\documentclass[aip,jcp, reprint, groupedaddress, floatfix]{revtex4-2}%
\usepackage{graphicx}
\usepackage{tabularx}
\usepackage{dcolumn}
\usepackage{longtable}
\usepackage{tensor}
\usepackage{color}
\usepackage{xcolor}
\usepackage{placeins}
\usepackage{bm}
\usepackage{amsmath}
\usepackage{amsfonts}
\usepackage{amssymb}
\usepackage{enumerate}
\usepackage[urlcolor=white]{hyperref}%
\usepackage{soul}
\providecommand{\U}[1]{\protect\rule{.1in}{.1in}}

\graphicspath{{C:/Users/moghadda/Desktop/VGApaper/Figures/}{C:/Users/Jiri/Dropbox/Papers/Chemistry_papers/2023/VGA_high-order/Figures/}}
\colorlet{RED}{red}
\begin{document}
\title{High-order geometric integrators for the variational Gaussian approximation}
\author{Roya Moghaddasi Fereidani}
\email{roya.moghaddasifereidani@epfl.ch}
\author{Ji\v{r}\'{\i} J. L. Van\'{\i}\v{c}ek}
\email{jiri.vanicek@epfl.ch}
\affiliation{Laboratory of Theoretical Physical Chemistry, Institut des Sciences et
Ing\'{e}nierie Chimiques, Ecole Polytechnique F\'{e}d\'{e}rale de Lausanne
(EPFL), CH-1015 Lausanne, Switzerland}
\date{\today}

\begin{abstract}
Among the single-trajectory Gaussian-based methods for solving the
time-dependent Schr\"{o}dinger equation, the variational Gaussian
approximation is the most accurate one. In contrast to Heller's original
thawed Gaussian approximation, it is symplectic, conserves energy exactly, and may partially account for tunneling. However, the variational method is
also much more expensive. To improve its efficiency, we symmetrically compose
the second-order symplectic integrator of Faou and Lubich and obtain geometric
integrators that can achieve an arbitrary even order of convergence in the
time step. We demonstrate that the high-order integrators can speed up
convergence drastically compared to the second-order algorithm and, in
contrast to the popular fourth-order Runge-Kutta method, are time-reversible
and conserve the norm and the symplectic structure exactly, regardless of the
time step. To show that the method is not restricted to low-dimensional
systems, we perform most of the analysis on a non-separable twenty-dimensional
model of coupled Morse oscillators. We also show that the variational method
may capture tunneling and, in general, improves accuracy over the
non-variational thawed Gaussian approximation.

\end{abstract}
\maketitle


\section{Introduction}

Nuclear quantum effects play an important role in many fundamental phenomena
in physics and
chemistry.~\cite{book_Nitzan:2006,book_Tannor:2007,book_Heller:2018} The idea
of using Gaussian wavepackets~\cite{Pattanayak_Schieve:1994,
Pattanayak_Schieve:1997, book_Tannor:2007,book_Lubich:2008,
Faou_Lubich:2009_v2, Richings_Lasorne:2015, book_Heller:2018,
garashchuk:2021,Edery:2021} for the semiclassical description of nuclei goes
back to the works of Heller~\cite{Heller:1975, Heller:1976a, Heller:1981} and
Hagedorn.~\cite{Hagedorn:1980_v2, Hagedorn:1998} In addition to many
convenient mathematical properties,~\cite{garashchuk:2021} the Gaussian
wavepacket is the exact solution of the Schr\"{o}dinger equation in a
many-dimensional harmonic potential, which is often used as a starting point
for modeling and discussing molecular vibrations. The localized nature of
Gaussians allows nuclear dynamics to be performed on the fly, without the need
to pre-compute a full potential energy surface. As a result, the
Gaussian-based methods can be easily
combined~\cite{Lasorne_Worth:2006,Ben-Nun_Martinez:1998a,Wehrle_Vanicek:2014_v2}
with ab initio evaluation of the potential. In addition, the Gaussian
wavepackets inherit a symplectic structure from the manifold of the
quantum-mechanical Hilbert space.~\cite{Faou_Lubich:2006, Ohsawa:2015a,
Ohsawa_Leok:2013}

Employing a superposition of Gaussian basis functions to represent the nuclear
wavepacket makes it possible to address more subtle quantum effects, including
interference, tunneling, diffraction, wavepacket splitting, and nonadiabatic
transitions. A number of multi-trajectory Gaussian-based approaches, such as
the full multiple spawning,~\cite{Martinez_Levine:1996a,
Ben-Nun_Martinez:2000, Curchod_Martinez:2018_v2} coupled coherent
states,~\cite{Shalashilin_Child:2004} minimum energy
method,~\cite{Sawada_Heather:1985, Sawada_Metiu:1986a, Sawada_Metiu:1986b}
variational multiconfigurational Gaussians,~\cite{Worth_Burghardt:2003,
Richings_Lasorne:2015} multiconfigurational Ehrenfest
method,~\cite{Shalashilin:2009, Shalashilin:2010} Gaussian dephasing
representation,~\cite{Sulc_Vanicek:2013} initial value
representation,~\cite{Miller:2001_v2} frozen Gaussian
approximation,~\cite{Heller:1981} Herman-Kluk
propagator,~\cite{Herman_Kluk:1984} hybrid dynamics,~\cite{Grossmann:2006}
multiple coherent states,\cite{Ceotto_Atahan:2009a} and
divide-and-conquer semiclassical
dynamics\cite{Ceotto_Conte:2017,Gabas_Ceotto:2019} were developed to capture
these effects. However, the use of multiple coupled or uncoupled Gaussians
makes these methods rather expensive and difficult to converge, especially in
combination with on-the-fly ab initio simulation of large systems.

To simulate systems with weak anharmonicity and mild quantum effects, it is
sometimes sufficient to use single-trajectory Gaussian-based methods. Because
they avoid the issue of convergence with respect to the number of
trajectories, the single-trajectory techniques preserve more geometric
properties. An original method in this family is Heller's thawed Gaussian
approximation (TGA),~\cite{Heller:1975,Lee_Heller:1982,book_Heller:2018} which
propagates a single Gaussian wavepacket in the local
harmonic approximation of the potential. The TGA is much more accurate than
the global harmonic approximations because it at least partially includes
anharmonicity.~\cite{Wehrle_Vanicek:2014_v2, Wehrle_Vanicek:2015_v2,
Patoz_Vanicek:2018} However, because  the TGA uses a classical trajectory, it
cannot describe quantum tunneling.~\cite{Begusic_Vanicek:2022} Here,
we explore the variational Gaussian approximation
(VGA),~\cite{Coalson_Karplus:1990,book_Lubich:2008,Lasser_Lubich:2020} which
evolves a single Gaussian wavepacket according to the Dirac-Frenkel-McLachlan
time-dependent variational principle.~\cite{Dirac:1930,
book_Frenkel:1934,McLachlan:1964,Heller:1976,
book_Kramer_Saraceno:1980,book_Lubich:2008} In contrast to the TGA, the VGA
conserves both the symplectic structure and
energy~\cite{Faou_Lubich:2006,book_Hairer_Wanner:2006,book_Lubich:2008} and,
in addition, may partially capture tunneling.~\cite{Buch:2002,
Hasegawa:2014}

The variational Gaussian wavepacket dynamics was introduced in the seminal
work of Heller.~\cite{Heller:1976} Heather and Metiu derived equations of
motion for the Gaussian's parameters using a variational \textquotedblleft
minimum error method\textquotedblright.~\cite{Heather_Metiu:1985} Coalson and
Karplus~\cite{Coalson_Karplus:1990} obtained a refined version of these
equations by applying the time-dependent variational principle to a
multi-dimensional Gaussian wavepacket ansatz. Poirier derived the equations of
the VGA using quantum trajectories.\cite{Poirier:DF_quantum_trajectories} The
non-canonical symplectic structure of these equations was found for a
spherical Gaussian wavepacket by Faou and Lubich~\cite{Faou_Lubich:2006} and
generalized to an arbitrary multi-dimensional Gaussian wavepacket by Ohsawa
and Leok.~\cite{Ohsawa_Leok:2013} The equations of the VGA contain expectation
values of the potential and its first two derivatives, which, in general,
cannot be evaluated analytically. Therefore, for practical applications, the
potential should be
approximated,\cite{Pattanayak_Schieve:1994,Ohsawa_Leok:2013,Moghaddasi_Vanicek:2023a}
which introduces further errors. To avoid these additional errors here, we
have designed a multi-dimensional nonseparable coupled Morse oscillator
potential, whose matrix elements can be computed exactly.

Faou and Lubich developed an integration method to numerically solve the
equations of motion for the VGA.~\cite{Faou_Lubich:2006} Their integrator is
symplectic, norm-conserving, time-reversible, and for sufficiently small time
steps, energy-conserving.~\cite{book_Lubich:2008} It is of second-order
accuracy in the time step.~\cite{book_Lubich:2008} Here, to make the
VGA more practical, we make their algorithm more efficient by increasing the
order of convergence using various recursive and non-recursive composition
techniques.~\cite{book_Hairer_Wanner:2006, book_Leimkuhler_Reich:2004,
Yoshida:1990, Suzuki:1990, McLachlan:1995, Wehrle_Vanicek:2011,
Sofroniou_Spaletta:2005, Choi_Vanicek:2019, Roulet_Vanicek:2019} We also generalize their method  from a spherical to a general
multi-dimensional Gaussian and demonstrate the geometric properties of the high-order integrators. 

The remainder of this paper is organized as follows. After reviewing the
variational Gaussian wavepacket dynamics in Sec.~\ref{sec:theory}, we discuss
its geometric properties in Sec.~\ref{sec:geom_prop}. Nearly all of these
geometric properties are preserved by the symplectic integrators, which are
described in Sec.~\ref{sec:sym_integrators}. In Sec.~\ref{sec:num_examples},
we provide numerical examples that confirm the improved accuracy of the VGA
over those of other single-trajectory Gaussian-based methods. We also use the multi-dimensional coupled Morse potential to numerically verify the convergence, geometric properties, and increased efficiency of the high-order integrators.
Section~\ref{sec:conclusion} concludes this paper.

\section{Variational Gaussian approximation}

\label{sec:theory}

Assuming the validity of the Born-Oppenheimer
approximation,~\cite{Born_Oppenheimer:1927,Heller:1981a} the motion of the
nuclei can be described by the time-dependent Schr\"{o}dinger equation
\begin{equation}
i\hbar\, d|\Psi_{t}\rangle/dt=\hat{H}|\Psi_{t}\rangle\label{eq:TDSE}%
\end{equation}
with a time-independent Hamiltonian operator
\begin{align}
\hat{H}=\hat{T}+\hat{V}=T(\hat{p})+V(\hat{q})
\label{eq:H}%
\end{align}
where $\hat{T}\equiv T(\hat{p}):=\hat{p}^{T}\cdot m^{-1}\cdot \hat{p}/2$ is the kinetic energy, depending only
on the momentum $p$,  $\hat{V} \equiv V(\hat{q})$ is the potential energy, depending only on the
position $q$, and $m$ is the real-symmetric mass matrix. Solving
Eq.~(\ref{eq:TDSE}) in high-dimensional systems is a formidable task, and
various approaches were developed to approximate the
solution.~\cite{book_Tannor:2007} Among these, the
VGA~\cite{Coalson_Karplus:1990,Lasser_Lubich:2020} is obtained by applying the
time-dependent variational
principle~\cite{Dirac:1930,book_Frenkel:1934,book_Lubich:2008}
\begin{equation}
\langle\delta\psi_{t}\,|\big(i\hbar\frac{d}{dt}-\hat{H}\big)|\,\psi_{t}%
\rangle=0\label{eq:DiracFrenkelVP1}%
\end{equation}
to the complex Gaussian ansatz~\cite{Heller:1976}
\begin{align}
\psi_{t}(q) &  =\text{exp}\big\{(i/\hbar)\big[(q-q_{t})^{T}\cdot A_{t}%
\cdot(q-q_{t})/2\nonumber\\
&  +p_{t}^{T}\cdot(q-q_{t})+\gamma_{t}\big]\big\}\label{eq:GWP}%
\end{align}
approximating the wavefunction $\Psi_{t}$. In Eq.~(\ref{eq:GWP}), $q_{t}$ and
$p_{t}$ are $D$-dimensional real vectors representing the position and
momentum of the Gaussian's center, $A_{t}=\mathcal{A}_{t}+i\mathcal{B}_{t}$ is
a $D\times D$ complex symmetric matrix whose real part $\mathcal{A}_{t}$
introduces a spatial chirp and whose positive-definite imaginary part
$\mathcal{B}_{t}$ determines the width of the Gaussian, and $\gamma_{t}=\phi
_{t}+i\delta_{t}$ is a complex number whose real part $\phi_{t}$ introduces a
time-dependent phase and whose imaginary part $\delta_{t}$ ensures
normalization at all times. The squared norm of $\psi_{t}$ is
\begin{equation}
I(\mathcal{B}_{t},\delta_{t}):=\lVert\psi_{t}\rVert^{2}=[\text{det}({\pi\hbar
}/{\mathcal{B}_{t}})]^{1/2}e^{-{2\delta_{t}}/{\hbar}}.\label{eq:Norm}%
\end{equation}
In Appendix~\ref{sec:symplectic_structure} we show that applying the
variational principle~(\ref{eq:DiracFrenkelVP1}) to the Gaussian
ansatz~(\ref{eq:GWP}) yields the system
\begin{align}
\dot{q}_{t} &  =m^{-1}\cdot p_{t},\label{eq:qEOM}\\
\dot{p}_{t} &  =-V_{1},\label{eq:pEOM}\\
\dot{A}_{t} &  =-A_{t}\cdot m^{-1}\cdot A_{t}-V_{2},\label{eq:AEOM}\\
\dot{\gamma}_{t} &  =T(p_{t})-V_{0}+(i\hbar/2)\,\text{Tr}\big(m^{-1}\cdot
A_{t}\big)\label{eq:gammaEOM}%
\end{align}
of ordinary differential equations for the parameters, where $T(p_{t}%
)=p_{t}^{T}\cdot m^{-1}\cdot p_{t}/2$ and%
\begin{equation}
V_{0}=\langle\hat{V}\rangle-\text{Tr}\big({\langle\hat{V}^{\prime\prime
}\rangle}\cdot\Sigma_{t}\big)/2,\quad V_{1}=\langle\hat{V}^{\prime}%
\rangle,\quad V_{2}=\langle\hat{V}^{\prime\prime}\rangle
.\label{eq:V0_V1_V2_VGA}%
\end{equation}
Here, $\hat{V}^{\prime}:=V^{\prime}(q)|_{q=\hat{q}}$ and $\hat{V}%
^{\prime\prime}:=V^{\prime\prime}(q)|_{q=\hat{q}}$ denote the gradient and
Hessian of the potential energy operator and
\begin{equation}
\Sigma_{t}:=\langle(\hat{q}-q_{t})\otimes(\hat{q}-q_{t})^{T}\rangle
=(\hbar/2)\,\mathcal{B}_{t}^{-1}%
\end{equation}
is the position covariance. Above and throughout this paper, we use a
shorthand notation $\langle\hat{O}\rangle:=\langle\psi_{t}|\hat{O}|\psi
_{t}\rangle$ for the expectation value of the operator $\hat{O}$ in the
normalized state $\psi_{t}$. Since by assumption the Gaussian wavepacket
retains its Gaussian form for all times, the VGA cannot describe wavepacket splitting.


Rewriting Gaussian~(\ref{eq:GWP}) in Hagedorn's
parametrization~\cite{Hagedorn:1980_v2,Lasser_Lubich:2020}
\begin{align}
\psi_{t}(q)  &  =(\pi\hbar)^{-D/4}(\text{det}\,Q_{t})^{-1/2}\text{exp}%
\big\{(i/\hbar)\big[(q-q_{t})^{T}\cdot\nonumber\\
&  P_{t}\cdot Q_{t}^{-1}\cdot(q-q_{t})/2+p_{t}^{T}\cdot(q-q_{t})+S_{t}%
\big]\big\} \label{eq:HGWP}%
\end{align}
leads to equivalent, yet more classical-like equations
\begin{align}
\dot{Q}_{t}  &  =m^{-1}\cdot P_{t},\label{eq:HQEOM}\\
\dot{P}_{t}  &  =-V_{2}\cdot Q_{t},\label{eq:HPEOM}\\
\dot{S}_{t}  &  =T(p_{t})-V_{0}. \label{eq:HSEOM}%
\end{align}
The new parameters $Q_{t}$ and $P_{t}$ are two $D\times D$ complex matrices,
related to the Gaussian's width via $A_{t}=P_{t}\cdot Q_{t}^{-1}$ and
satisfying the relations
\begin{align}
&  Q_{t}^{T}\cdot P_{t}-P_{t}^{T}\cdot Q_{t}=0,\\
&  Q_{t}^{\dagger}\cdot P_{t}-P_{t}^{\dagger}\cdot Q_{t}=2iI_{D},
\label{eq:QPrels}%
\end{align}
where $I_{D}$ is the $D\times D$ identity matrix. $S_{t}$ is a real scalar
generalizing the classical action. The norm of the Gaussian (\ref{eq:HGWP})
is
\begin{equation}
\lVert\psi(t)\rVert=\text{det}\big[\text{Im}\,(P_{t}\cdot Q_{t}^{-1})\cdot
Q_{t}\cdot Q_{t}^{\dagger}\big]^{-1/4}. \label{eq:HNorm}%
\end{equation}


\section{Geometric properties of the VGA}

\label{sec:geom_prop}

If $\hat{P}(\psi_{t})$ denotes the orthogonal projection onto the tangent
space at $\psi_{t}$ of the approximation manifold $M$ of complex Gaussians,
the variational principle~(\ref{eq:DiracFrenkelVP1}) is equivalent to the
nonlinear Schr\"{o}dinger equation~\cite{book_Lubich:2008,Roulet_Vanicek:2021,
Vanicek:2023}
\begin{equation}
i\hbar\,d|\psi_{t}\rangle/dt=\hat{H}_{\text{eff}}(\psi_{t})\,|\psi_{t}%
\rangle\label{eq:Nonlin_TDSE}%
\end{equation}
with an effective, state-dependent Hamiltonian
\begin{equation}
\hat{H}_{\text{eff}}(\psi_{t}):=\hat{P}(\psi_{t})\,\hat{H}=\hat{T}+\hat
{P}(\psi_{t})\,\hat{V}. 
\label{eq:Heff}%
\end{equation}
In the position representation, the effective potential $\hat{V}_{\text{eff}}(\psi_{t}):=\hat{P}(\psi_{t})\,\hat{V}$ is a quadratic
function~\cite{Lasser_Lubich:2020,Vanicek:2023}
\begin{align}
V_{\text{eff}}(q;\psi_{t})  &  =V_{0}+V_{1}^{T}\cdot(q-q_{t})\nonumber\\
&  \quad+(q-q_{t})^{T}\cdot V_{2}\cdot(q-q_{t})/2, \label{eq:Veff}%
\end{align}
where $V_{0}$, $V_{1}$, and $V_{2}$ are given in Eq.~(\ref{eq:V0_V1_V2_VGA}).
The time evolution operator of the effective Hamiltonian (\ref{eq:Heff}) is
also nonlinear and can be expressed as
\begin{equation}
\hat{U}_{\text{eff}}(t,t_{0};\psi):=\mathcal{T}\text{exp}\bigg[-\frac{i}%
{\hbar}\int_{t_{0}}^{t}\hat{H}_{\text{eff}}({\psi_{t^{\prime}}})\,dt^{\prime
}\bigg], \label{eq:U}%
\end{equation}
where $\mathcal{T}$ denotes the time-ordering operator. Next, we discuss the
geometric properties of the linear Schr\"{o}dinger equation that are preserved
by the VGA.

\begin{figure}
[!htbp]
\includegraphics[width=0.45\textwidth]{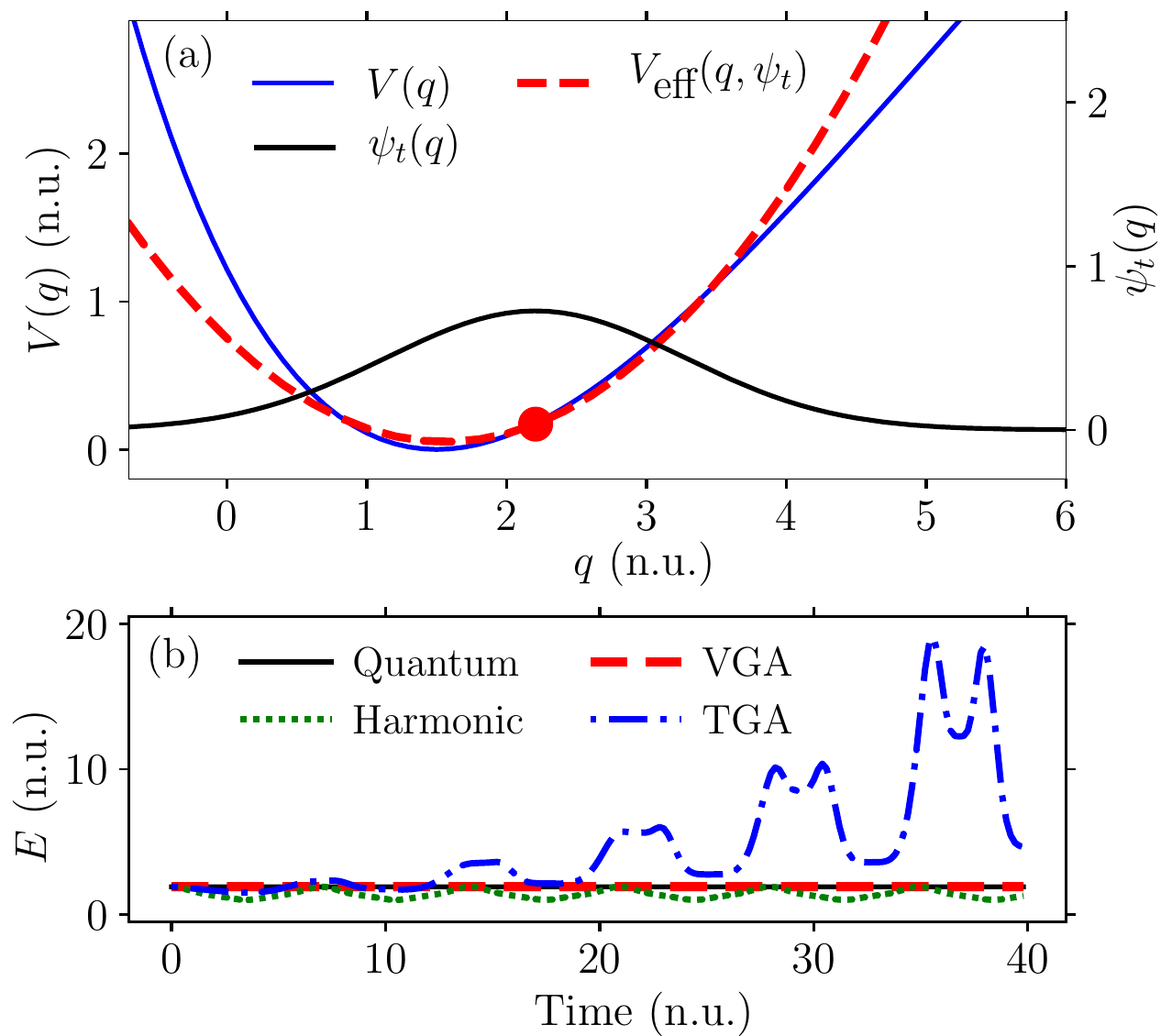}
\caption{Energy conservation by the VGA. (a) Gaussian wavepacket $\psi_{t}$ propagated in a Morse potential $V$ using the VGA with an effective potential $V_{\textrm{eff}}$, (b) energy of the wavepacket propagated with various methods.
	} \label{Fig1}
\end{figure}

\subsection{Energy conservation}

Although a nonlinear evolution does not generally conserve
energy,~\cite{Roulet_Vanicek:2021,Vanicek_Begusic:2021} the energy is
conserved in the VGA as in any other method derived from the time-dependent
variational principle.~\cite{Kramer_Saraceno:1981,Coalson_Karplus:1990,Beck_Meyer:2000, book_Lubich:2008,Habershon:2012,Joubert-Doriol_Izmaylov:2015,Hackl_Cirac:2020,Lasser_Lubich:2020} To see this, note that the arbitrary infinitesimal
change $\delta\psi_{t}$ in the variational principle~(\ref{eq:DiracFrenkelVP1}%
) can be chosen to be proportional to $\dot{\psi}_{t}$. Therefore,
\begin{align}
\dot{E}  &  =d\langle\hat{H}\rangle/dt=2\,\text{Re}\langle\dot{\psi}_{t}%
|\hat{H}|\psi_{t}\rangle\nonumber\\
&  =2\,\text{Re}\langle\dot{\psi}_{t}|i\hbar\,\dot{\psi}_{t}\rangle
=2\,\text{Re}\big[i\hbar\,\lVert\dot{\psi}_{t}\rVert^{2}\big]=0,
\end{align}
which proves conservation of energy $E=\langle\hat{H}\rangle$.

Panel (a) of Fig.~\ref{Fig1} shows an example of a Gaussian wavepacket
propagated in a Morse potential. The effective potential $V_{\text{eff}}$ of
the VGA differs from the local harmonic approximation since $V_{\text{eff}}$
is not tangent to $V$. Panel (b) compares the energies of the wavepacket
propagated with various methods. Unlike the two non-variational semiclassical
methods (the TGA and harmonic approximation), the VGA conserves energy
exactly. Note that we use the natural units (n.u.) with $\hbar=m=1$. For other simulation details, see the supplementary material.


\subsection{Effective energy conservation}

The VGA also conserves the effective energy $E_{\text{eff}}=\langle\hat
{H}_{\text{eff}} \rangle=\langle\hat{T} \rangle+\langle\hat{V}_{\text{eff}}
\rangle$,~\cite{Vanicek:2023} because the effective energy is equal to the
energy $(\langle\hat{H}_{\text{eff}} \rangle=\langle\hat{H}\rangle)$, and the
energy is conserved. The equality $\langle\hat{H}_{\text{eff}} \rangle
=\langle\hat{H}\rangle$ follows because
\begin{align}
\langle\hat{V}_{\text{eff}} \rangle &  =\langle\hat{V}\rangle-\text{Tr}%
\big({\langle\hat{V}^{\prime\prime} \rangle} \cdot\Sigma_{t}\big)/2+\langle
\hat{V}^{\prime}\rangle^{T} \cdot\langle\hat{q}-q_{t} \rangle\nonumber\\
&  \quad+\text{Tr}[{\langle\hat{V}^{\prime\prime} \rangle} \cdot\langle
(\hat{q}-q_{t})\otimes(\hat{q}-q_{t})^{T}\rangle]/2\nonumber\\
&  =\langle\hat{V}\rangle. \label{eq:Veff_exp_value}%
\end{align}

\subsection{Norm conservation}

The manifold $M$ of unnormalized complex Gaussian wavepackets contains
rays,~\cite{book_Lubich:2008,Vanicek:2023} i.e., for each $\psi_{t}\in M$ and
each complex number $\lambda$, we have $\lambda\,\psi_{t}\in M$. Therefore,
the variation $\delta\psi_{t}\propto\psi_{t}$ is permitted; invoking the
variational principle~(\ref{eq:DiracFrenkelVP1}) with $\delta\psi_{t}%
\propto\psi_{t}$ implies that~\cite{Lasser_Lubich:2020}
\begin{align}
d\lVert\psi_{t}\rVert^{2}/dt  &  =2\,\text{Re}\langle\psi_{t}|\dot{\psi}%
_{t}\rangle=2\,\text{Re}\langle\psi_{t}|(i\hbar)^{-1}\hat{H}\psi_{t}%
\rangle\nonumber\\
&  =2\,\text{Re}\big[(i\hbar)^{-1}\langle\hat{H}\rangle\big]=0.
\end{align}
Thus, the VGA, as well as other, more general Gaussian wavepacket methods,
conserve the norm $\lVert\psi_{t}\rVert$ of the propagated
Gaussian.~\cite{Vanicek:2023,Coalson_Karplus:1990, book_Lubich:2008,Lasser_Lubich:2020}

\subsection{Non-conservation of the inner product and distance}

Due to nonlinearity, the VGA generally does not conserve the inner product
between states $\psi_{1}$ and $\psi_{2}$:~\cite{Vanicek:2023}
\begin{align}
d \langle\psi_{1}|\psi_{2}\rangle/dt &  =\langle\dot{\psi}_{1}|\psi
_{2}\rangle+\langle\psi_{1}|\dot{\psi}_{2}\rangle\nonumber\\
&  =(i\hbar)^{-1}\langle\psi_{1}|\hat{H}_{\text{eff}}(\psi_{1})-\hat
{H}_{\text{eff}}(\psi_{2})|\psi_{2}\rangle\neq0.
\end{align}
Although the VGA conserves the norm, the non-conservation of the inner product
leads to the non-conservation of the distance between $\psi_{1}$ and $\psi
_{2}$:
\begin{align}
d(\psi_{1},\psi_{2})  &  :=\lVert\psi_{1}-\psi_{2}\rVert=\langle\psi_{1}%
-\psi_{2}|\psi_{1}-\psi_{2}\rangle^{1/2}\nonumber\\
&  \,\,=\big(\lVert\psi_{1}\rVert^{2}+\lVert\psi_{2}\rVert^{2}-2\,\text{Re}%
\,\langle\psi_{1}|\psi_{2}\rangle\big)^{1/2}. \label{eq:distance}%
\end{align}

\subsection{Time reversibility}

The nonlinear evolution~(\ref{eq:U}) is reversible
since~\cite{Roulet_Vanicek:2021,Vanicek:2023}
\begin{equation}
\psi_{t,\text{FB}}:=\hat{U}_{\text{eff}}(t_{0},t;\psi)\hat{U}_{\text{eff}%
}(t,t_{0};\psi)\psi_{0}=\psi_{0}, \label{eq:TR}%
\end{equation}
where $\psi_{t,\text{FB}}$ is the state obtained by propagating the initial
state $\psi_{0}$ first forward in time with the evolution operator $\hat
{U}_{\text{eff}}(t,t_{0};\psi)$ and then backward in time with the reverse
evolution operator
\begin{align}
\hat{U}_{\text{eff}}(t_{0},t;\psi):  &  =\tilde{\mathcal{T}}\text{exp}%
\bigg[-\frac{i}{\hbar}\int_{t}^{t_{0}}\hat{H}_{\text{eff}}({\psi_{t^{\prime}}%
})\,dt^{\prime}\bigg]\nonumber\\
&  =\tilde{\mathcal{T}}\text{exp}\bigg[\frac{i}{\hbar}\int_{t_{0}}^{t}\hat
{H}_{\text{eff}}({\psi_{t^{\prime}}})\,dt^{\prime}\bigg]\nonumber\\
&  =\hat{U}_{\text{eff}}(t,t_{0};\psi)^{-1}; \label{eq:Ureverse}%
\end{align}
$\tilde{\mathcal{T}}$ is the reverse time-ordering operator.

\subsection{Symplecticity}

Manifold $M$ of Gaussian wavepackets can be endowed with a non-canonical
symplectic structure.~\cite{book_Hairer_Wanner:2006,Ohsawa_Leok:2013} Faou and
Lubich showed that the spherical Gaussian wavepacket inherits this symplectic
structure from the infinite-dimensional Hilbert space manifold by the
variational principle.~\cite{Faou_Lubich:2006} Ohsawa and Leok used the
symplectic structure of this manifold to derive the variational Gaussian
wavepacket dynamics as a non-canonical Hamiltonian system with the Hamiltonian
function $h(q_{t}, p_{t}, A_{t}, \gamma_{t}):=\langle\psi_{t} |\hat{H} |
\psi_{t} \rangle$.~\cite{Ohsawa_Leok:2013} Employing a combination of their
approaches, in Appendix~\ref{sec:symplectic_structure}, we find the
non-canonical symplectic structure of the more general non-spherical Gaussian
wavepacket and rederive the equations of motion~(\ref{eq:qEOM}%
)-(\ref{eq:gammaEOM}) for the Gaussian's parameters.


\section{Geometric integrators for the VGA}

\label{sec:sym_integrators}

\subsection{Second-order symplectic integrator}

Faou and Lubich proposed a symplectic algorithm for the numerical time
integration of the differential equations of the
VGA.~\cite{Faou_Lubich:2006,book_Hairer_Wanner:2006, book_Lubich:2008} The
integrator is based on the splitting of the Hamiltonian into the kinetic and
potential energy terms. We have generalized their method for scalar mass $m$
and width $A_{t}$ to non-diagonal, symmetric matrices $m$ and $A_{t}%
$.~\cite{Vanicek:2023} During the kinetic propagation $[\hat{H}_{\text{eff}%
}=T(\hat{p})]$, Eqs. (\ref{eq:qEOM})-(\ref{eq:gammaEOM}) have the analytical
solution
\begin{align}
q_{t}  &  =q_{0}+t\,m^{-1}\cdot p_{0},\\
p_{t}  &  =p_{0},\\
A_{t}  &  =(A_{0}^{-1}+t\,m^{-1})^{-1},\label{eq:Tstep_AEOM}\\
\gamma_{t}  &  =\gamma_{0}+t\,T(p_{0})+(i\hbar/2)\,\text{ln}\,[\,\text{det}\,( I_{D}+t\,m^{-1}\cdot A_{0})], \label{eq:Tstep_gamma_EOM}%
\end{align}
and, during the potential propagation $[\hat{H}_{\text{eff}}=V_{\text{eff}%
}(\hat{q})]$, they have the analytical solution
\begin{align}
q_{t}  &  =q_{0},\\
p_{t}  &  =p_{0}-t\,V_{1}(q_{0},\text{Im}\,A_{0}),\\
A_{t}  &  =A_{0}-t\,V_{2}(q_{0},\text{Im}\,A_{0}),\\
\gamma_{t}  &  =\gamma_{0}-t\,V_{0}(q_{0},\text{Im}\,A_{0}). \label{eq:VstepSO}%
\end{align}
Applying the potential propagation for time $\Delta t/2$, kinetic propagation
for time $\Delta t$, and potential propagation for time $\Delta t/2$ in
sequence yields a \textquotedblleft
potential-kinetic-potential\textquotedblright(VTV) algorithm that is of the
second order in the time step $\Delta t$. Another second-order algorithm is
the \textquotedblleft kinetic-potential-kinetic\textquotedblright(TVT)
algorithm, which is obtained by swapping the potential and kinetic
propagations in the VTV algorithm. Each of these two numerical algorithms
gives the state $\psi_{t+\Delta t}$ at time $t+\Delta t$ from the state
$\psi_{t}$ at time $t$:
\begin{equation}
|\psi_{t+\Delta t}\rangle=\hat{U}_{2}(t+\Delta t,t;\psi)|\psi_{t}\rangle,
\label{eq:U2}%
\end{equation}
where $\hat{U}_{2}$ is the approximate second-order evolution operator
associated with the VTV or TVT algorithm.

In Hagedorn's parametrization, the flow $\Phi_{\text{T},t}$ associated with
the kinetic propagation is
\begin{align}
q_{t}  &  =q_{0}+t\, m^{-1} \cdot p_{0},\label{eq:Hf_T_qEOM}\\
p_{t}  &  =p_{0},\label{eq:Hf_T_pEOM}\\
Q_{t}  &  =Q_{0}+t\, m^{-1} \cdot P_{0},\label{eq:Hf_T_QEOM}\\
P_{t}  &  =P_{0},\label{eq:Hf_T_PEOM}\\
S_{t}  &  =S_{0}+t\,T(p_{0}), \label{eq:Hf_T_SEOM}%
\end{align}
and the potential flow $\Phi_{\text{V},t}$ is
\begin{align}
q_{t}  &  =q_{0},\label{eq:Hf_v_qEOM}\\
p_{t}  &  =p_{0}-t\,V_{1}(q_{0},Q_{0}),\label{eq:Hf_v_pEOM}\\
Q_{t}  &  =Q_{0},\label{eq:Hf_v_QEOM}\\
P_{t}  &  =P_{0}-t\, V_{2} (q_{0},Q_{0}) \cdot Q_{0},\label{eq:Hf_v_PEOM}\\
S_{t}  &  =S_{0}-t\,V_{0}(q_{0},Q_{0}).
\end{align}


\subsection{High-order symplectic integrators}

\label{sec-Compositions}

High-order integrators can be obtained by composing  the second-order (VTV or TVT) algorithm~(\ref{eq:U2}). More precisely, any
symmetric algorithm $\hat{U}_{p}$ of even order $p$ can generate an evolution
operator $\hat{U}_{p+2}$ of order $p+2$ if it is symmetrically composed as
\begin{multline}
\hat{U}_{p+2}(t+\Delta t,t;\psi):=\hat{U}_{p}(t+\xi_{M}\Delta t,t+\xi
_{M-1}\Delta t;\psi)\nonumber\\
\quad\cdots\hat{U}_{p}(t+\xi_{1}\Delta t,t;\psi),
\end{multline}
where $M$ is the total number of composition steps and $\xi_{n}:=\sum
_{j=1}^{n}\gamma_{j}$ denotes the sum of the first $n$ real composition
coefficients $\gamma_{j}$, which satisfy the relations $\sum_{j=1}^{M}%
\gamma_{j}=1$ (consistency), $\gamma_{M+1-j}=\gamma_{j}$ (symmetry), and
$\sum_{j=1}^{M}\gamma_{j}^{p+1}=0$ (order increase
guarantee).~\cite{book_Hairer_Wanner:2006} The most common composition methods
are the recursive triple-jump~\cite{Yoshida:1990} ($M=3$) and Suzuki's
fractal~\cite{Suzuki:1990} ($M=5$). Although both methods can generate
high-order integrators, the number of composition steps grows exponentially
with the order of convergence. To further increase the efficiency, we mainly
use \textquotedblleft optimal\textquotedblright\ nonrecursive
methods\cite{Kahan_Li:1997, Sofroniou_Spaletta:2005} to obtain integrators of
sixth-, eighth-, and tenth-order. We refer to them as \textquotedblleft
optimal\textquotedblright composition methods because they minimize the
magnitudes of composition steps defined as $\sum_{j=1}^{M}|\gamma_{j}|$ or
$\max_{j}|\gamma_{j}|$. Suzuki's fractal gives the optimal fourth-order
scheme.~\cite{Choi_Vanicek:2019} For more details on these composition
schemes, see Ref.~\onlinecite{Choi_Vanicek:2019}. We compare the efficiencies
and numerically verify the predicted order of convergence of these integrators
for the VGA in Sec.~\ref{sec:num_examples} and in the supplementary material.


\subsection{Geometric properties of the symplectic integrators}

Each kinetic or potential step of the symplectic integrators is the exact
solution of the nonlinear Schr\"{o}dinger equation (\ref{eq:Nonlin_TDSE}) with
$\hat{H}_{\text{eff}}=\hat{T}$ or $\hat{H}_{\text{eff}}=\hat{V}_{\text{eff}}$,
and thus has all the geometric properties of the VGA. All symplectic
integrators that are obtained by symmetric composition of the kinetic and
potential steps are time-reversible, norm-conserving, and
symplectic.~\cite{Faou_Lubich:2006, book_Lubich:2008,Vanicek:2023} However,
due to the splitting, they are only approximately energy-conserving, with an
error $\mathcal{O}(\Delta t^{M})$ where $M$ is greater than or equal to the
order of the integrator.~\cite{Faou_Lubich:2006, book_Lubich:2008,
Roulet_Vanicek:2019, Choi_Vanicek:2021b,Vanicek:2023}


\section{Numerical examples}

\label{sec:num_examples}

In what follows, we investigate the VGA and the proposed high-order
integrators in different model systems. For the numerical experiments, we have
specifically chosen the quartic double-well and coupled Morse potentials, for
both of which the expectation values of the potential energy, gradient, and
Hessian, needed in the VGA, can be computed analytically. We also compare the
VGA with two other Gaussian wavepacket methods, the TGA and harmonic
approximation, defined in Appendix~\ref{sec:TGA_harmonic}.

\subsection{Over-the-barrier motion and tunneling in a double well}

\label{sec-DW}

Double-well systems are ubiquitous in chemistry, physics, and
biology.~\cite{Thorwart_Hanggi:2001} Well-known molecular examples of
double-well systems include the inversion of ammonia, phosphine, and
arsine.~\cite{Begusic_Vanicek:2022} The most remarkable phenomenon
in double-well potentials is the quantum tunneling,~\cite{Book_Razavy:2003}
which allows hopping between its two minima through a classically forbidden
region. Here, we consider a one-dimensional symmetric double-well potential
\begin{equation}
V(q) = a-b\,q^{2}+ c\,q^{4} \label{eq:Double-Well}%
\end{equation}
with positive $a$, $b$, and $c$. This potential is a special case of the
quartic potential, described by Eq.~(\ref{eq:Quartic}) in
Appendix~\ref{subsec:Quartic} with parameters $V(q_{\text{eq}})=a$,
$V^{\prime}(q_{\text{ref}})=V^{\prime\prime\prime}(q_{\text{ref}})=0$,
$V^{\prime\prime}(q_{\text{ref}})=-2\,b$, and $V^{(4)}(q_{\text{ref}})=24
\,c$. The expectation values of the quartic potential, its gradient and
Hessian are derived in Appendix~\ref{subsec:Quartic}.

Figure~\ref{Fig2} analyzes the dynamics of a wavepacket propagated in the
double-well potential~(\ref{eq:Double-Well}) with $a=1$, $b=5$, and $c=2.5$.
The initial state was a real Gaussian with width \textquotedblleft
matrix\textquotedblright$A_{0}=4\,i$ and zero momentum. Depending on the
initial position of the Gaussian, its energy was above or below the potential
barrier. The grid for the exact quantum dynamics consisted of $512$ points
between $-10$ and $10$. Time step $\Delta t=0.001$ and the second-order (TVT)
symplectic integrator were used in all simulations.


\begin{figure}
[htb]\centering
\includegraphics[width=0.45\textwidth]{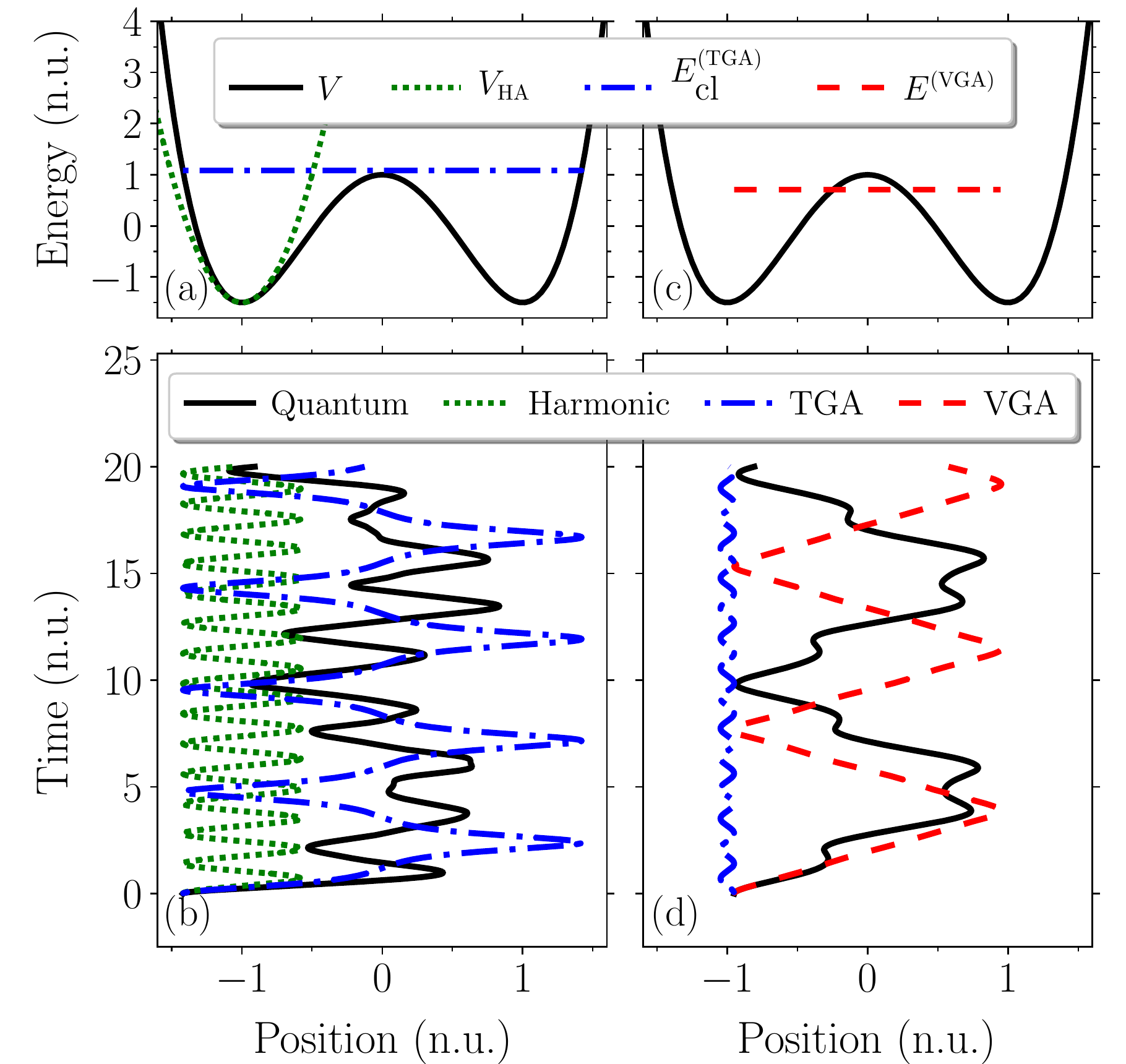}
\caption{ Thawed Gaussian approximation (TGA) can describe the classical  over-the-barrier motion in double-well systems, and the variational Gaussian approximation (VGA) can also describe quantum tunneling. Panel (b) associated with panel (a) shows that the TGA with classical energy $E_{\textrm{cl}}$ above the barrier can cross the barrier, whereas the harmonic approximation with approximated harmonic potential $V_{\textrm{HA}}$ cannot. Panel (d) associated with panel (c) shows that the VGA, unlike the TGA, can even tunnel through the barrier. }
\label{Fig2}
\end{figure}



The left panels of Fig.~\ref{Fig2} show the \textquotedblleft
over-the-barrier\textquotedblright motion of a wavepacket with initial
position $q_{0}=-1.42$ and energy $E \approx5.36$. Panel (a) shows the
double-well potential, its harmonic approximation $V_{\text{HA}}$ at the
minimum of the left well, and the conserved ``classical'' energy
$E_{\text{cl}}=p^{2}_{t}/2m+V(q_{t})\approx1.083$ of the TGA calculated at the
Gaussian's center, which evolves according to Hamilton's equations of
motion~[Eqs.~(\ref{eq:qEOM}) and~(\ref{eq:pEOM}) with coefficients
(\ref{eq:V0_V1_V2_TGA})]. Because the classical energy is slightly above the
potential barrier, the wavepacket passes the barrier, which is confirmed in
panel (b) by the TGA and the exact quantum results. In contrast, the same
wavepacket propagated with the harmonic approximation cannot cross the barrier
because the harmonic potential confines it to the left well [see panels (a)
and (b)].


Several studies~\cite{Buch:2002, Hasegawa:2014} found that the VGA
may realize tunneling in double-well systems. Our results, shown in
the right-hand panels of Fig.~\ref{Fig2}, confirm this observation for a
wavepacket with initial position $q_{0}=-0.95$ and energy $E\approx0.71$,
which is below the potential barrier [panel (c)]. Panel (d) shows that unlike
the TGA, which shows small oscillations around the minimum of the left well,
the VGA captures quantum tunneling at least qualitatively, since the VGA
wavepacket clearly moves back and forth between the two wells.


\begin{figure}
[!htbp]
\includegraphics[width=0.45\textwidth]{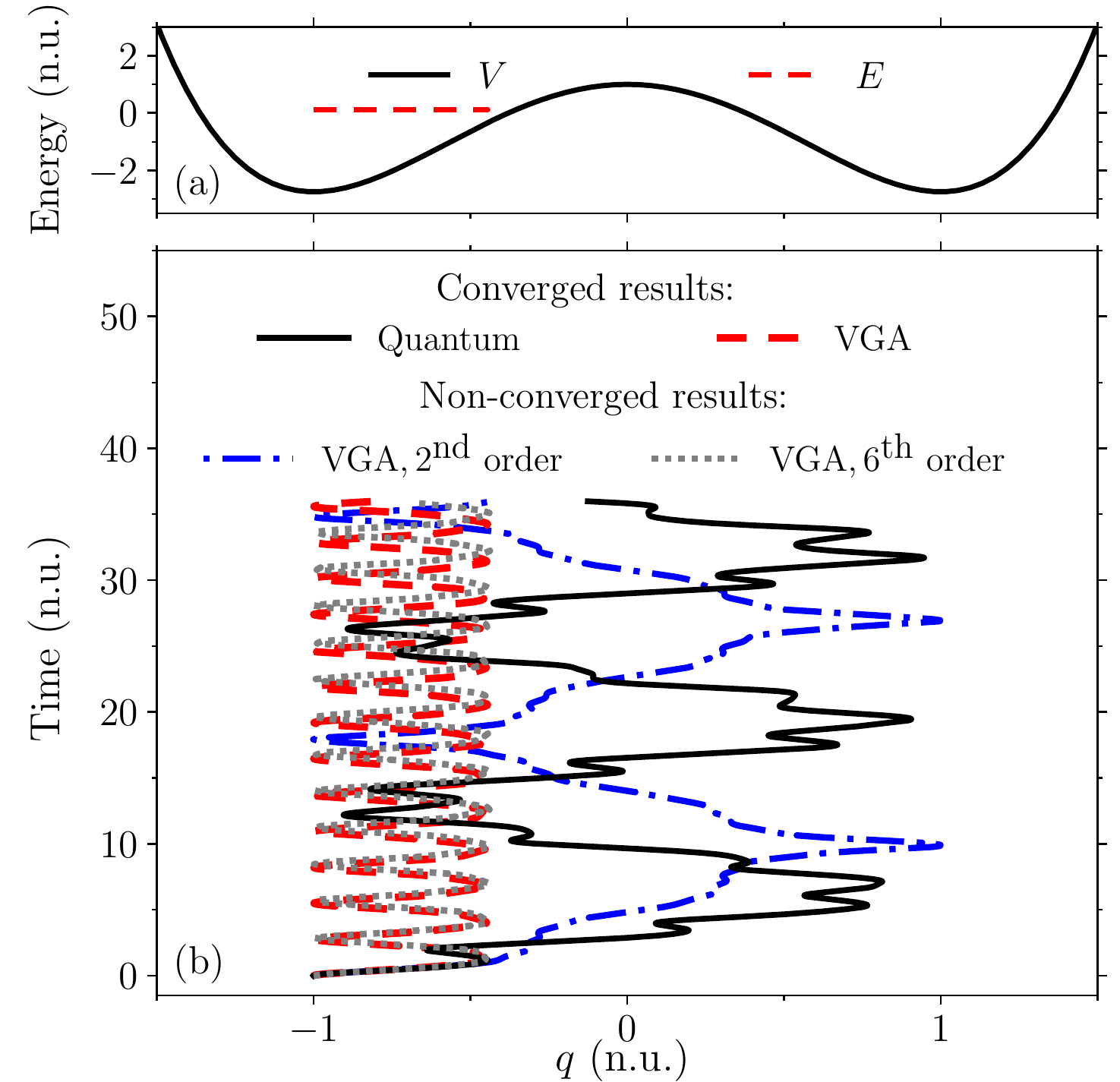}
\caption{ Failure of the VGA to detect tunneling in highly anharmonic systems and the importance of using high-order integrators. (a) The double well potential and the wavepacket’s energy.  (b) The difference between the exact quantum and fully converged VGA results shows that the VGA cannot capture tunneling in this case. However, due to cancellation of errors, the numerically non-converged results of the VGA, obtained with the second-order integrator, recover the tunneling. Interestingly, with approximately the same computational effort, the sixth-order integrator calculations already agree with the fully converged VGA result.
} \label{Fig3}
\end{figure}


However, sometimes the VGA can fail to capture tunneling (see Fig.~\ref{Fig3}), especially when the double well is highly anharmonic or the barrier is extremely high. Figure~\ref{Fig3} shows the
dynamics of an initially Gaussian wavepacket with $q_{0}=-1$, $p_{0}=0$, and width
$A_{0}=(20/9)^{2}\,i$ in the
double well~(\ref{eq:Double-Well}) with parameters $a=1$, $b=7.5$, and
$c=3.75$. The energy of the wavepacket, $E\sim0.12$, is below the energy of
the barrier. Comparison of the exact quantum calculation with the fully converged VGA result  obtained using the sixth-order integrator with a small time step of $\Delta t=0.02$ implies that the VGA fails to capture tunneling in this system.

Let us demonstrate the importance of high-order integrators in situations at
the border of tunneling and non-tunneling regimes. Panel (b) of Fig.~\ref{Fig3} also shows less converged results of the VGA obtained by a low- and a high-order integrators. Due to the low accuracy, a simulation by the
second-order integrator with a time step of $\Delta t=0.08$ and computational
cost of $\sim297.7\,\text{s}$, measured in central processing unit (CPU) time, incorrectly
shows tunneling. Interestingly, even with a much larger time step of $\Delta
t=0.18$ and a slightly lower computational cost of $\sim275.4\,\text{s}$, the
sixth-order integrator gives the correct results. To make the computational
cost of the initialization and finalization negligible, we considered the CPU
time corresponding to a longer simulation time $t=10^{4}=10000$.

\subsection{Multi-dimensional coupled Morse potential}

\label{subsec:Coupled_Morse}

To study multi-dimensional systems without having to approximate $\langle
\hat{V}\rangle$, $\langle\hat{V}^{\prime}\rangle$, and $\langle\hat{V}%
^{\prime\prime}\rangle$, we have designed a non-separable,
arbitrary-dimensional anharmonic potential with analytical expectation values.
This $D$-dimensional coupled Morse potential,
\begin{equation}
V(q)=V_{\text{eq}}+\sum_{j=1}^{D}\,V_{j}(q_{j})+V_{\text{cpl}}(q),
\label{eq:Coupled-Morse}%
\end{equation}
consists of $D$ standard one-dimensional Morse potentials $V_{j}(q_{j})$ for
all its vibrational modes $q_{j}$, which are, in addition, mutually coupled
with a somewhat artificial, non-separable multi-dimensional Morse coupling
$V_{\text{cpl}}(q)$. In Eq.~(\ref{eq:Coupled-Morse}), $V_{\text{eq}}$ is the
potential at the equilibrium position $q_{\text{eq}}$, and each
one-dimensional Morse potential
\begin{equation}
V_{j}(q_{j}):=d_{e}^{\prime}\big[1-y_{j}(a_{j}^{\prime},q_{j})\big]^{2}
\label{eq:1D_Morse}%
\end{equation}
depends on the dissociation energy $d_{e}^{\prime}$, decay parameter
$a_{j}^{\prime}$, and one-dimensional Morse variable
\begin{equation}
y_{j}(a_{j}^{\prime},q_{j}):=\exp\big[-a_{j}^{\prime}\,(q_{j}-q_{\text{eq}%
,j})\big]. \label{eq:Morse_var_1D}%
\end{equation}
The $D$-dimensional Morse coupling
\begin{equation}
V_{\text{cpl}}(q):=d_{e}\,\big[1-y(a,q)\big]^{2} \label{eq:cpl_Morse}%
\end{equation}
depends on the dissociation energy $d_{e}$, decay vector $a$, and $D$-dimensional Morse variable
\begin{equation}
y(a,q):=\exp\big[-a^{T}\cdot(q-q_{\text{eq}})\big]. \label{eq:Morse_var}%
\end{equation}
The coupling $V_{\text{cpl}}(q)$ results in non-separability. The decay
parameter $a_{j}^{\prime}$, dissociation energy $d_{e}^{\prime}$, and
dimensionless anharmonicity $\chi_{j}^{\prime}$ are related by the
equation~\cite{Begusic_Vanicek:2019}
\begin{equation}
a_{j}^{\prime}=\chi_{j}^{\prime}\,\sqrt{8\,d_{e}^{\prime}},
\label{eq:anhar_rel_1}%
\end{equation}
and, similarly, the decay vector $a$, dissociation energy $d_{e}$, and
dimensionless anharmonicity vector $\chi$ are related via
\begin{equation}
a=\chi\,\sqrt{8\,d_{e}}. \label{eq:anhar_rel_2}%
\end{equation}
The expectation values $\langle\hat{V}\rangle$, $\langle\hat{V}^{\prime
}\rangle$, and $\langle\hat{V}^{\prime\prime}\rangle$ in a Gaussian wavepacket
are derived in Appendix \ref{subsec:Morse}.

Next, we report the results of several simulations that demonstrate: (i) the
better accuracy of the VGA over other single-trajectory Gaussian-based
methods, (ii) the preservation of the geometric properties of the VGA by the
symplectic integrators, and (iii) the efficiency of high-order integrators.
After inspecting a low-dimensional ($2$D) system, for which the grid-based
quantum calculations are available as a benchmark, we analyze the convergence
and geometric properties of the symplectic integrators in a high-dimensional
($20$D) system.


\begin{figure}
[htbp]
\includegraphics[width=0.45\textwidth]{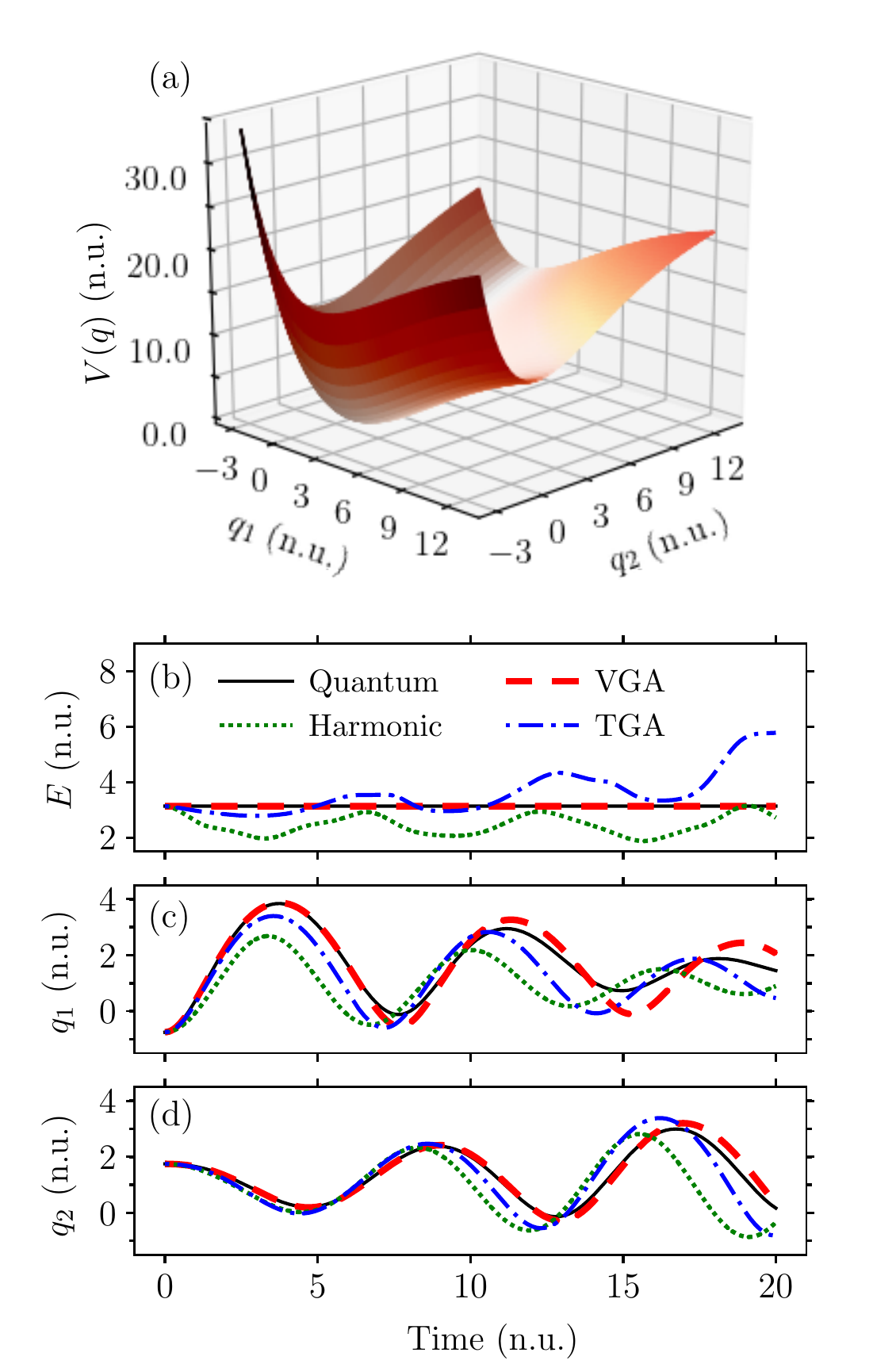}
\caption{Dynamics of an initially Gaussian wavepacket in a two-dimensional coupled Morse potential shown in panel (a). (b) Energy. (c) and (d): Expectation values of coordinates. }
\label{Fig4}
\end{figure}


Panel (a) of Fig.~\ref{Fig4} shows a two-dimensional coupled Morse
potential~(\ref{eq:Coupled-Morse}) with energy $V_{\text{eq}}=0$ at the
equilibrium position $q_{\text{eq}}=(1,1)$. It is composed of two
one-dimensional Morse potentials with the same dissociation energy
$d_{e}^{\prime}=11.25$ and different anharmonicities $\chi_{1}^{\prime}=0.02$
and $\chi_{2}^{\prime}=0.017$. The parameters of the coupling term are
$d_{e}=5.75$ and $\chi=(0.014,0.017)$. The initial state was a real Gaussian
with $q_{0}=(-0.75,1.75)$, $p_{0}=(0,0)$, and a diagonal width matrix $A_{0}$
with non-zero elements $A_{0,11}=A_{0,22}=i$. This wavepacket was then
propagated for $20000$ steps of $\Delta t=0.001$ with the second-order
symplectic integrator. The position grid for the exact quantum dynamics
consisted of $256$ points between $-3$ and $13$ in both directions.

Panel (b) of Fig.~\ref{Fig4} indicates that the VGA conserves energy, while neither the TGA nor the harmonic approximation are energy-conserving. Panels (c) and (d) show that, for very short times, all approximate
methods recover the exact quantum results, but their accuracies decrease with
increasing time. However, the VGA remains accurate for longer than the TGA,
which in turn remains accurate for longer than the harmonic approximation.


\begin{figure}
[htbp]
\includegraphics[width=0.45\textwidth]{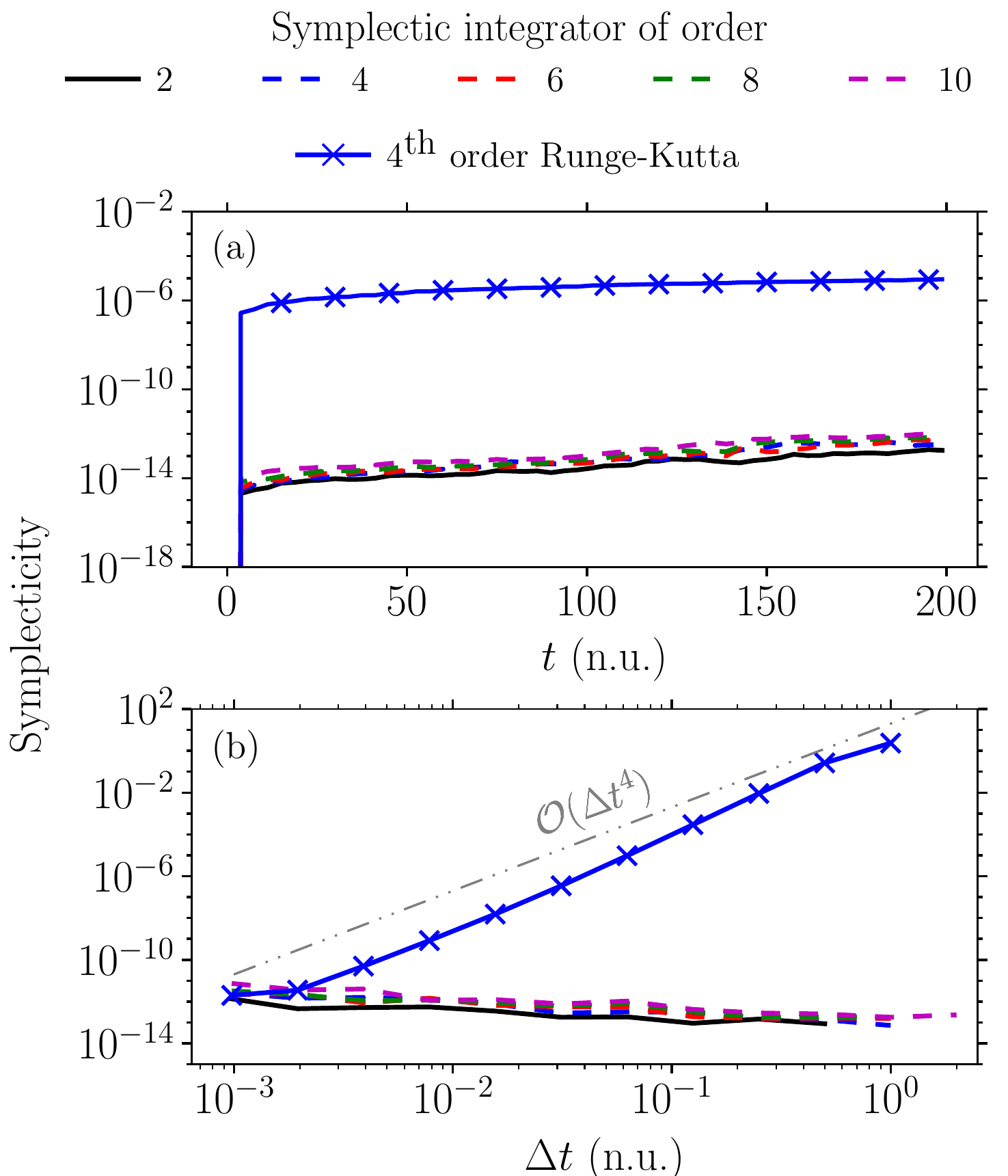}
\caption{Conservation of the symplectic structure of the Gaussian wavepackets by the symplectic integrators and its non-conservation by the fourth-order Runge-Kutta method. The system is the same two-dimensional coupled Morse potential as in Fig.~\ref{Fig4}. The symplecticity~(\ref{eq:symplecticity}) is shown: (a) as a function of time $t$ for a fixed time step $\Delta t= 2^{-4}\, \textrm{n.u.}$ and (b) as a function of time step $\Delta t$ at the final time $t_{f}=200\, \textrm{n.u.}$ To avoid clutter, only the high-order symplectic integrators obtained with the optimal composition are displayed.}
\label{Fig5}
\end{figure}

We also used the two-dimensional system to numerically verify the
symplecticity of the integrators for the VGA. We have developed a numerical procedure to check the symplecticity of the VGA by measuring the distance
\begin{align}
&  d\big(\Phi_{t}^{\prime}(z_{0})^{T}\cdot B(z_{t})\cdot\Phi_{t}^{\prime
}(z_{0}),B(z_{0})\big)\nonumber\\
&  \,\,=\lVert\Phi_{t}^{\prime}(z_{0})^{T}\cdot B(z_{t})\cdot\Phi_{t}^{\prime
}(z_{0})-B(z_{0})\rVert\label{eq:symplecticity}%
\end{align}
between the \textquotedblleft initial\textquotedblright\ and \textquotedblleft
final\textquotedblright\ symplectic structure matrices $B(z_{0})$ and
$B(z_{t})$. Here, vector $z_{t}$ contains components of the parameters $q_{t}%
$, $p_{t}$, $Q_{t}$, and $P_{t}$, $B(z_{t})$ is a skew-symmetric matrix
representing the symplectic two-form of Gaussian
wavepackets,~\cite{Ohsawa:2015a} and $\Phi_{t}^{\prime}(z_{0})$ is the
Jacobian of the VGA evolution $z_{t}=\Phi_{t}(z_{0})$. We chose to work in
Hagedorn's parametrization because the evaluation of the Jacobian is simpler
and the symplectic structure matrix $B$ is independent of $z_{t}$; see
Appendix~\ref{sec:num_symplectic_structure} for details.

Figure~\ref{Fig5} shows the symplecticity~(\ref{eq:symplecticity}) of the
Gaussian wavepackets propagated with the VGA in the two-dimensional potential
shown in Fig.~\ref{Fig4}(a). Although the propagation time in Fig.~\ref{Fig5}
is ten times longer than that in Fig.~\ref{Fig4}, all symplectic integrators
conserve the symplectic structure as a function of both time and time step. In
contrast, Fig.~\ref{Fig5} shows that the popular fourth-order Runge-Kutta
approach is not symplectic.


To show that, unlike grid-based quantum methods, the VGA is feasible in
high-dimensional models, we have constructed a twenty-dimensional coupled
Morse potential~(\ref{eq:Coupled-Morse}), composed of twenty Morse
potentials~(\ref{eq:1D_Morse}) with the same dissociation energy
$d_{e}^{\prime}=0.1$ and anharmonicity parameters $\chi_{j}^{\prime
},\,\,j=1,\dots,20$ uniformly varying in the range between $0.001$ and
$0.005$. The parameters of the coupling term are $d_{e}=0.075$ and $\chi
_{j}=(3/4)\,\chi_{j}^{\prime}$. The initial Gaussian was real, had zero
position and momentum and a diagonal width matrix with non-zero elements
$A_{0,jj}=4\,d_{e}\chi_{j}\,i$. The wavepacket was propagated for
$2^{17}=131072$ steps of $\Delta t=0.125$ with the second-order symplectic integrator.


\begin{figure}
[htbp]
\includegraphics[width=0.45\textwidth]{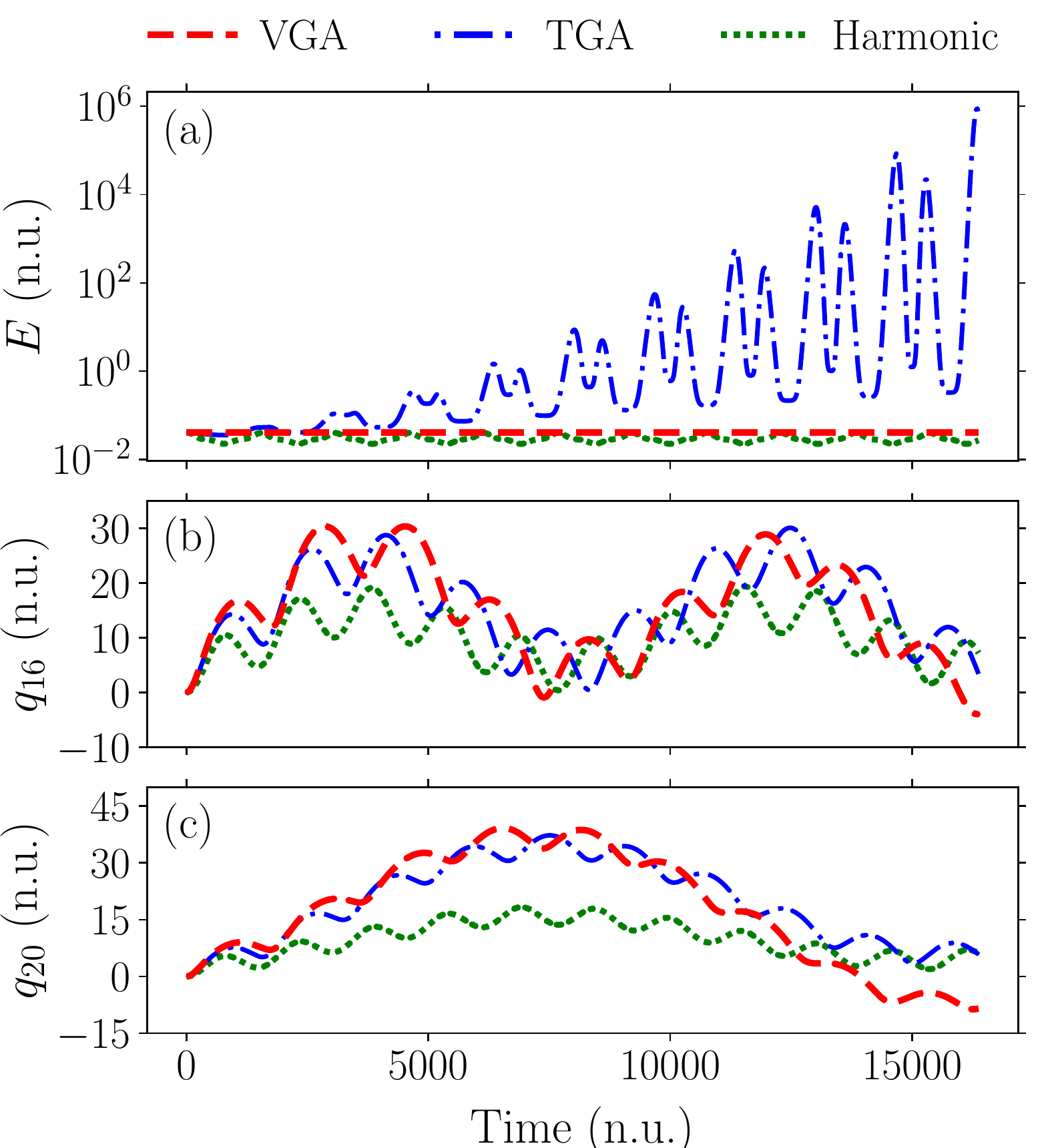}
\caption{Dynamics of a Gaussian wavepacket propagated in a twenty-dimensional coupled Morse potential with the variational Gaussian (VGA), thawed Gaussian (TGA), and harmonic approximations. Energy [panel (a)] and displacement along two coordinates [panels (b) and (c)] are shown.	
} \label{Fig6}
\end{figure}


Figure \ref{Fig6} compares the dynamics of the Gaussian wavepacket propagated
with different methods in this twenty-dimensional system. Initially, results
of all methods overlap almost perfectly. However, after a short time, first
the harmonic approximation and later the TGA start to deviate from the VGA.

\begin{figure}
[!htbp]
\includegraphics[width=0.45\textwidth]{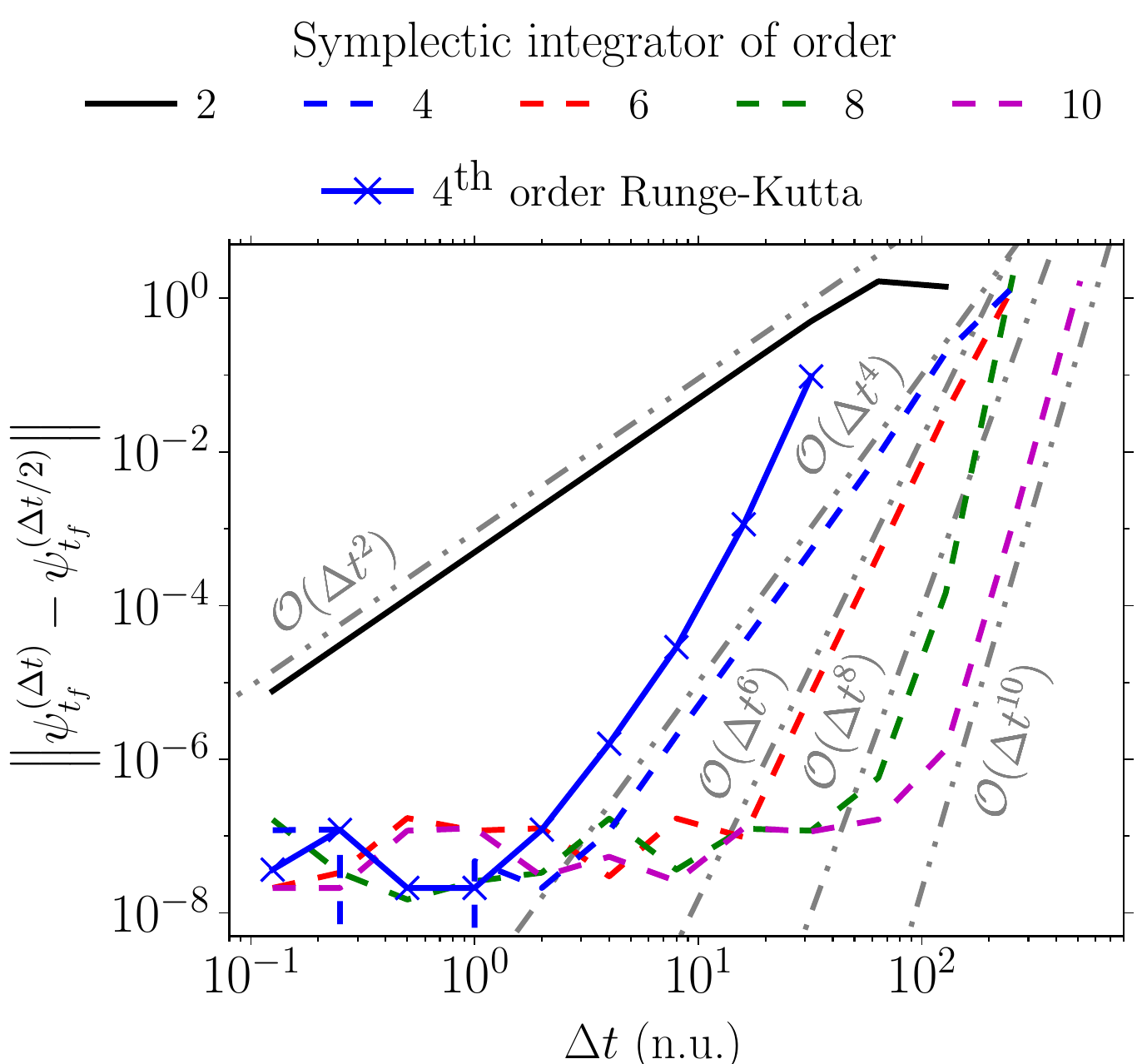}
\caption{Convergence of the symplectic integrators and of the fourth-order Runge-Kutta method for the VGA, measured by the convergence error at the final time $t_{f} = 2^{16} \, \textrm{n.u.}=65536 \, \textrm{n.u.}$ as a function of the time step $\Delta t$. The convergence error is defined as the distance $d(\psi^{(\Delta t)}_{t},\psi^{(\Delta t/2)}_{t})\equiv \lVert \psi^{(\Delta t)}_{t}-\psi^{(\Delta t/2)}_{t} \rVert$, where $\psi^{(\Delta t)}_{t}$ denotes the state at time $t$ obtained after propagation with the time step $\Delta t$.
	} \label{Fig7}
\end{figure}

To analyze the convergence and geometric properties of the integrators, we
repeated the VGA simulation with several high-order symplectic integrators and
with the fourth-order Runge-Kutta method. Figure \ref{Fig7} compares the
convergence of various methods as a function of the time step. For all
methods, the obtained orders of convergence agree with the predicted ones,
indicated by the gray straight lines.


\begin{figure}
[!htbp]
\includegraphics[width=0.45\textwidth]{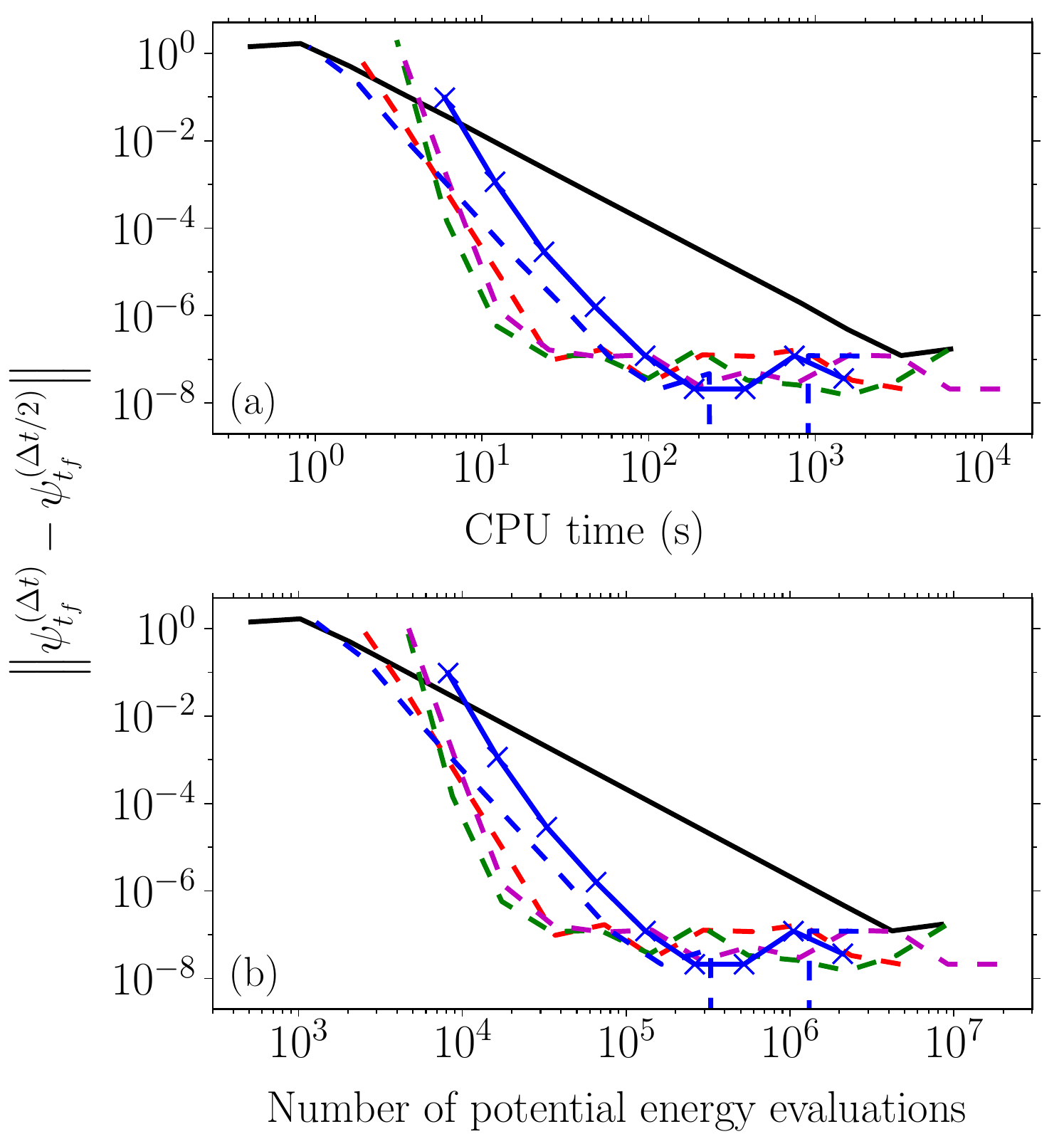}
\caption{Efficiency of the symplectic integrators and of the fourth-order Runge-Kutta method for the VGA. The efficiency is measured by plotting the convergence error, defined in the caption of Fig.~\ref{Fig7}, as a function of (a) the computational cost (CPU time) or (b) the number of potential energy evaluations. The line labels are the same as those in Fig. \ref{Fig7}.
	} \label{Fig8}
\end{figure}
Since the high-order methods require many composition substeps to be performed
at each time step $\Delta t$, the higher efficiency is not guaranteed solely
by a higher order of convergence. Therefore, in Fig.~\ref{Fig8}, we provide
two direct ways to measure the efficiency: one plotting the convergence error
as a function of the CPU time, and the other plotting the error as a function
of the number of potential energy evaluations. The similarity between panels
(a) and (b) confirms that the potential propagation substeps are the most
time-consuming parts of the simulation. In addition, Fig.~\ref{Fig8} shows
that high-order optimal integrators are more efficient than both the
second-order symplectic integrator and the fourth-order Runge-Kutta method.
For example, below a rather large error of $10^{-1}$, the fourth-order
symplectic integrator is already more efficient than the second-order
algorithm. The efficiency gain increases when high accuracy is desired.
Indeed, for a moderate error of $10^{-6}$, the eighth-order method is $100$
times faster than the second-order symplectic method and more than $5$ times
faster than the fourth-order Runge-Kutta approach. The plateau indicates the
machine precision error.


\begin{figure}
[ht]
\includegraphics[width=0.45\textwidth]{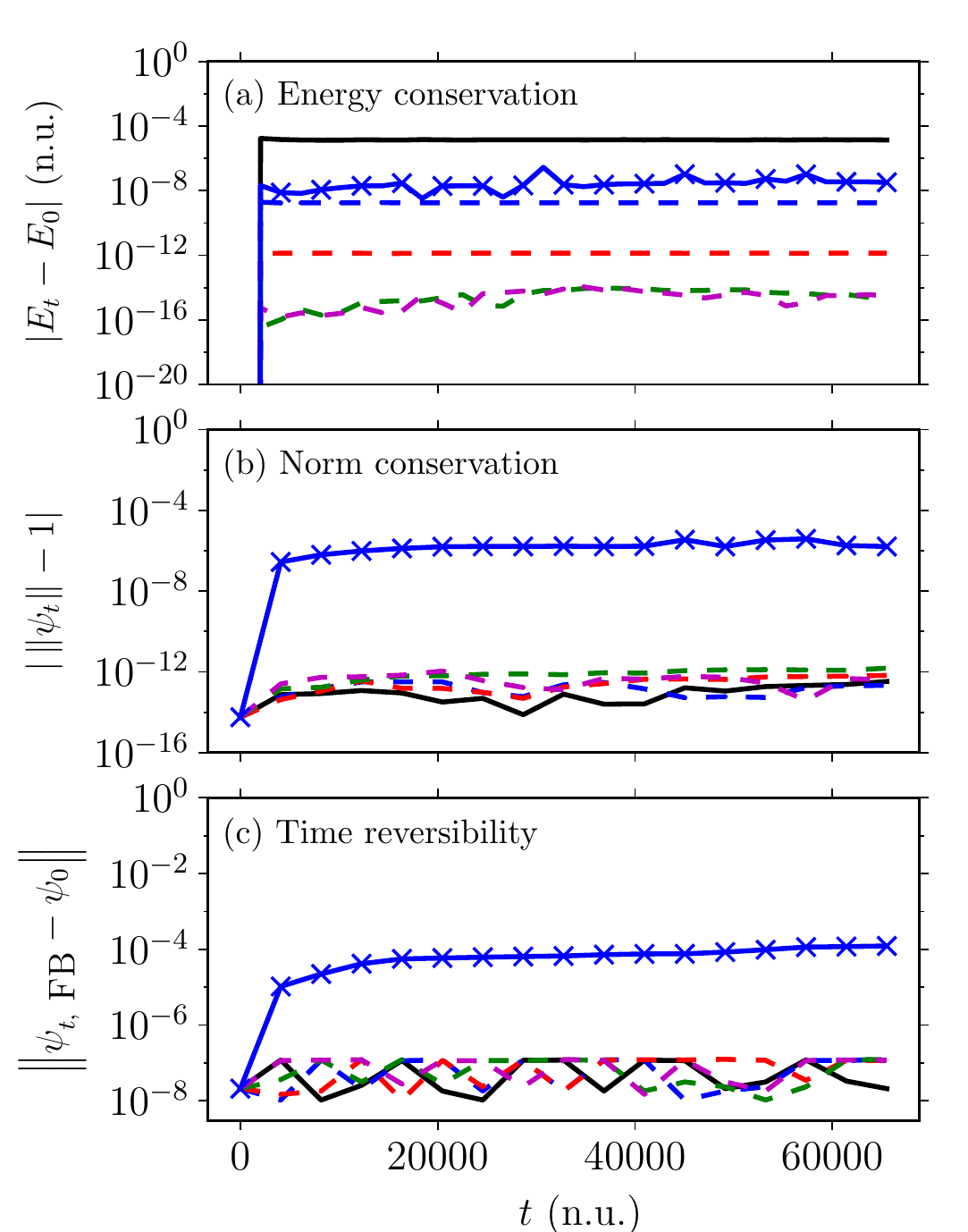}
\caption{Geometric properties of various integrators for the VGA as a function of time $t$ for a large time step $\Delta t = 8 \, \textrm{n.u.}$ (a) Energy, (b) norm, and (c) time reversibility [Eq. (\ref{eq:TR_distance})] are shown. The line labels are the same as those in Fig. \ref{Fig7}.
} \label{Fig9}
\end{figure}



\begin{figure}
[!htbp]
\includegraphics[width=0.45\textwidth]{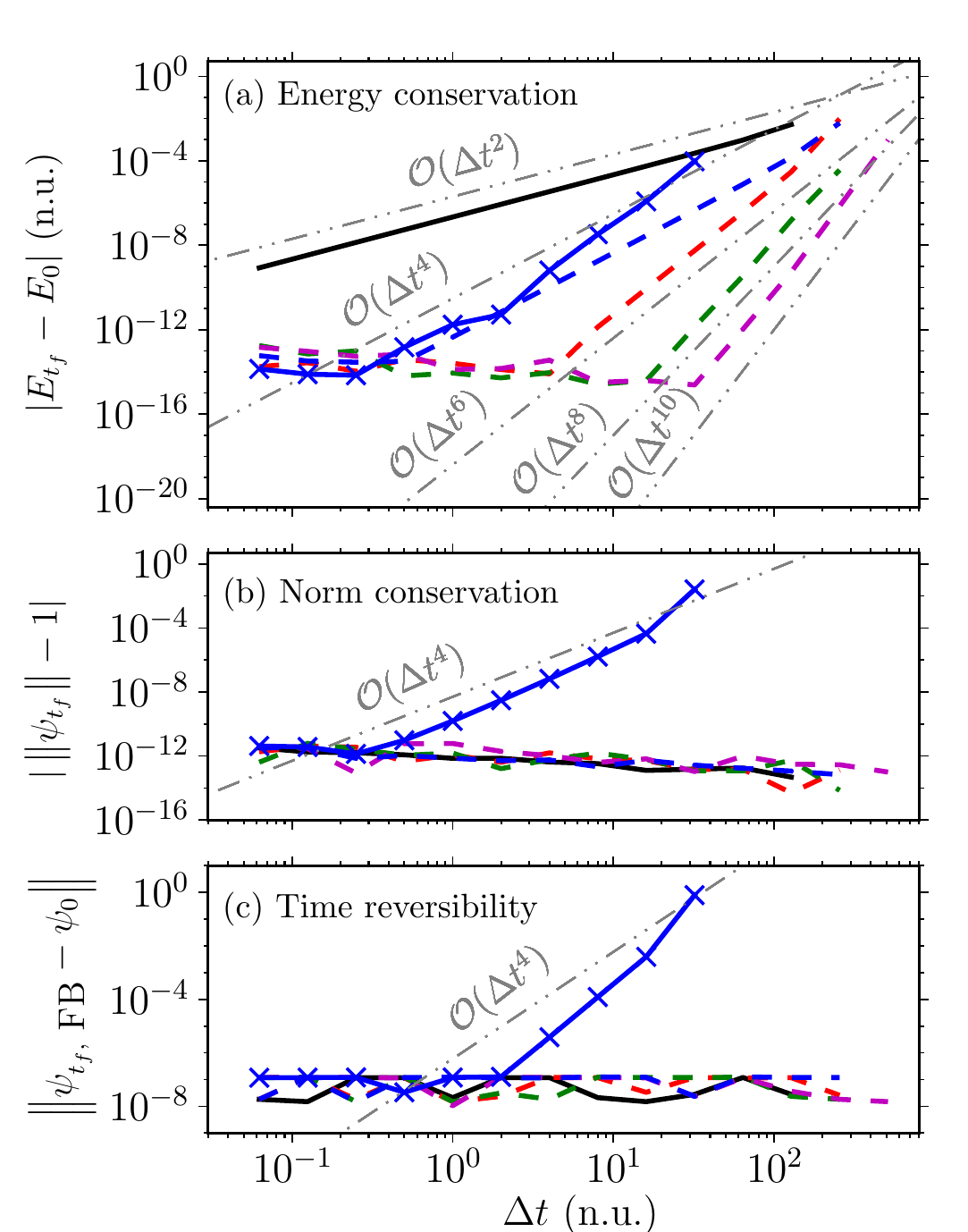}
\caption{Geometric properties of various integrators for the VGA as a function of the time step $\Delta t$ measured at the final time $t_{f} = 2^{16} \, \textrm{n.u.}=65536 \, \textrm{n.u.}$ (a) Energy, (b) norm, and (c) time reversibility [Eq.~(\ref{eq:TR_distance})] are shown. The line labels are the same as those in Fig. \ref{Fig7}.
} \label{Fig10}
\end{figure}

Figure \ref{Fig9} shows the time dependence of energy, norm, and time
reversibility, while Fig.~\ref{Fig10} shows how these geometric properties
depend on the time step. To analyze the conservation of norm and energy, we
compute $|\lVert\psi_{t}\rVert-1|$ and $|E_{t}-E_{0}|$ respectively. The
energy is calculated from Eqs.~(\ref{eq:Energy}),~(\ref{eq:KinEnergy}),
and~(\ref{eq:exp_value_coupled_Morse_pot}) in the Appendix. Time reversibility
is checked by measuring the distance
\begin{equation}
d\big(\psi_{t,\text{FB}},\psi_{0}\big)=\lVert\psi_{t,\text{FB}}-\psi_{0}%
\rVert\label{eq:TR_distance}%
\end{equation}
between the \textquotedblleft forward-backward\textquotedblright propagated
state $\psi_{t,\text{FB}}$ defined in Eq.~(\ref{eq:TR}) and the initial state
$\psi_{0}$. Due to the unnecessarily large computational cost, we did not
analyze the symplecticity (\ref{eq:symplecticity}) for this twenty-dimensional
system; the conservation of the symplectic structure by the symplectic
integrators was already verified for a two-dimensional potential in Fig.
\ref{Fig5}.

Panels (a) of Figs. \ref{Fig9} and \ref{Fig10} show near-conservation of
energy by the symplectic integrators. Our symplectic integrators cannot
conserve energy exactly since the alternation between kinetic and potential
propagations makes the effective Hamiltonian time-dependent. However, since
the VGA is energy-conserving, energy conservation is seen for time steps
$\Delta t$ that are small enough that the numerical errors become negligible.
The gray lines in Fig. \ref{Fig10}(a) indicate that the energy conservation
follows the order of convergence of the symplectic integrators. Panels (b) and
(c) confirm that all symplectic integrators are exactly norm-conserving and
time-reversible, regardless of the size of the time step. Furthermore, all
three panels show that very small time steps would be required for the
fourth-order Runge-Kutta method to conserve norm and energy and to be
reversible. Note that the convergence of energy and reversibility by the
Runge-Kutta method appears somewhat faster than $\mathcal{O}(\Delta t^{4})$.

\begin{figure}
[!htbp]
\includegraphics[width=0.45\textwidth]{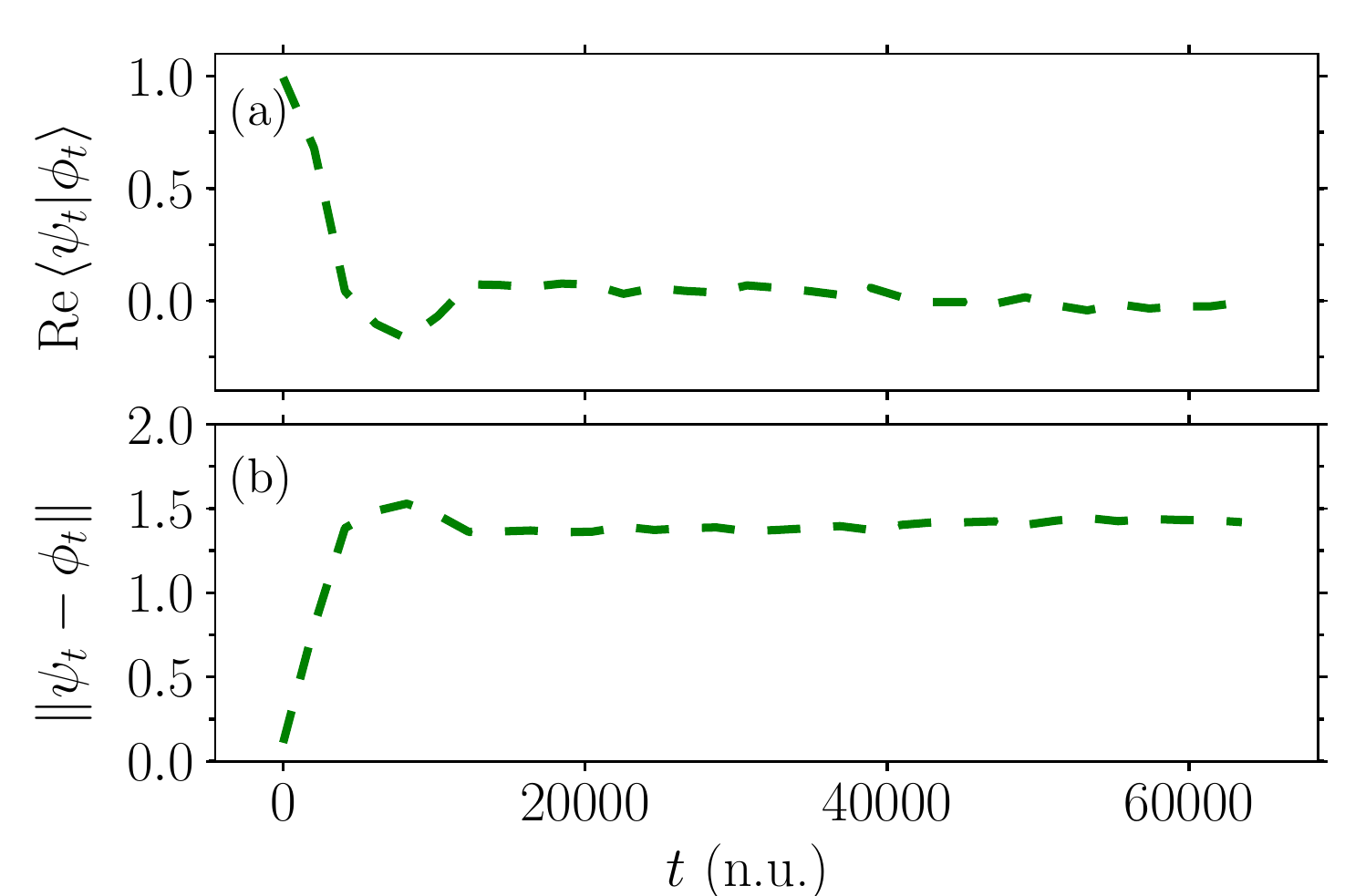}
\caption{Non-conservation of the (a) inner product [Eq. (\ref{eq:GWPs_overlap})] and (b) distance between two states [Eq. (\ref{eq:distance})] by the VGA. The system is the twenty-dimensional coupled Morse potential~(\ref{eq:Coupled-Morse}), and state $\psi_{0}$ is the Gaussian wavepacket defined in the text introducing Fig.~\ref{Fig6}. State $\phi_{0}$ is $\psi_{0}$ displaced by $1\, \textrm{n.u}$ along all its twenty modes.
	} \label{Fig11}
\end{figure}
In Figs. \ref{Fig5}, \ref{Fig9}, and \ref{Fig10}, we showed the geometric
properties conserved exactly by the VGA and investigated whether the
symplectic integrators can preserve them. Figure~\ref{Fig11} displays two
properties, the inner product and distance~(\ref{eq:distance}), which are not
conserved even by the VGA itself. The analytical expression for the inner
product of two Gaussians is given in Appendix.~\ref{subsec:inner_prod}. To
eliminate numerical errors, we propagated the wavepackets using the
eighth-order integrator with a time step of $\Delta t=0.5$ since, as shown in
Fig~\ref{Fig7}, this integrator gives highly accurate results with this time step.


\section{Conclusion}

\label{sec:conclusion}

In this paper, we have revisited the VGA and analyzed its accuracy,
efficiency, and geometric properties. Our results confirm that the VGA is an
efficient semiclassical method for simulating weakly anharmonic
high-dimensional systems, which are very expensive or beyond reach of exact quantum methods. Furthermore, by comparing the results of the VGA, TGA, and
harmonic approximation with exact quantum calculations in several
low-dimensional systems, we have confirmed that the VGA, although more
computationally expensive, is the most accurate single-trajectory
Gaussian-based method. We also verified that the VGA conserves energy and may
approximately capture tunneling. The reader should, however, keep in mind that the single Gaussian ansatz is a very restrictive approximation and that, despite conserving the energy and symplectic structure, even a fully converged result of the VGA can be far from the exact quantum solution of a problem.

To reduce the computational cost of the VGA, in Sec.~\ref{sec:sym_integrators}%
, we derived efficient high-order geometric integrators by symmetrically
composing the second-order symplectic integrator. These high-order integrators
are symplectic, norm-conserving, time-reversible, and for small time steps,
energy-conserving. Using the VGA to simulate multi-dimensional coupled Morse
potentials, we numerically demonstrated the geometric properties of the
symplectic integrators. In particular, we transformed the analytical technique
used by Faou and Lubich~\cite{Faou_Lubich:2006} and by Ohsawa and
Leok~\cite{Ohsawa_Leok:2013} into a practical numerical method for checking the symplecticity of the geometric integrators.

Although the VGA appears, at first sight, to be an uncontrollable approximation, it is possible to systematically improve VGA using Hagedorn wave packets.\cite{Lasser_Lubich:2020} The equations of motion of the VGA require expectation values of the potential
energy, its gradient, and the Hessian, which are not always analytically
available. Although various numerical quadrature techniques exist, they would be too expensive in ab initio real-world applications. One way to reduce this cost is to approximate the potential by its Taylor expansion around the Gaussian’s center. Applying
the VGA to the local harmonic approximation of the potential yields the TGA,
which conserves neither the symplectic structure nor the effective
energy.~\cite{Vanicek:2023} However, the TGA combined with the ab initio evaluation of the potential has already produced reasonably accurate spectra of a number of polyatomic molecules, such as oligothiophenes,~\cite{Wehrle_Vanicek:2014_v2,Begusic_Vanicek:2019} tetrafluorobenzene, ammonia, phosphine, and arsine.~\cite{Begusic_Vanicek:2022} By applying the VGA to the local cubic approximation of the potential, one obtains a method that not only improves the accuracy over the TGA, but also preserves both the symplectic structure and the effective energy. ~\cite{Ohsawa_Leok:2013, Pattanayak_Schieve:1994,Vanicek:2023} The detailed discussion of this approach, which should still be practical for ab initio calculations of spectra of medium-sized molecules, is deferred to our forthcoming paper.~\cite{Moghaddasi_Vanicek:2023a}


\section*{Supplementary material}

See the supplementary material for the convergence and efficiency of the
high-order symplectic integrators obtained using the triple-jump and
Suzuki-fractal composition schemes, for the analysis of the VGA in Hagedorn's
parametrization [Eq.~(\ref{eq:HGWP})], and for the separate convergence of
individual parameters of the Gaussian wavepacket in both Heller's
[Eq.~(\ref{eq:GWP})] and Hagedorn's parametrizations.

\begin{acknowledgments}
The authors thank Christian Lubich and Tomoki Ohsawa for useful discussions
and acknowledge the financial support from the European Research Council (ERC)
under the European Union's Horizon 2020 research and innovation program (grant
agreement No. 683069 -- MOLEQULE).
\end{acknowledgments}

\section*{Author declarations}

\subsection*{Conflict of interest}

The authors have no conflicts to disclose.

\section*{Data availability}

The data that support the findings of this study are available within the article and its supplementary material.

\appendix

\section{Two non-variational Gaussian wavepacket methods}

\label{sec:TGA_harmonic}

\subsection{Heller's original thawed Gaussian approximation (TGA)}

\label{subsubsec:TGA}

In the TGA,~\cite{Heller:1975} the potential energy $V$ is replaced by its
local harmonic approximation (LHA)
\begin{align}
V_{\text{LHA}}(q,\psi_{t}) &  =V(q_{t})+V^{\prime}(q_{t})^{T}\cdot
(q-q_{t})\nonumber\\
&  \quad+(q-q_{t})^{T}\cdot V^{\prime\prime}(q_{t})\cdot(q-q_{t}%
)/2\label{eq:V_LHA}%
\end{align}
about the Gaussian's center $q_{t}$. In Eq.~(\ref{eq:V_LHA}), $V(q_{t})$,
$V^{\prime}(q_{t})$, and $V^{\prime\prime}(q_{t})$ represent the potential,
gradient, and Hessian at $q_{t}$. Replacing $V$ with $V_{\text{LHA}}$ and
inserting the thawed Gaussian ansatz~(\ref{eq:GWP}) [or~(\ref{eq:HGWP})] into
the time-dependent Schr\"{o}dinger equation~(\ref{eq:TDSE})
gives~\cite{Heller:1975, Begusic_Vanicek:2019,
Vanicek_Begusic:2021,Vanicek:2023} the equations of motion~(\ref{eq:qEOM}%
)-(\ref{eq:gammaEOM})[or~(\ref{eq:HQEOM})-(\ref{eq:HSEOM}) in Hagedorn's
parametrization~\cite{Begusic_Vanicek:2022}] with coefficients
\begin{equation}
V_{0}=V(q_{t}),\,V_{1}=V^{\prime}(q_{t}),\,V_{2}=V^{\prime\prime}%
(q_{t}).\label{eq:V0_V1_V2_TGA}%
\end{equation}


\subsection{Harmonic approximation}

\label{subsubsec:HA}

In the harmonic approximation (HA),~\cite{Tannor_Heller:1982,Vanicek:2023} the
potential $V$ is replaced by its second-order Taylor expansion
\begin{align}
V_{\text{HA}}(q,\psi_{t}) &  =V(q_{\text{ref}})+V^{\prime}(q_{\text{ref}}%
)^{T}\cdot(q-q_{\text{ref}})\nonumber\\
&  \quad+(q-q_{\text{ref}})^{T}\cdot V^{\prime\prime}(q_{\text{ref}}%
)\cdot(q-q_{\text{ref}})/2\label{eq:V_HA}%
\end{align}
about a reference geometry $q_{\text{ref}}$. Replacing $V$ with $V_{\text{HA}%
}$ and inserting the thawed Gaussian ansatz~(\ref{eq:GWP}) [or~(\ref{eq:HGWP}%
)] into the time-dependent Schr\"{o}dinger equation~(\ref{eq:TDSE}), we obtain
the equations of motion~(\ref{eq:qEOM})-(\ref{eq:gammaEOM})
[or~(\ref{eq:HQEOM})-(\ref{eq:HSEOM}) in Hagedorn's parametrization] with
coefficients~\cite{Tannor_Heller:1982,Vanicek:2023}
\begin{equation}
V_{0}=V_{\text{HA}}(q_{t}),\,V_{1}=V_{\text{HA}}^{\prime}(q_{t}),\,V_{2}%
=V_{\text{HA}}^{\prime\prime}(q_{t}).\label{eq:V0_V1_V2_HA}%
\end{equation}

\section{Various properties of the Gaussian wavepacket}

\label{sec:useful_eqns} Here, we derive several expressions needed to obtain
the equations of motion [Eqs.~(\ref{eq:qEOM})-(\ref{eq:gammaEOM}) with
coefficients~(\ref{eq:V0_V1_V2_VGA})] and the geometric properties of the VGA.

\subsection{Position and momentum covariances}

\label{sec:Gaussian_integrals}

The position and momentum covariance matrices of a Gaussian wavepacket in
either Heller's [Eq.~(\ref{eq:GWP})] or Hagedorn's [Eq.~(\ref{eq:HGWP})]
parametrization are
\begin{align}
\Sigma_{t} &  :=\langle(\hat{q}-q_{t})\otimes(\hat{q}-q_{t})^{T}%
\rangle\nonumber\\
&  \,\,=(\hbar/2)\,\mathcal{B}_{t}^{-1}=(\hbar/2)\,Q_{t}\cdot Q_{t}^{\dagger
},\label{eq:Covq}\\
\Pi_{t} &  :=\langle(\hat{p}-p_{t})\otimes(\hat{p}-p_{t})^{T}\rangle
\nonumber\\
&  \,\,=(\hbar/2)\,A_{t}\cdot\mathcal{B}_{t}^{-1}\cdot A_{t}^{\dag}%
=(\hbar/2)\,P_{t}\cdot P_{t}^{\dagger}.\label{eq:Covp}%
\end{align}
The Isserlis theorem~\cite{Isserlis:1918} implies that the fourth moment of
position is~\cite{Brookes:2011}
\begin{equation}
\langle x_{j}\,x_{k}\,x_{l}\,x_{m}\rangle=\Sigma_{t,jk}\,\Sigma_{t,lm}%
+\Sigma_{t,jl}\,\Sigma_{t,km}+\Sigma_{t,jm}\,\Sigma_{t,kl},\label{eq:Isserli}%
\end{equation}
where $x:=q-q_{t}$ is a vector of difference coordinates.

\subsection{Overlap of Gaussian wavepackets}

\label{subsec:inner_prod}

The overlap of two Gaussian wavepackets is
\begin{align}
\langle\psi_{1t}|  &  \psi_{2t} \rangle=(\text{det}\,Z)^{-1/2}\nonumber\\
&  \times\text{exp}\big\{(i/\hbar) \big[-\delta\lambda^{T} \cdot(\delta
W)^{-1} \cdot\delta\lambda/2 + \delta\eta\big]\big\}, \label{eq:GWPs_overlap}%
\end{align}
where the tensor difference $\delta\Lambda:=\Lambda_{2}-\Lambda^{*}_{1}$ is a
shorthand notation used for a particular scalar $\eta$, vector $\lambda$, or
matrix $W$, which depend on the Gaussian's parametrization. For Heller's
parametrization~(\ref{eq:GWP}), we have~\cite{Begusic_Vanicek:2020a}
\begin{align}
Z  &  = (1/2i\pi\hbar)\,\delta W,\\
W  &  = A_{t},\\
\lambda &  = p_{t} - A_{t} \cdot q_{t},\\
\eta &  = \gamma_{t}- (\lambda+p_{t})^{T} \cdot q_{t}/2,
\label{eq:GWPs_overlap_comp}%
\end{align}
whereas for Hagedorn's parametrization~(\ref{eq:HGWP}), we have
\begin{align}
Z  &  = \big(Q_{1t}^{\dagger} \cdot P_{2t} - P_{1t}^{\dagger} \cdot
Q_{2t}\big)/(2i),\\
W  &  = P_{t} \cdot Q^{-1}_{t},\\
\lambda &  = p_{t} - P_{t} \cdot Q^{-1}_{t}\cdot q_{t},\\
\eta &  = S_{t}-(\lambda+p_{t})^{T} \cdot q_{t}/2.
\label{eq:HGWPs_overlap_comp}%
\end{align}

\subsection{Energy of the Gaussian wavepacket}

The energy of a normalized Gaussian wavepacket $\psi_{t}$ is computed as the
expectation value
\begin{equation}
E_{t}:=\langle\hat{H}\rangle=\langle\hat{T}\rangle+\langle\hat{V}%
\rangle.\label{eq:Energy}%
\end{equation}
The first term is the kinetic energy
\begin{align}
\langle\hat{T}\rangle &  =\langle\hat{p}^{T}\cdot m^{-1}\cdot\hat{p}%
\rangle/2=\text{Tr}\big(m^{-1}\cdot\langle\,\hat{p}\otimes\hat{p}^{T}%
\rangle\big)/2\nonumber\\
&  =p_{t}^{T}\cdot m^{-1}\cdot p_{t}/2+\text{Tr}\big(m^{-1}\cdot\Pi
_{t}\big)/2,\label{eq:KinEnergy}%
\end{align}
and the second term is the potential energy
\begin{equation}
\langle\hat{V}\rangle=\int V(q)\,\rho_{t}(q)\,d^{D}q,\label{eq:V_exp_value}%
\end{equation}
where $\rho_{t}$ is the normalized position density
\begin{equation}
\rho_{t}(q):=|\psi_{t}(q)|^{2}=[\text{det}({2\pi\,\Sigma_{t}})]^{-1/2}%
\,e^{-x^{T}\cdot\Sigma_{t}^{-1}\cdot x/2},\label{eq:GWP_density}%
\end{equation}
with $x:=q-q_{t}$ and $\Sigma_{t}$ the position covariance~(\ref{eq:Covq}).
Unlike the kinetic energy, the potential energy cannot generally be evaluated analytically.

In the rest of this appendix, we differentiate the energy of the Gaussian
wavepacket with respect to its parameters. The derived expressions are used in
Appendix~\ref{sec:symplectic_structure} to obtain the equations of motion for
the VGA and to demonstrate numerically the
preservation of the symplectic structure by the symplectic integrators
designed for the VGA.


\subsection{Partial derivatives of the kinetic energy}

\label{subsec:dT}

Inserting Eq.~(\ref{eq:Covp}) for the momentum covariance into
Eq.~(\ref{eq:KinEnergy}) for the kinetic energy gives
\begin{align}
\langle\hat{T}\rangle &  =p_{t}^{T}\cdot m^{-1}\cdot p_{t}/2\nonumber\\
&  +(\hbar/4)\,\text{Tr}\big(m^{-1}\cdot\mathcal{A}_{t}\cdot\mathcal{B}%
_{t}^{-1}\cdot\mathcal{A}_{t}+m^{-1}\cdot\mathcal{B}_{t}\big).
\end{align}
Partial derivatives of $\langle\hat{T}\rangle$ with respect to the parameters
$q_{t},\phi_{t}$, and $\gamma_{t}$ vanish. The other derivatives are
\begin{align}
\partial\langle\hat{T}\rangle/\partial p_{t} &  =m^{-1}\cdot p_{t},\\
\partial\langle\hat{T}\rangle/\partial\mathcal{A}_{t} &  =(\hbar
/4)\,\big(m^{-1}\cdot\mathcal{A}_{t}\cdot\mathcal{B}_{t}^{-1}+\mathcal{B}%
_{t}^{-1}\cdot\mathcal{A}_{t}\cdot m^{-1}\big),\label{eq:dTdA}\\
\partial\langle\hat{T}\rangle/\partial\mathcal{B}_{t} &  =(\hbar
/4)\,\big(m^{-1}-\mathcal{B}_{t}^{-1}\cdot\mathcal{A}_{t}\cdot m^{-1}%
\cdot\mathcal{A}_{t}\cdot\mathcal{B}_{t}^{-1}\big).\label{eq:dTdB}%
\end{align}
To derive Eqs.~(\ref{eq:dTdA}) and~(\ref{eq:dTdB}), we used the relation
\begin{equation}
\partial\,\text{Tr}[F(X)]/\partial X=f(X)^{T}\label{eq:Trace_derivative}%
\end{equation}
for the derivative of the trace of a general function $F(X)$ of a square
matrix $X$, and applied it to $X=\mathcal{A}_{t}$ and $\mathcal{B}_{t}$. In
Eq.~(\ref{eq:Trace_derivative}), $f(\cdot)$ is the scalar derivative of
$F(\cdot)$.~\cite{Petersen_Pedersen:2012}

\subsection{Partial derivatives of the potential energy}

\label{subsec:dV}

Partial derivatives of the potential energy (\ref{eq:V_exp_value}) with
respect to the parameters $p_{t},\mathcal{A}_{t},\phi_{t}$, and $\gamma_{t}$
vanish. The other derivatives are
\begin{align}
\partial\langle\hat{V}\rangle/\partial q_{t} &  =\langle\hat{V}^{\prime
}\rangle,\label{eq:V_q}\\
\partial\langle\hat{V}\rangle/\partial\mathcal{B}_{t} &  =-(\hbar
/4)\,\mathcal{B}_{t}^{-1}\cdot\langle\hat{V}^{\prime\prime}\rangle
\cdot\mathcal{B}_{t}^{-1}.\label{eq:V_B}%
\end{align}
To derive Eq.~(\ref{eq:V_q}), we integrated by parts. Equation~(\ref{eq:V_B})
follows from Eq.~(\ref{eq:V_exp_value}) by substituting the derivative
\begin{align}
\partial\rho_{t}(q)/\partial\mathcal{B}_{t} &  =\partial\big[\text{det}%
({\mathcal{B}_{t}/\pi\hbar})^{1/2}\,e^{-x^{T}\cdot\mathcal{B}_{t}\cdot
x/\hbar}\big]/\partial\mathcal{B}_{t}\nonumber\\
&  =\big(\mathcal{B}_{t}^{-1}/2-x\otimes x^{T}/\hbar\big)\rho_{t}%
(q)\label{eq:drho_dB}%
\end{align}
of the density~(\ref{eq:GWP_density}) and integrating twice by parts. Finally,
to derive Eq.~(\ref{eq:drho_dB}), we used Eq.~(\ref{eq:Covq}) for the position
covariance and the relation
\begin{equation}
\partial\,\text{det}\mathcal{B}_{t}/\partial\mathcal{B}_{t}=\big(\text{det}%
\mathcal{B}_{t}\big)\,(\mathcal{B}_{t}^{-1})^{T}\label{eq:deriv_det}%
\end{equation}
for the derivative of a determinant.~\cite{Petersen_Pedersen:2012}


\label{subsec:dVdQ}

To find the partial derivatives of the potential energy~(\ref{eq:V_exp_value})
with respect to Hagedorn's parameters $Q_{t}$ and $P_{t}$, one needs to
express the density (\ref{eq:GWP_density}) in terms of these parameters. To do
this, we use Eq.~(\ref{eq:Covq}) to write
\begin{equation}
\mathcal{B}_{t}^{-1}=Q_{t}\cdot Q_{t}^{\dagger}=\sum_{r=1}^{2}Q_{t}^{(r)}\cdot
Q_{t}^{(r)\,T},\label{eq:B_Q_rel}%
\end{equation}
where $Q_{t}^{(1)}$ and $Q_{t}^{(2)}$ are the real and imaginary parts of
$Q_{t}$. Therefore, the density depends only on $Q_{t}$, implying that
$\partial\rho_{t}(q)/\partial P_{t}=0$ and
\begin{equation}
\partial\langle\hat{V}\rangle/\partial\,P_{t}=0.\label{eq:dV_P}%
\end{equation}
Furthermore, using Eq.~(\ref{eq:B_Q_rel}) and the chain rule, we have
\begin{align}
\partial\rho_{t}(q)/\partial Q_{t}^{(r)} &  =2\,[\partial\rho_{t}%
(q)/{\partial\mathcal{B}_{t}^{-1}}]\cdot Q_{t}^{(r)}\nonumber\\
&  =-2\,\mathcal{B}_{t}\cdot\lbrack\partial\rho_{t}(q)/{\partial
\mathcal{B}_{t}}]\cdot\mathcal{B}_{t}\cdot Q_{t}^{(r)},\label{eq:drho_dQ}%
\end{align}
which with Eq.~(\ref{eq:drho_dB}) gives
\begin{equation}
\partial\langle\hat{V}\rangle/\partial\,Q_{t}^{(r)}=(\hbar/2)\,\langle\hat
{V}^{\prime\prime}\rangle\cdot Q_{t}^{(r)},\quad r=1,2.\label{eq:dV_Q}%
\end{equation}
Similarly, if we differentiate the $n$th potential derivative $\hat{V}^{(n)}$,
which is a tensor of rank $n$, we will get a tensor of rank $n+2$ with
components
\begin{equation}
\partial\langle\hat{V}^{(n)}\rangle_{k_{1}\dots k_{n}}/\partial Q_{ij}%
^{(r)}=\frac{\hbar}{2}\,\sum_{l=1}^{D}\,\langle\hat{V}^{(n+2)}\rangle
_{k_{1}\dots k_{n}il}\,Q_{lj}^{(r)}.\label{eq:V_Q}%
\end{equation}


\section{Symplectic wavepacket dynamics}

\label{sec:symplectic_structure}

To reveal the symplectic structure of the Schr\"{o}dinger equation
(\ref{eq:TDSE}), one can identify the complex wavefunction $\Psi_{t}%
(q)=v_{t}(q)+iw_{t}(q)$ with a real pair $\Psi_{t}=(v_{t},w_{t})$, where
$v_{t}:=\text{Re}\,\Psi_{t}$ and $w_{t}:=\text{Im}\,\Psi_{t}$%
.~\cite{book_Marsden_Ratiu:1999} Since the Hamiltonian $\hat{H}$ is a real
operator, Eq.~({\ref{eq:TDSE}}) can be written as the canonical Hamiltonian
system~\cite{book_Marsden_Ratiu:1999,Faou_Lubich:2006,book_Hairer_Wanner:2006}
\begin{equation}
\hbar\,\dot{\Psi}_{t}=J\,\nabla_{\Psi_{t}}H(\Psi_{t}),\label{eq:H_Sys}%
\end{equation}
where $H\big(\Psi_{t}\big):=\langle\Psi_{t}|\hat{H}|\Psi_{t}\rangle/2$, and
\begin{equation}
J=%
\begin{pmatrix}
0 & 1\\
-1 & 0
\end{pmatrix}
\label{eq:J}%
\end{equation}
is the canonical symplectic matrix. From the symplectic point of view, the
time-dependent variational principle can be expressed by the real inner
product~\cite{book_Lubich:2008}
\begin{equation}
\langle\delta\psi_{t}\,|\hbar\,\dot{\psi}_{t}-J\,\nabla_{\psi_{t}}H(\psi
_{t})\rangle=0,\label{eq:DiracFrenkelVP}%
\end{equation}
where $\psi_{t}$ is an approximation to the solution of Eq. (\ref{eq:H_Sys}).
This is equivalent to requiring that the residual of the Schr\"{o}dinger
equation is always orthogonal to the tangent space $T_{\psi_{t}}M$ of the
approximation manifold $M$ at the point $\psi_{t}$. If one maps $\psi_{t}$ to
a new coordinate $z_{t}$ with a function $\chi$, i.e., $\psi_{t}=\chi
(z_{t})\in M$, Eq.~(\ref{eq:DiracFrenkelVP})
becomes~\cite{Faou_Lubich:2006,book_Hairer_Wanner:2006}
\begin{equation}
B(z_{t})\,\dot{z}_{t}=\nabla_{z_{t}}H\big(\chi(z_{t}%
)\big).\label{eq:Non_Con_H_Sys}%
\end{equation}
In Eq.~(\ref{eq:Non_Con_H_Sys}), $B(z_{t})=\hbar\,X(z_{t})^{T}J^{-1}X(z_{t})$
is the non-canonical symplectic matrix, where $X=(V,W)$ is the real pair of
the complex derivative $X_{\mathbb{C}}(z_{t})=\chi^{\prime}(z_{t}%
)=V(z_{t})+iW(z_{t})$. Similar to the canonical symplectic matrix~(\ref{eq:J}%
), the non-canonical symplectic matrix $B(z_{t})$ is skew-symmetric, but, in
general, depends on $z_{t}$.~\cite{Faou_Lubich:2006,book_Hairer_Wanner:2006}

\subsection{Derivations of the VGA equations of motion}

In the VGA, the manifold $M$ consists of unnormalized complex Gaussian
wavepackets $\chi(z_{t})$ [Eq.~(\ref{eq:GWP})] with the squared norm
coefficient $I(\mathcal{B}_{t},\delta_{t})$ [Eq.~(\ref{eq:Norm})] and
parameters
\begin{equation}
z_{t}:=\big(q_{t}^{T},p_{t}^{T},\Tilde{\mathcal{A}}_{t}^{T},\Tilde
{\mathcal{B}}_{t}^{T},\phi_{t},\delta_{t}\big)^{T}\in\mathbb{R}^{2D+2D^{2}%
+2},\label{eq:local_coord}%
\end{equation}
where $\Tilde{\mathcal{A}}_{t}$ and $\Tilde{\mathcal{B}}_{t}$ are $D^{2}%
$-dimensional column vectors containing elements of the real and imaginary
parts of the width matrix $A_{t}$ in a column-wise manner, i.e.,
$\Tilde{\mathcal{A}}_{j+D(k-1)}=\mathcal{A}_{jk}$ and $\Tilde{\mathcal{B}%
}_{j+D(k-1)}=\mathcal{B}_{jk}$. The tangent space $T_{\chi}M$ consists of
vector derivatives
\begin{equation}
\chi^{\prime}(z_{t})=\bigg(\frac{\partial\chi}{\partial q_{t}},\frac
{\partial\chi}{\partial p_{t}},\frac{\partial\chi}{\partial\Tilde{\mathcal{A}%
}_{t}},\frac{\partial\chi}{\partial\Tilde{\mathcal{B}}_{t}},\frac{\partial
\chi}{\partial\phi_{t}},\frac{\partial\chi}{\partial\delta_{t}}\bigg),
\end{equation}
with
\begin{align}
&  \partial\chi/\partial q_{t}=-(i/\hbar)\,(A_{t}\cdot x+p_{t})\,\chi,\\
&  \partial\chi/\partial p_{t}=(i/\hbar)\,x\,\chi,\\
&  \partial\chi/\partial\Tilde{\mathcal{A}}_{j+D(k-1)}=\partial\chi
/\partial\mathcal{A}_{jk}=(i/2\hbar)\,x_{j}\,x_{k}\,\chi,\\
&  \partial\chi/\partial\Tilde{\mathcal{B}}_{j+D(k-1)}=\partial\chi
/\partial\mathcal{B}_{jk}=-(1/2\hbar)\,x_{j}\,x_{k}\,\chi,\\
&  \partial\chi/\partial\phi_{t}=(i/\hbar)\,\chi,\\
&  \partial\chi/\partial\delta_{t}=-(1/\hbar)\,\chi,
\end{align}
where $x:=q-q_{t}$ is the difference coordinate vector. Using the relation
$X^{T}J^{-1}X=-\text{Im}\,X_{\mathbb{C}}^{\dagger}X_{\mathbb{C}}%
$,~\cite{Faou_Lubich:2006,book_Hairer_Wanner:2006} we compute the
non-canonical symplectic structure matrix of (\ref{eq:GWP}) as
\begin{align}
&  B(z_{t})=I(\mathcal{B}_{t},\delta_{t})/2\nonumber\\
\times &
\begin{pmatrix}
0 & -I_{D} & 0 & \frac{1}{2}p_{t}\otimes\beta_{t}^{T} & 0 & \frac{2}{\hbar
}p_{t}\\
I_{D} & 0 & 0 & 0 & 0 & 0\\
0 & 0 & 0 & -\frac{\hbar}{8}\Gamma_{t} & 0 & -\frac{1}{2}\beta_{t}\\
-\frac{1}{2}\beta_{t}\otimes p_{t}^{T} & 0 & \frac{\hbar}{8}\Gamma_{t} & 0 &
\frac{1}{2}\beta_{t} & 0\\
0 & 0 & 0 & -\frac{1}{2}\beta_{t}^{T} & 0 & -\frac{2}{\hbar}\\
-\frac{2}{\hbar}p_{t}^{T} & 0 & \frac{1}{2}\beta_{t}^{T} & 0 & \frac{2}{\hbar}
& 0
\end{pmatrix}
,\label{eq:omega}%
\end{align}
where $\beta_{t}$ is a $D^{2}$-dimensional vector with components
\begin{equation}
\beta_{j+D(k-1)}:=(2/\hbar)\Sigma_{t,jk}=\big(\mathcal{B}^{-1}\big)_{jk},
\end{equation}
and $\Gamma_{t}$ is a $D^{2}\times D^{2}$ real matrix with components
\begin{align}
&  \Gamma_{j+D(k-1),\,l+D(m-1)}=(2/\hbar)^{2}\langle x_{j}\,x_{k}%
\,x_{l}\,x_{m}\rangle\nonumber\\
&  \qquad=\big(\mathcal{B}^{-1}\big)_{jk}\big(\mathcal{B}^{-1}\big)_{lm}%
+\big(\mathcal{B}^{-1}\big)_{jl}\big(\mathcal{B}^{-1}\big)_{km}\nonumber\\
&  \qquad\quad+\big(\mathcal{B}^{-1}\big)_{jm}\big(\mathcal{B}^{-1}\big)_{kl},
\end{align}
obtained from Eqs.~(\ref{eq:Covq}) and~(\ref{eq:Isserli}). Using the relations
from Appendices~\ref{subsec:dT} and~\ref{subsec:dV}, the energy gradient with
respect to the coordinates $z_{t}$ is
\begin{align}
\nabla_{z_{t}}H\big(\chi(z_{t})\big)=I(\mathcal{B}_{t} &  ,\delta
_{t})/2\,\big(\langle\hat{V}^{\prime}\rangle,m^{-1}\cdot p_{t},(\hbar
/4)\,\epsilon_{t},\nonumber\\
&  (\hbar/4)\,\zeta_{t},0,-(2/\hbar)\,\langle\hat{H}\rangle
\big),\label{eq:grad_E}%
\end{align}
where $\langle\hat{H}\rangle$ is defined in Eq.~(\ref{eq:Energy}) and
$\epsilon_{t}$ and $\zeta_{t}$ are $D^{2}$-dimensional vectors with elements
\begin{align}
\epsilon_{j+D(k-1)} &  =\big(m^{-1}\cdot\mathcal{A}_{t}\cdot\mathcal{B}%
_{t}^{-1}+\mathcal{B}_{t}^{-1}\cdot\mathcal{A}_{t}\cdot m^{-1}\big)_{jk},\\
\zeta_{j+D(k-1)} &  =\big(m^{-1}-\mathcal{B}_{t}^{-1}\cdot\mathcal{A}_{t}\cdot
m^{-1}\cdot\mathcal{A}_{t}\cdot\mathcal{B}_{t}^{-1}\nonumber\\
-(2/\hbar) &  \,\langle\hat{H}\rangle\,\mathcal{B}_{t}^{-1}-\mathcal{B}%
_{t}^{-1}\cdot\langle\hat{V}^{\prime\prime}\rangle\cdot\mathcal{B}_{t}%
^{-1}\big)_{jk}.
\end{align}
Substituting Eq.~(\ref{eq:local_coord}) for $z_{t}$, Eq.~(\ref{eq:omega}) for
the symplectic form $B(z_{t})$, and Eq.~(\ref{eq:grad_E}) for the gradient
$\nabla_{z_{t}}H\big(\chi(z_{t})\big)$ into Eq. (\ref{eq:Non_Con_H_Sys})
yields the equations of motion for parameters $q_{t},p_{t},\Tilde{\mathcal{A}%
}_{t},\Tilde{\mathcal{B}}_{t},\phi_{t},\text{and}\,\delta_{t}$. Combining the
real and imaginary parts of $A_{t}$ and those of $\gamma_{t}$ into single
equations, these equations are given by Eqs.~(\ref{eq:qEOM}%
)-(\ref{eq:gammaEOM}) with coefficients~(\ref{eq:V0_V1_V2_VGA}).


\subsection{Symplecticity of the geometric integrators}

\label{sec:num_symplectic_structure}

The geometric integrators designed for the VGA preserve the symplectic
structure of the Gaussian wavepackets.~\cite{book_Hairer_Wanner:2006} To
verify this analytically, one should show that the Jacobian $\Phi_{t}^{\prime
}=\Phi_{t}^{\prime}(z_{0})$ of the flow $z_{t}=\Phi_{t}(z_{0})$ of an
integrator satisfies the relation~\cite{book_Hairer_Wanner:2006}
\begin{equation}
\Phi_{t}^{\prime}(z_{0})^{T}\cdot B(z_{t})\cdot\Phi_{t}^{\prime}%
(z_{0})=B(z_{0}).
\label{eq:Hairer}
\end{equation}
[Note that matrix $B$ in Eq.~(\ref{eq:Hairer}) is the inverse of matrix $B$ appearing in Eq.(4.2) of Chapter VII of Ref.~\onlinecite{book_Hairer_Wanner:2006}).]
The flow of a symplectic integrator is composed of a sequence of kinetic and
potential flows, and thus its Jacobian is equal to the matrix product of the
Jacobians of the elementary flows. The kinetic flow
equations~(\ref{eq:Tstep_AEOM}) and (\ref{eq:Tstep_gamma_EOM}) contain the
inverse and determinant of the complex matrix $A_{t}$, and thus it is very
difficult to decouple these equations into separate equations for the real and
imaginary parts of $A_{t}$ and $\gamma_{t}$. It is much simpler to find the
Jacobian of the flow in Hagedorn's parametrization. Ohsawa found the reduced
symplectic structure which corresponds to the Gaussian wavepacket without a
phase in the equivalent manifold with phase symmetry.~\cite{Ohsawa:2015a} For
simplicity, we choose the reduced symplectic form, which is the constant
block-diagonal matrix
\begin{equation}
B(z_{t})=%
\begin{pmatrix}
J_{2D} & 0 & 0\\
0 & \frac{\hbar}{2}J_{2D^{2}} & 0\\
0 & 0 & \frac{\hbar}{2}J_{2D^{2}}%
\end{pmatrix}
,\label{eq:HOmega}%
\end{equation}
with
\begin{equation}
z_{t}:=\bigg(q_{t}^{T},p_{t}^{T},\widetilde{,Q_{t}^{(1)}}^{T},\widetilde{P_{t}%
^{(1)}}^{T},\widetilde{Q_{t}^{(2)}}^{T},\widetilde{P_{t}^{(2)}}^{T}%
\bigg)^{T}\in\mathbb{R}^{2D+4D^{2}},\label{eq:z_t}%
\end{equation}
where $\widetilde{\Lambda}$, which is used for $\Lambda=Q_{t}^{(1)}%
,P_{t}^{(1)},Q_{t}^{(2)}$, and $P_{t}^{(2)}$ is a $D^{2}$-dimensional vector
containing elements of the $D\times D$ matrix $\Lambda$ in a column-wise
manner, i.e., $\widetilde{\Lambda}_{j+D(k-1)}=\Lambda_{jk}$. Also,
$J_{2D}=J\otimes I_{D}$, where $J$ is the two-dimensional symplectic
matrix~(\ref{eq:J}), and $J_{2D^{2}}=J_{2D}\otimes I_{D}$.
Equations~(\ref{eq:Hf_T_qEOM})-(\ref{eq:Hf_T_PEOM}) yield the Jacobian of the
kinetic flow $\Phi_{\text{T},t}$, which is the block-diagonal matrix
\begin{equation}
\Phi_{\text{T},t}^{\prime}(z_{0})=%
\begin{pmatrix}
M_{2D} & 0 & 0\\
0 & M_{2D^{2}} & 0\\
0 & 0 & M_{2D^{2}}%
\end{pmatrix}
,\label{eq:HJT}%
\end{equation}
where
\begin{equation}
M_{2D}=%
\begin{pmatrix}
I_{D} & t\,m^{-1}\\
0 & I_{D}%
\end{pmatrix}
\label{eq:Stability}%
\end{equation}
is the stability matrix and $M_{2D^{2}}=M_{2D}\otimes I_{D}$. The kinetic flow
with Jacobian~(\ref{eq:HJT}) is symplectic, i.e., $\Phi_{\text{T},t}^{\prime
}(z_{0})^{T}\cdot B(z_{t})\cdot\Phi_{\text{T},t}^{\prime}(z_{0})=B(z_{0}),$
because
\begin{equation}
M_{2D}^{T}\cdot J_{2D}\cdot M_{2D}=%
\begin{pmatrix}
I_{D} & 0\\
t\,m^{-1} & I_{D}%
\end{pmatrix}%
\begin{pmatrix}
0 & -I_{D}\\
I_{D} & t\,m^{-1}%
\end{pmatrix}
=J_{2D},
\end{equation}
and using matrix and tensor multiplication
\begin{align}
M_{2D^{2}}^{T} &  \cdot J_{2D^{2}}\cdot M_{2D^{2}}\nonumber\\
&  =\big(M_{2D}^{T}\otimes I_{D}\big)\cdot\big(J_{2D}\otimes I_{D}%
\big)\cdot\big(M_{2D}\otimes I_{D}\big)\nonumber\\
&  =\big(M_{2D}^{T}\cdot J_{2D}\cdot M_{2D}\big)\otimes I_{D}=J_{2D^{2}%
}.\label{eq:Stability_is_symplectic_2}%
\end{align}
Similarly, Eqs. (\ref{eq:Hf_v_qEOM})-(\ref{eq:Hf_v_PEOM}) imply that the
Jacobian of the potential flow $\Phi_{\text{V},t}$ is the $(2D+4D^{2}%
)$-dimensional matrix
\begin{equation}
\Phi_{\text{V},t}^{\prime}(z_{0})=I_{2D+4D^{2}}-t%
\begin{pmatrix}
0 & 0 & 0 & 0 & 0 & 0\\
a & 0 & b^{(1)} & 0 & b^{(2)} & 0\\
0 & 0 & 0 & 0 & 0 & 0\\
c^{(1)} & 0 & d^{(11)} & 0 & d^{(12)} & 0\\
0 & 0 & 0 & 0 & 0 & 0\\
c^{(2)} & 0 & d^{(21)} & 0 & d^{(22)} & 0
\end{pmatrix},
\label{eq:HJV}%
\end{equation}
where
\begin{align}
a_{j,k} &  =\langle\hat{V}^{\prime\prime}\rangle_{jk},\label{eq:a}\\
b_{j,D(k-1)+l}^{(r)} &  =(\hbar/2)\,\langle\hat{V}^{\prime\prime\prime}%
\rangle_{jkm}\,Q_{t,ml}^{(r)},\label{eq:b}\\
c_{D(j-1)+k,l}^{(r)} &  =\langle\hat{V}^{\prime\prime\prime}\rangle
_{jml}\,Q_{t,mk}^{(r)},\label{eq:c}\\
d_{D(j-1)+k,\,D(l-1)+m}^{(rs)} &  =(\hbar/2)\,\langle\hat{V}^{(4)}%
\rangle_{jnlp}\,Q_{t,nk}^{(r)}\,Q_{t,pm}^{(s)}\nonumber\\
&  \quad+\langle\hat{V}^{\prime\prime}\rangle_{jn}\,\delta_{nl}\,\delta
_{km}\,\delta_{rs}\label{eq:d}%
\end{align}
specify the components of the $D\times D$ matrix $a$, $D\times D^{2}$ matrix
$b^{(r)}$, $D^{2}\times D$ matrix $c^{(r)}$, and $D^{2}\times D^{2}$ matrix
$d^{(rs)}$ for all $r,s\in\{1,2\}$. In Eqs.~(\ref{eq:b})-(\ref{eq:d}),
$\hat{V}^{\prime\prime\prime}:=V^{\prime\prime\prime}(q)|_{q=\hat{q}}$ and
$\hat{V}^{(4)}:=V^{(4)}(q)|_{q=\hat{q}}$. It is easy to show that the
potential flow with Jacobian~(\ref{eq:HJV}) is symplectic, i.e.,
$\Phi_{\text{V},t}^{\prime}(z_{0})^{T}\cdot B(z_{t})\cdot\Phi_{\text{V}%
,t}^{\prime}(z_{0})=B(z_{0})$, if and only if
\begin{align}
a^{T} &  =a,\label{eq:condition_1}\\
\big(b^{(r)}\big)^{T} &  =(\hbar/2)\,c^{(r)},\label{eq:condition_2}\\
\big(d^{(rs)}\big)^{T} &  =d^{(sr)}.\label{eq:condition_3}%
\end{align}
Because $\langle\hat{V}^{\prime\prime}\rangle$, $\langle\hat{V}^{\prime
\prime\prime}\rangle$, and $\langle\hat{V}^{(4)}\rangle$ are totally
symmetric, Eqs. (\ref{eq:a})-(\ref{eq:d}) for $a$, $b^{(r)}$, $c^{(r)}$ and
$d^{(rs)}$ imply that conditions~(\ref{eq:condition_1})-(\ref{eq:condition_3})
hold for the Jacobian~(\ref{eq:HJV}), and thus the
potential flow is symplectic. Since both kinetic and potential flows are
symplectic, any composition of them is also symplectic. This proves the
conservation of the symplectic structure by the geometric integrators. In Sec.\ref{subsec:Coupled_Morse}, we verified their symplecticity numerically by measuring the accuracy with which Eq.~(\ref{eq:Hairer}) is satisfied if $\Phi_{t}(z_{0})$ denotes the composed flow consisting many steps, each of which, in turn, is composed by several potential and kinetic substeps. For that, the Jacobian $\Phi^{\prime}_{t}(z_{0})$ of the composed flow appearing in Eq.~(\ref{eq:Hairer}) was obtained by matrix multiplication of the Jacobians~(\ref{eq:HJT}) and~(\ref{eq:HJV}) of all kinetic and potential steps.


\section{Expectation values in a Gaussian wavepacket}

\label{sec:exp_value_pot}

In this appendix, we derive the expectation values in a Gaussian wavepacket of
the potential energy, gradient, and Hessian of the multi-dimensional quartic
and coupled Morse potentials.

\subsection{Quartic potential and its derivatives}

\label{subsec:Quartic}

An important class of potentials consists of the quartic polynomials, which
have applications in many areas of research, including molecular force-field
design~\cite{Linde_Hase:1990,Yagi_Gordon:2004} and tunneling
theory.~\cite{Mandelli_Ceotto:2022,Masoumi_Wachter:2017} We consider the most
general form of a $D$-dimensional quartic potential
\begin{align}
V(q)  &  =V(q_{\text{ref}}) + V^{\prime}(q_{\text{ref}})^{T} \cdot x +x^{T}
\cdot V^{\prime\prime}(q_{\text{ref}}) \cdot x/2\nonumber\\
&  \quad+ V^{\prime\prime\prime}(q_{\text{ref}})_{ijk} \, x_{i} \, x_{j} \,
x_{k}/3!\nonumber\\
&  \quad+ V^{(4)}(q_{\text{ref}})_{ijkl} \, x_{i} \, x_{j} \, x_{k} \,
x_{l}/4!, \label{eq:Quartic}%
\end{align}
where the scalar $V(q_{\text{ref}})$, vector $V^{\prime}(q_{\text{ref}})$,
matrix $V^{\prime\prime}(q_{\text{ref}})$, rank-$3$ tensor $V^{\prime
\prime\prime}(q_{\text{ref}})$, and rank-$4$ tensor $V^{(4)}(q_{\text{ref}})$
are the potential energy and its first four derivatives at a reference
position $q_{\text{ref}}$. Here, $x:=q-q_{\text{ref}}$ is a $D$-dimensional
vector with elements $x_{i}$. The gradient and Hessian of the quartic
potential are
\begin{align}
V^{\prime}(q)_{i}  &  =V^{\prime}(q_{\text{ref}})_{i} + V^{\prime\prime
}(q_{\text{ref}})_{ij} \, x_{j}\nonumber\\
&  \quad+V^{\prime\prime\prime}(q_{\text{ref}})_{ijk} \, x_{j} \,
x_{k}/2\nonumber\\
&  \quad+ V^{(4)}(q_{\text{ref}})_{ijkl} \, x_{j} \, x_{k} \, x_{l}%
/3!,\label{eq:QuarticGrad}\\
V^{\prime\prime}(q)_{ij}  &  = V^{\prime\prime}(q_{\text{ref}})_{ij}+
V^{\prime\prime\prime}(q_{\text{ref}})_{ijk} \, x_{k}\nonumber\\
&  \quad+ V^{(4)}(q_{\text{ref}})_{ijkl} \, x_{k} \, x_{l}/2.
\label{eq:QuarticHess}%
\end{align}

Using Eqs.~(\ref{eq:Covq}) and~(\ref{eq:Isserli}) and the fact that
$V^{\prime\prime}(q_{\text{eq}})$, $V^{\prime\prime\prime}(q_{\text{eq}})$,
and $V^{(4)}(q_{\text{eq}})$ are totally symmetric tensors, the expectation
values of the quartic potential and its first two derivatives in a Gaussian
wavepacket are
\begin{align}
\langle\hat{V} \rangle &  =V(q_{\text{ref}}) +\text{Tr}[V^{\prime\prime
}(q_{\text{ref}}) \cdot\Sigma_{t}] /2\nonumber\\
&  \quad+ V^{(4)}(q_{\text{ref}})_{ijkl} \, \Sigma_{t,ij}\, \Sigma
_{t,kl}/8,\label{eq:QuarticExp}\\
\langle\hat{V}^{\prime}\rangle_{i}  &  =V^{\prime}(q_{\text{ref}})_{i} +
V^{\prime\prime\prime}(q_{\text{ref}})_{ijk}\, \Sigma_{t,jk}%
/2,\label{eq:QuarticGradExp}\\
\langle\hat{V}^{\prime\prime}\rangle_{ij}  &  = V^{\prime\prime}%
(q_{\text{ref}})_{ij}+ V^{(4)}(q_{\text{ref}})_{ijkl} \, \Sigma_{t,kl}/2.
\label{eq:QuarticHessExp}%
\end{align}


\subsection{Coupled Morse potential and its derivatives}

\label{subsec:Morse}

The coupled Morse potential~(\ref{eq:Coupled-Morse}) is introduced in the main
text. The expectation values of this potential and its first four derivatives
are
\begin{align}
\langle\hat{V}\rangle &  =V_{\text{eq}}+\sum_{j=1}^{D}\langle V_{j}(\hat
{q}_{j})\rangle+\langle V_{\text{cpl}}(\hat{q})\rangle
,\label{eq:exp_value_coupled_Morse_pot}\\
\langle\hat{V}^{(k)}\rangle &  =\sum_{j=1}^{D}\langle V_{j}^{(k)}(\hat{q}%
_{j})\rangle+\langle V_{\text{cpl}}^{(k)}(\hat{q})\rangle,
\label{eq:exp_value_coupled_Morse_fourth-der}%
\end{align}
where $k\in\{1,2,3,4\}$,
\begin{align}
V_{j}^{(k)}(q_{j}) = (-1)^{k-1} 2\,d_{e}^{\prime}\big(a_{j}^{\prime}\big)^{k}
\,\big[y(a_{j}^{\prime},q_{j})-2^{k-1}y(a_{j}^{\prime},q_{j})^{2}
\big]
\label{eq:Der_Morse_1D}
\end{align}
is the $k$th derivative of the one-dimensional Morse
potential~(\ref{eq:1D_Morse}), and
\begin{align}
V_{\text{cpl}}^{(k)}(q)_{l_{1}\dots l_{k}} &= (-1)^{k-1}\,2\,d_{e}\,a_{l_{1}} \dots a_{l_{k}}\,\notag\\& \qquad \big[y(a,q)-2^{k-1}\,y(a,q)^{2}\big]
\label{eq:Der_cpl-Morse}
\end{align}
is the $k$th derivatives of the $D$-dimensional coupling
term~(\ref{eq:cpl_Morse}).

Expectation value of the Morse variable~(\ref{eq:Morse_var}) in the Gaussian
wavepacket can be evaluated analytically as
\begin{align}
\langle y(a,q)\rangle &  =\int e^{-a^{T}\cdot(q-q_{\text{eq}})}\,\rho
_{t}(q)\,d^{D}q\nonumber\\
&  =\big[\text{det}\big({2\pi\,\Sigma_{t}^{-1}}\big)\big]^{-1/2}%
\,e^{-a^{T}\cdot(q_{t}-q_{\text{eq}})}\nonumber\\
&  \quad\times\int e^{-(q-q_{t})^{T}\cdot\Sigma_{t}^{-1}\cdot(q-q_{t}%
)/2-a^{T}\cdot(q-q_{\text{t}})}\,d^{D}q\nonumber\\
&  =e^{-a^{T}\cdot(q_{t}-q_{\text{eq}})+a^{T}\cdot\Sigma_{t}\cdot
a/2}\nonumber\\
&  =y(a,q_{t})\,z(a),\label{eq:MorseVarIntegral}%
\end{align}
where we defined the \textquotedblleft Morse Gaussian
variable\textquotedblright%
\begin{equation}
z(a):=\exp\big(a^{T}\cdot\Sigma_{t}\cdot a/2\big).
\end{equation}
Likewise,
\begin{equation}
\langle y(a,q)^{2}\rangle=y(a,q_{t})^{2}\,z(a)^{4}%
.\label{eq:MorseVar2Integral}%
\end{equation}
To find the expectation value of a one-dimensional Morse variable $y_{j}%
(a_{j}^{\prime},q_{j})$ in a multi-dimensional Gaussian wavepacket, we
considered an auxiliary vector $\tilde{a}$ with $D$ components $\tilde{a}%
_{k}=a_{j}^{\prime}\,\delta_{jk}$. Noting that
\begin{equation}
y_{j}(a_{j}^{\prime},q_{j})=e^{-a_{j}^{\prime}\,(q_{j}-q_{\text{eq},j}%
)}=e^{-\tilde{a}\cdot(q-q_{\text{eq}})}=y(\tilde{a},q),
\end{equation}
defining a one-dimensional Morse Gaussian variable
\begin{equation}
z_{j}(a_{j}^{\prime}):=e^{\Sigma_{t,jj}\,{a_{j}^{\prime}}^{2}/2}=e^{\tilde
{a}^{T}\cdot\Sigma_{t}\cdot\tilde{a}/2}=z(\tilde{a}),
\end{equation}
and using the results~(\ref{eq:MorseVarIntegral})
and~(\ref{eq:MorseVar2Integral}), we find that
\begin{align}
\langle y_{j}(a_{j}^{\prime},q_{j})\rangle &  =\langle y(\tilde{a}%
,q)\rangle=y(\tilde{a},q_{t})\,z(\tilde{a})\nonumber\\
&  =y_{j}(a_{j}^{\prime},q_{t,j})\,z_{j}(a_{j}^{\prime}%
),\label{eq:1D_MorseVarIntegral}\\
\langle y_{j}(a_{j}^{\prime},q_{j})^{2}\rangle &  =\langle y(\tilde{a}%
,q)^{2}\rangle=y(\tilde{a},q_{t})^{2}\,z(\tilde{a})^{4}\nonumber\\
&  =y_{j}(a_{j}^{\prime},q_{t,j})^{2}\,z_{j}(a_{j}^{\prime})^{4}%
.\label{eq:1D_MorseVar2Integral}%
\end{align}
From Eqs.~(\ref{eq:1D_MorseVarIntegral}) and (\ref{eq:1D_MorseVar2Integral}),
we obtain expectation values of the one-dimensional Morse
potential~(\ref{eq:1D_Morse}) and its derivative~(\ref{eq:Der_Morse_1D}):
\begin{align}
\langle V_{j}(\hat{q}_{j})\rangle &  =d_{e}^{\prime}\,(1-2\,m_{j}%
+n_{j}),\label{eq:1D_Morse_exp_value}\\
\langle V_{j}^{(k)}(\hat{q}_{j})\rangle &  = (-1)^{k-1} 2 \,d_{e}^{\prime}\,\big(a_{j}^{\prime}\big)^{k} \,(m_{j}-2^{k-1}\,\,n_{j}), 
\label{eq:Der_1D_Morse_exp_value}
\end{align}
where we introduced $m_{j}:=y_{j}(a_{j}^{\prime},q_{t,j})\,z_{j}(a_{j}%
^{\prime})$ and $n_{j}:=y_{j}(a_{j}^{\prime},q_{t,j})^{2}\,z_{j}(a_{j}%
^{\prime})^{4}$ to simplify the notation. Similarly,
Eqs.(\ref{eq:MorseVarIntegral}) and (\ref{eq:MorseVar2Integral}) give
expectation values of the coupling term~(\ref{eq:cpl_Morse}) and its
derivative~(\ref{eq:Der_cpl-Morse}):
\begin{align}
\langle V_{\text{cpl}}(\hat{q})\rangle &  =d_{e}%
\,(1-2\,M+N),\label{eq:cpl_Morse_exp_value}\\
 \langle V_{\text{cpl}}^{(k)}(\hat{q}) \rangle_{l_{1}\dots l_{k}}  &  = (-1)^{k-1}\,2\,d_{e}\,a_{l_{1}} \dots a_{l_{k}}\,(M-2^{k-1}N),
\label{eq:Der_cpl_Morse_exp_value}
\end{align}
where $M:=y(a,q_{t})\,z(a)$ and $N:=y(a,q_{t})^{2}\,z(a)^{4}$.


\bibliographystyle{aipnum4-2}
\bibliography{VGA_high-order}

\begin{thebibliography}{88}%
\makeatletter
\providecommand \@ifxundefined [1]{%
 \@ifx{#1\undefined}
}%
\providecommand \@ifnum [1]{%
 \ifnum #1\expandafter \@firstoftwo
 \else \expandafter \@secondoftwo
 \fi
}%
\providecommand \@ifx [1]{%
 \ifx #1\expandafter \@firstoftwo
 \else \expandafter \@secondoftwo
 \fi
}%
\providecommand \natexlab [1]{#1}%
\providecommand \enquote  [1]{``#1''}%
\providecommand \bibnamefont  [1]{#1}%
\providecommand \bibfnamefont [1]{#1}%
\providecommand \citenamefont [1]{#1}%
\providecommand \href@noop [0]{\@secondoftwo}%
\providecommand \href [0]{\begingroup \@sanitize@url \@href}%
\providecommand \@href[1]{\@@startlink{#1}\@@href}%
\providecommand \@@href[1]{\endgroup#1\@@endlink}%
\providecommand \@sanitize@url [0]{\catcode `\\12\catcode `\$12\catcode
  `\&12\catcode `\#12\catcode `\^12\catcode `\_12\catcode `\%12\relax}%
\providecommand \@@startlink[1]{}%
\providecommand \@@endlink[0]{}%
\providecommand \url  [0]{\begingroup\@sanitize@url \@url }%
\providecommand \@url [1]{\endgroup\@href {#1}{\urlprefix }}%
\providecommand \urlprefix  [0]{URL }%
\providecommand \Eprint [0]{\href }%
\providecommand \doibase [0]{https://doi.org/}%
\providecommand \selectlanguage [0]{\@gobble}%
\providecommand \bibinfo  [0]{\@secondoftwo}%
\providecommand \bibfield  [0]{\@secondoftwo}%
\providecommand \translation [1]{[#1]}%
\providecommand \BibitemOpen [0]{}%
\providecommand \bibitemStop [0]{}%
\providecommand \bibitemNoStop [0]{.\EOS\space}%
\providecommand \EOS [0]{\spacefactor3000\relax}%
\providecommand \BibitemShut  [1]{\csname bibitem#1\endcsname}%
\let\auto@bib@innerbib\@empty
\bibitem [{\citenamefont {Nitzan}(2006)}]{book_Nitzan:2006}%
  \BibitemOpen
  \bibfield  {author} {\bibinfo {author} {\bibfnamefont {A.}~\bibnamefont
  {Nitzan}},\ }\href@noop {} {\emph {\bibinfo {title} {Chemical dynamics in
  condensed phases: relaxation, transfer and reactions in condensed molecular
  systems}}}\ (\bibinfo  {publisher} {Oxford university press},\ \bibinfo
  {year} {2006})\BibitemShut {NoStop}%
\bibitem [{\citenamefont {Tannor}(2007)}]{book_Tannor:2007}%
  \BibitemOpen
  \bibfield  {author} {\bibinfo {author} {\bibfnamefont {D.~J.}\ \bibnamefont
  {Tannor}},\ }\href@noop {} {\emph {\bibinfo {title} {Introduction to Quantum
  Mechanics: A Time-Dependent Perspective}}}\ (\bibinfo  {publisher}
  {University Science Books},\ \bibinfo {address} {Sausalito},\ \bibinfo {year}
  {2007})\BibitemShut {NoStop}%
\bibitem [{\citenamefont {Heller}(2018)}]{book_Heller:2018}%
  \BibitemOpen
  \bibfield  {author} {\bibinfo {author} {\bibfnamefont {E.~J.}\ \bibnamefont
  {Heller}},\ }\href@noop {} {\emph {\bibinfo {title} {The semiclassical way to
  dynamics and spectroscopy}}}\ (\bibinfo  {publisher} {Princeton University
  Press},\ \bibinfo {address} {Princeton, NJ},\ \bibinfo {year}
  {2018})\BibitemShut {NoStop}%
\bibitem [{\citenamefont {Pattanayak}\ and\ \citenamefont
  {Schieve}(1994)}]{Pattanayak_Schieve:1994}%
  \BibitemOpen
  \bibfield  {author} {\bibinfo {author} {\bibfnamefont {A.~K.}\ \bibnamefont
  {Pattanayak}}\ and\ \bibinfo {author} {\bibfnamefont {W.~C.}\ \bibnamefont
  {Schieve}},\ }\href {https://doi.org/10.1103/PhysRevE.50.3601} {\bibfield
  {journal} {\bibinfo  {journal} {Phys.\ Rev.~E}\ }\textbf {\bibinfo {volume}
  {50}},\ \bibinfo {pages} {3601} (\bibinfo {year} {1994})}\BibitemShut
  {NoStop}%
\bibitem [{\citenamefont {Pattanayak}\ and\ \citenamefont
  {Schieve}(1997)}]{Pattanayak_Schieve:1997}%
  \BibitemOpen
  \bibfield  {author} {\bibinfo {author} {\bibfnamefont {A.~K.}\ \bibnamefont
  {Pattanayak}}\ and\ \bibinfo {author} {\bibfnamefont {W.~C.}\ \bibnamefont
  {Schieve}},\ }\href {https://doi.org/10.1103/PhysRevE.56.278} {\bibfield
  {journal} {\bibinfo  {journal} {Phys.\ Rev.~E}\ }\textbf {\bibinfo {volume}
  {56}},\ \bibinfo {pages} {278} (\bibinfo {year} {1997})}\BibitemShut
  {NoStop}%
\bibitem [{\citenamefont {Lubich}(2008)}]{book_Lubich:2008}%
  \BibitemOpen
  \bibfield  {author} {\bibinfo {author} {\bibfnamefont {C.}~\bibnamefont
  {Lubich}},\ }\href@noop {} {\emph {\bibinfo {title} {From Quantum to
  Classical Molecular Dynamics: Reduced Models and Numerical Analysis}}},\
  \bibinfo {edition} {12th}\ ed.\ (\bibinfo  {publisher} {European Mathematical
  Society},\ \bibinfo {address} {Z\"{u}rich},\ \bibinfo {year}
  {2008})\BibitemShut {NoStop}%
\bibitem [{\citenamefont {Faou}, \citenamefont {Gradinaru},\ and\ \citenamefont
  {Lubich}(2009)}]{Faou_Lubich:2009_v2}%
  \BibitemOpen
  \bibfield  {author} {\bibinfo {author} {\bibfnamefont {E.}~\bibnamefont
  {Faou}}, \bibinfo {author} {\bibfnamefont {V.}~\bibnamefont {Gradinaru}},\
  and\ \bibinfo {author} {\bibfnamefont {C.}~\bibnamefont {Lubich}},\ }\href
  {https://doi.org/10.1137/080729724} {\bibfield  {journal} {\bibinfo
  {journal} {SIAM J.\ Sci.\ Comp.}\ }\textbf {\bibinfo {volume} {31}},\
  \bibinfo {pages} {3027} (\bibinfo {year} {2009})}\BibitemShut {NoStop}%
\bibitem [{\citenamefont {Richings}\ \emph {et~al.}(2015)\citenamefont
  {Richings}, \citenamefont {Polyak}, \citenamefont {Spinlove}, \citenamefont
  {Worth}, \citenamefont {Burghardt},\ and\ \citenamefont
  {Lasorne}}]{Richings_Lasorne:2015}%
  \BibitemOpen
  \bibfield  {author} {\bibinfo {author} {\bibfnamefont {G.~W.}\ \bibnamefont
  {Richings}}, \bibinfo {author} {\bibfnamefont {I.}~\bibnamefont {Polyak}},
  \bibinfo {author} {\bibfnamefont {K.~E.}\ \bibnamefont {Spinlove}}, \bibinfo
  {author} {\bibfnamefont {G.~A.}\ \bibnamefont {Worth}}, \bibinfo {author}
  {\bibfnamefont {I.}~\bibnamefont {Burghardt}},\ and\ \bibinfo {author}
  {\bibfnamefont {B.}~\bibnamefont {Lasorne}},\ }\href
  {https://doi.org/10.1080/0144235X.2015.1051354} {\bibfield  {journal}
  {\bibinfo  {journal} {Int. \ Rev. \ Phys. \ Chem.}\ }\textbf {\bibinfo
  {volume} {34}},\ \bibinfo {pages} {269} (\bibinfo {year} {2015})}\BibitemShut
  {NoStop}%
\bibitem [{\citenamefont {Garashchuk}(2021)}]{garashchuk:2021}%
  \BibitemOpen
  \bibfield  {author} {\bibinfo {author} {\bibfnamefont {S.}~\bibnamefont
  {Garashchuk}},\ }in\ \href@noop {} {\emph {\bibinfo {booktitle} {Basis Sets
  in Computational Chemistry}}}\ (\bibinfo  {publisher} {Springer},\ \bibinfo
  {year} {2021})\ pp.\ \bibinfo {pages} {215--252}\BibitemShut {NoStop}%
\bibitem [{\citenamefont {Edery}(2021)}]{Edery:2021}%
  \BibitemOpen
  \bibfield  {author} {\bibinfo {author} {\bibfnamefont {A.}~\bibnamefont
  {Edery}},\ }\href {https://doi.org/10.1103/PhysRevD.104.125015} {\bibfield
  {journal} {\bibinfo  {journal} {Phys.\ Rev.~D}\ }\textbf {\bibinfo {volume}
  {104}},\ \bibinfo {pages} {125015} (\bibinfo {year} {2021})}\BibitemShut
  {NoStop}%
\bibitem [{\citenamefont {Heller}(1975)}]{Heller:1975}%
  \BibitemOpen
  \bibfield  {author} {\bibinfo {author} {\bibfnamefont {E.~J.}\ \bibnamefont
  {Heller}},\ }\href {https://doi.org/10.1063/1.430620} {\bibfield  {journal}
  {\bibinfo  {journal} {J.~Chem.\ Phys.}\ }\textbf {\bibinfo {volume} {62}},\
  \bibinfo {pages} {1544} (\bibinfo {year} {1975})}\BibitemShut {NoStop}%
\bibitem [{\citenamefont {Heller}(1976{\natexlab{a}})}]{Heller:1976a}%
  \BibitemOpen
  \bibfield  {author} {\bibinfo {author} {\bibfnamefont {E.~J.}\ \bibnamefont
  {Heller}},\ }\href {https://doi.org/10.1063/1.432974} {\bibfield  {journal}
  {\bibinfo  {journal} {J.~Chem.\ Phys.}\ }\textbf {\bibinfo {volume} {65}},\
  \bibinfo {pages} {4979} (\bibinfo {year} {1976}{\natexlab{a}})}\BibitemShut
  {NoStop}%
\bibitem [{\citenamefont {Heller}(1981{\natexlab{a}})}]{Heller:1981}%
  \BibitemOpen
  \bibfield  {author} {\bibinfo {author} {\bibfnamefont {E.~J.}\ \bibnamefont
  {Heller}},\ }\href {https://doi.org/10.1063/1.442382} {\bibfield  {journal}
  {\bibinfo  {journal} {J.~Chem.\ Phys.}\ }\textbf {\bibinfo {volume} {75}},\
  \bibinfo {pages} {2923} (\bibinfo {year} {1981}{\natexlab{a}})}\BibitemShut
  {NoStop}%
\bibitem [{\citenamefont {Hagedorn}(1980)}]{Hagedorn:1980_v2}%
  \BibitemOpen
  \bibfield  {author} {\bibinfo {author} {\bibfnamefont {G.~A.}\ \bibnamefont
  {Hagedorn}},\ }\href {https://doi.org/10.1007/BF01230088} {\bibfield
  {journal} {\bibinfo  {journal} {Commun.\ Math.\ Phys.}\ }\textbf {\bibinfo
  {volume} {71}},\ \bibinfo {pages} {77} (\bibinfo {year} {1980})}\BibitemShut
  {NoStop}%
\bibitem [{\citenamefont {Hagedorn}(1998)}]{Hagedorn:1998}%
  \BibitemOpen
  \bibfield  {author} {\bibinfo {author} {\bibfnamefont {G.~A.}\ \bibnamefont
  {Hagedorn}},\ }\href {https://doi.org/10.1006/aphy.1998.5843} {\bibfield
  {journal} {\bibinfo  {journal} {Ann.\ Phys.\ (NY)}\ }\textbf {\bibinfo
  {volume} {269}},\ \bibinfo {pages} {77} (\bibinfo {year} {1998})}\BibitemShut
  {NoStop}%
\bibitem [{\citenamefont {Lasorne}\ \emph {et~al.}(2006)\citenamefont
  {Lasorne}, \citenamefont {Bearpark}, \citenamefont {Robb},\ and\
  \citenamefont {Worth}}]{Lasorne_Worth:2006}%
  \BibitemOpen
  \bibfield  {author} {\bibinfo {author} {\bibfnamefont {B.}~\bibnamefont
  {Lasorne}}, \bibinfo {author} {\bibfnamefont {M.~J.}\ \bibnamefont
  {Bearpark}}, \bibinfo {author} {\bibfnamefont {M.~A.}\ \bibnamefont {Robb}},\
  and\ \bibinfo {author} {\bibfnamefont {G.~A.}\ \bibnamefont {Worth}},\ }\href
  {https://doi.org/https://doi.org/10.1016/j.cplett.2006.10.099} {\bibfield
  {journal} {\bibinfo  {journal} {Chem.\ Phys.\ Lett.}\ }\textbf {\bibinfo
  {volume} {432}},\ \bibinfo {pages} {604 } (\bibinfo {year}
  {2006})}\BibitemShut {NoStop}%
\bibitem [{\citenamefont {Ben-Nun}\ and\ \citenamefont
  {Mart\'{\i}nez}(1998)}]{Ben-Nun_Martinez:1998a}%
  \BibitemOpen
  \bibfield  {author} {\bibinfo {author} {\bibfnamefont {M.}~\bibnamefont
  {Ben-Nun}}\ and\ \bibinfo {author} {\bibfnamefont {T.~J.}\ \bibnamefont
  {Mart\'{\i}nez}},\ }\href
  {https://doi.org/https://doi.org/10.1016/S0009-2614(98)01115-4} {\bibfield
  {journal} {\bibinfo  {journal} {Chem.\ Phys.\ Lett.}\ }\textbf {\bibinfo
  {volume} {298}},\ \bibinfo {pages} {57} (\bibinfo {year} {1998})}\BibitemShut
  {NoStop}%
\bibitem [{\citenamefont {Wehrle}, \citenamefont {\v{S}ulc},\ and\
  \citenamefont {Van\'{i}\v{c}ek}(2014)}]{Wehrle_Vanicek:2014_v2}%
  \BibitemOpen
  \bibfield  {author} {\bibinfo {author} {\bibfnamefont {M.}~\bibnamefont
  {Wehrle}}, \bibinfo {author} {\bibfnamefont {M.}~\bibnamefont {\v{S}ulc}},\
  and\ \bibinfo {author} {\bibfnamefont {J.~J.~L.}\ \bibnamefont
  {Van\'{i}\v{c}ek}},\ }\href {https://doi.org/10.1063/1.4884718} {\bibfield
  {journal} {\bibinfo  {journal} {J.~Chem.\ Phys.}\ }\textbf {\bibinfo {volume}
  {140}},\ \bibinfo {pages} {244114} (\bibinfo {year} {2014})}\BibitemShut
  {NoStop}%
\bibitem [{\citenamefont {Faou}\ and\ \citenamefont
  {Lubich}(2006)}]{Faou_Lubich:2006}%
  \BibitemOpen
  \bibfield  {author} {\bibinfo {author} {\bibfnamefont {E.}~\bibnamefont
  {Faou}}\ and\ \bibinfo {author} {\bibfnamefont {C.}~\bibnamefont {Lubich}},\
  }\href {https://doi.org/10.1007/s00791-006-0019-8} {\bibfield  {journal}
  {\bibinfo  {journal} {Comput.\ Visual.\ Sci.}\ }\textbf {\bibinfo {volume}
  {9}},\ \bibinfo {pages} {45} (\bibinfo {year} {2006})}\BibitemShut {NoStop}%
\bibitem [{\citenamefont {Ohsawa}(2015)}]{Ohsawa:2015a}%
  \BibitemOpen
  \bibfield  {author} {\bibinfo {author} {\bibfnamefont {T.}~\bibnamefont
  {Ohsawa}},\ }\href {https://doi.org/10.1007/s11005-015-0780-z} {\bibfield
  {journal} {\bibinfo  {journal} {Lett.\ Math.\ Phys.}\ }\textbf {\bibinfo
  {volume} {105}},\ \bibinfo {pages} {1301} (\bibinfo {year}
  {2015})}\BibitemShut {NoStop}%
\bibitem [{\citenamefont {Ohsawa}\ and\ \citenamefont
  {Leok}(2013)}]{Ohsawa_Leok:2013}%
  \BibitemOpen
  \bibfield  {author} {\bibinfo {author} {\bibfnamefont {T.}~\bibnamefont
  {Ohsawa}}\ and\ \bibinfo {author} {\bibfnamefont {M.}~\bibnamefont {Leok}},\
  }\href {https://doi.org/10.1088/1751-8113/46/40/405201} {\bibfield  {journal}
  {\bibinfo  {journal} {J.~Phys.~A}\ }\textbf {\bibinfo {volume} {46}},\
  \bibinfo {pages} {405201} (\bibinfo {year} {2013})}\BibitemShut {NoStop}%
\bibitem [{\citenamefont {Mart\'{\i}nez}, \citenamefont {Ben-Nun},\ and\
  \citenamefont {Levine}(1996)}]{Martinez_Levine:1996a}%
  \BibitemOpen
  \bibfield  {author} {\bibinfo {author} {\bibfnamefont {T.~J.}\ \bibnamefont
  {Mart\'{\i}nez}}, \bibinfo {author} {\bibfnamefont {M.}~\bibnamefont
  {Ben-Nun}},\ and\ \bibinfo {author} {\bibfnamefont {R.~D.}\ \bibnamefont
  {Levine}},\ }\href {https://doi.org/10.1021/jp953105a} {\bibfield  {journal}
  {\bibinfo  {journal} {J.~Phys.~C}\ }\textbf {\bibinfo {volume} {100}},\
  \bibinfo {pages} {7884} (\bibinfo {year} {1996})}\BibitemShut {NoStop}%
\bibitem [{\citenamefont {Ben-Nun}, \citenamefont {Quenneville},\ and\
  \citenamefont {Mart\'{\i}nez}(2000)}]{Ben-Nun_Martinez:2000}%
  \BibitemOpen
  \bibfield  {author} {\bibinfo {author} {\bibfnamefont {M.}~\bibnamefont
  {Ben-Nun}}, \bibinfo {author} {\bibfnamefont {J.}~\bibnamefont
  {Quenneville}},\ and\ \bibinfo {author} {\bibfnamefont {T.~J.}\ \bibnamefont
  {Mart\'{\i}nez}},\ }\href {https://doi.org/10.1021/jp994174i} {\bibfield
  {journal} {\bibinfo  {journal} {J.~Phys.\ Chem.~A}\ }\textbf {\bibinfo
  {volume} {104}},\ \bibinfo {pages} {5161} (\bibinfo {year}
  {2000})}\BibitemShut {NoStop}%
\bibitem [{\citenamefont {Curchod}\ and\ \citenamefont
  {Mart\'{\i}nez}(2018)}]{Curchod_Martinez:2018_v2}%
  \BibitemOpen
  \bibfield  {author} {\bibinfo {author} {\bibfnamefont {B.~F.~E.}\
  \bibnamefont {Curchod}}\ and\ \bibinfo {author} {\bibfnamefont {T.~J.}\
  \bibnamefont {Mart\'{\i}nez}},\ }\href
  {https://doi.org/10.1021/acs.chemrev.7b00423} {\bibfield  {journal} {\bibinfo
   {journal} {Chem.\ Rev.}\ }\textbf {\bibinfo {volume} {118}},\ \bibinfo
  {pages} {3305} (\bibinfo {year} {2018})}\BibitemShut {NoStop}%
\bibitem [{\citenamefont {Shalashilin}\ and\ \citenamefont
  {Child}(2004)}]{Shalashilin_Child:2004}%
  \BibitemOpen
  \bibfield  {author} {\bibinfo {author} {\bibfnamefont {D.~V.}\ \bibnamefont
  {Shalashilin}}\ and\ \bibinfo {author} {\bibfnamefont {M.~S.}\ \bibnamefont
  {Child}},\ }\href {https://doi.org/10.1063/1.1776111} {\bibfield  {journal}
  {\bibinfo  {journal} {J.~Chem.\ Phys.}\ }\textbf {\bibinfo {volume} {121}},\
  \bibinfo {pages} {3563} (\bibinfo {year} {2004})}\BibitemShut {NoStop}%
\bibitem [{\citenamefont {Sawada}\ \emph {et~al.}(1985)\citenamefont {Sawada},
  \citenamefont {Heather}, \citenamefont {Jackson},\ and\ \citenamefont
  {Metiu}}]{Sawada_Heather:1985}%
  \BibitemOpen
  \bibfield  {author} {\bibinfo {author} {\bibfnamefont {S.}~\bibnamefont
  {Sawada}}, \bibinfo {author} {\bibfnamefont {R.}~\bibnamefont {Heather}},
  \bibinfo {author} {\bibfnamefont {B.}~\bibnamefont {Jackson}},\ and\ \bibinfo
  {author} {\bibfnamefont {H.}~\bibnamefont {Metiu}},\ }\href
  {https://doi.org/10.1063/1.449204} {\bibfield  {journal} {\bibinfo  {journal}
  {J.~Chem.\ Phys.}\ }\textbf {\bibinfo {volume} {83}},\ \bibinfo {pages}
  {3009} (\bibinfo {year} {1985})}\BibitemShut {NoStop}%
\bibitem [{\citenamefont {Sawada}\ and\ \citenamefont
  {Metiu}(1986{\natexlab{a}})}]{Sawada_Metiu:1986a}%
  \BibitemOpen
  \bibfield  {author} {\bibinfo {author} {\bibfnamefont {S.}~\bibnamefont
  {Sawada}}\ and\ \bibinfo {author} {\bibfnamefont {H.}~\bibnamefont {Metiu}},\
  }\href {https://doi.org/10.1063/1.450774} {\bibfield  {journal} {\bibinfo
  {journal} {J.~Chem.\ Phys.}\ }\textbf {\bibinfo {volume} {84}},\ \bibinfo
  {pages} {6293} (\bibinfo {year} {1986}{\natexlab{a}})}\BibitemShut {NoStop}%
\bibitem [{\citenamefont {Sawada}\ and\ \citenamefont
  {Metiu}(1986{\natexlab{b}})}]{Sawada_Metiu:1986b}%
  \BibitemOpen
  \bibfield  {author} {\bibinfo {author} {\bibfnamefont {S.}~\bibnamefont
  {Sawada}}\ and\ \bibinfo {author} {\bibfnamefont {H.}~\bibnamefont {Metiu}},\
  }\href {https://doi.org/10.1063/1.450175} {\bibfield  {journal} {\bibinfo
  {journal} {J.~Chem.\ Phys.}\ }\textbf {\bibinfo {volume} {84}},\ \bibinfo
  {pages} {227} (\bibinfo {year} {1986}{\natexlab{b}})}\BibitemShut {NoStop}%
\bibitem [{\citenamefont {Worth}\ and\ \citenamefont
  {Burghardt}(2003)}]{Worth_Burghardt:2003}%
  \BibitemOpen
  \bibfield  {author} {\bibinfo {author} {\bibfnamefont {G.~A.}\ \bibnamefont
  {Worth}}\ and\ \bibinfo {author} {\bibfnamefont {I.}~\bibnamefont
  {Burghardt}},\ }\href {https://doi.org/10.1016/S0009-2614(02)01920-6}
  {\bibfield  {journal} {\bibinfo  {journal} {Chem.\ Phys.\ Lett.}\ }\textbf
  {\bibinfo {volume} {368}},\ \bibinfo {pages} {502} (\bibinfo {year}
  {2003})}\BibitemShut {NoStop}%
\bibitem [{\citenamefont {Shalashilin}(2009)}]{Shalashilin:2009}%
  \BibitemOpen
  \bibfield  {author} {\bibinfo {author} {\bibfnamefont {D.~V.}\ \bibnamefont
  {Shalashilin}},\ }\href {https://doi.org/10.1063/1.3153302} {\bibfield
  {journal} {\bibinfo  {journal} {J.~Chem.\ Phys.}\ }\textbf {\bibinfo {volume}
  {130}},\ \bibinfo {pages} {244101} (\bibinfo {year} {2009})}\BibitemShut
  {NoStop}%
\bibitem [{\citenamefont {Shalashilin}(2010)}]{Shalashilin:2010}%
  \BibitemOpen
  \bibfield  {author} {\bibinfo {author} {\bibfnamefont {D.~V.}\ \bibnamefont
  {Shalashilin}},\ }\href {https://doi.org/10.1063/1.3442747} {\bibfield
  {journal} {\bibinfo  {journal} {J.~Chem.\ Phys.}\ }\textbf {\bibinfo {volume}
  {132}},\ \bibinfo {eid} {244111} (\bibinfo {year} {2010})}\BibitemShut
  {NoStop}%
\bibitem [{\citenamefont {{\v{S}}ulc}\ \emph {et~al.}(2013)\citenamefont
  {{\v{S}}ulc}, \citenamefont {Hern{\'{a}}ndez}, \citenamefont
  {Mart{\'{i}}nez},\ and\ \citenamefont
  {Van{\'{i}}{\v{c}}ek}}]{Sulc_Vanicek:2013}%
  \BibitemOpen
  \bibfield  {author} {\bibinfo {author} {\bibfnamefont {M.}~\bibnamefont
  {{\v{S}}ulc}}, \bibinfo {author} {\bibfnamefont {H.}~\bibnamefont
  {Hern{\'{a}}ndez}}, \bibinfo {author} {\bibfnamefont {T.~J.}\ \bibnamefont
  {Mart{\'{i}}nez}},\ and\ \bibinfo {author} {\bibfnamefont {J.~J.~L.}\
  \bibnamefont {Van{\'{i}}{\v{c}}ek}},\ }\href
  {https://doi.org/10.1063/1.4813124} {\bibfield  {journal} {\bibinfo
  {journal} {J.~Chem.\ Phys.}\ }\textbf {\bibinfo {volume} {139}},\ \bibinfo
  {pages} {034112} (\bibinfo {year} {2013})}\BibitemShut {NoStop}%
\bibitem [{\citenamefont {Miller}(2001)}]{Miller:2001_v2}%
  \BibitemOpen
  \bibfield  {author} {\bibinfo {author} {\bibfnamefont {W.~H.}\ \bibnamefont
  {Miller}},\ }\href {https://doi.org/10.1021/jp003712k} {\bibfield  {journal}
  {\bibinfo  {journal} {J.~Phys.\ Chem.~A}\ }\textbf {\bibinfo {volume}
  {105}},\ \bibinfo {pages} {2942} (\bibinfo {year} {2001})}\BibitemShut
  {NoStop}%
\bibitem [{\citenamefont {Herman}\ and\ \citenamefont
  {Kluk}(1984)}]{Herman_Kluk:1984}%
  \BibitemOpen
  \bibfield  {author} {\bibinfo {author} {\bibfnamefont {M.~F.}\ \bibnamefont
  {Herman}}\ and\ \bibinfo {author} {\bibfnamefont {E.}~\bibnamefont {Kluk}},\
  }\href {https://doi.org/10.1016/0301-0104(84)80039-7} {\bibfield  {journal}
  {\bibinfo  {journal} {Chem.\ Phys.}\ }\textbf {\bibinfo {volume} {91}},\
  \bibinfo {pages} {27} (\bibinfo {year} {1984})}\BibitemShut {NoStop}%
\bibitem [{\citenamefont {Grossmann}(2006)}]{Grossmann:2006}%
  \BibitemOpen
  \bibfield  {author} {\bibinfo {author} {\bibfnamefont {F.}~\bibnamefont
  {Grossmann}},\ }\href {https://doi.org/10.1063/1.2213255} {\bibfield
  {journal} {\bibinfo  {journal} {J.~Chem.\ Phys.}\ }\textbf {\bibinfo {volume}
  {125}},\ \bibinfo {pages} {014111} (\bibinfo {year} {2006})}\BibitemShut
  {NoStop}%
\bibitem [{\citenamefont {Ceotto}\ \emph {et~al.}(2009)\citenamefont {Ceotto},
  \citenamefont {Atahan}, \citenamefont {Tantardini},\ and\ \citenamefont
  {Aspuru-Guzik}}]{Ceotto_Atahan:2009a}%
  \BibitemOpen
  \bibfield  {author} {\bibinfo {author} {\bibfnamefont {M.}~\bibnamefont
  {Ceotto}}, \bibinfo {author} {\bibfnamefont {S.}~\bibnamefont {Atahan}},
  \bibinfo {author} {\bibfnamefont {G.~F.}\ \bibnamefont {Tantardini}},\ and\
  \bibinfo {author} {\bibfnamefont {A.}~\bibnamefont {Aspuru-Guzik}},\ }\href
  {https://doi.org/10.1063/1.3155062} {\bibfield  {journal} {\bibinfo
  {journal} {J.~Chem.\ Phys.}\ }\textbf {\bibinfo {volume} {130}},\ \bibinfo
  {eid} {234113} (\bibinfo {year} {2009})}\BibitemShut {NoStop}%
\bibitem [{\citenamefont {Ceotto}, \citenamefont {Di~Liberto},\ and\
  \citenamefont {Conte}(2017)}]{Ceotto_Conte:2017}%
  \BibitemOpen
  \bibfield  {author} {\bibinfo {author} {\bibfnamefont {M.}~\bibnamefont
  {Ceotto}}, \bibinfo {author} {\bibfnamefont {G.}~\bibnamefont {Di~Liberto}},\
  and\ \bibinfo {author} {\bibfnamefont {R.}~\bibnamefont {Conte}},\ }\href
  {https://doi.org/10.1103/PhysRevLett.119.010401} {\bibfield  {journal}
  {\bibinfo  {journal} {Phys.\ Rev.\ Lett.}\ }\textbf {\bibinfo {volume}
  {119}},\ \bibinfo {pages} {010401} (\bibinfo {year} {2017})}\BibitemShut
  {NoStop}%
\bibitem [{\citenamefont {Gabas}, \citenamefont {{Di Liberto}},\ and\
  \citenamefont {Ceotto}(2019)}]{Gabas_Ceotto:2019}%
  \BibitemOpen
  \bibfield  {author} {\bibinfo {author} {\bibfnamefont {F.}~\bibnamefont
  {Gabas}}, \bibinfo {author} {\bibfnamefont {G.}~\bibnamefont {{Di
  Liberto}}},\ and\ \bibinfo {author} {\bibfnamefont {M.}~\bibnamefont
  {Ceotto}},\ }\href {https://doi.org/10.1063/1.5100503} {\bibfield  {journal}
  {\bibinfo  {journal} {J.~Chem.\ Phys.}\ }\textbf {\bibinfo {volume} {150}},\
  \bibinfo {pages} {224107} (\bibinfo {year} {2019})}\BibitemShut {NoStop}%
\bibitem [{\citenamefont {Lee}\ and\ \citenamefont
  {Heller}(1982)}]{Lee_Heller:1982}%
  \BibitemOpen
  \bibfield  {author} {\bibinfo {author} {\bibfnamefont {S.~Y.}\ \bibnamefont
  {Lee}}\ and\ \bibinfo {author} {\bibfnamefont {E.~J.}\ \bibnamefont
  {Heller}},\ }\href {https://doi.org/10.1063/1.443342} {\bibfield  {journal}
  {\bibinfo  {journal} {J.~Chem.\ Phys.}\ }\textbf {\bibinfo {volume} {76}},\
  \bibinfo {pages} {3035} (\bibinfo {year} {1982})}\BibitemShut {NoStop}%
\bibitem [{\citenamefont {Wehrle}, \citenamefont {Oberli},\ and\ \citenamefont
  {Van\'{i}\v{c}ek}(2015)}]{Wehrle_Vanicek:2015_v2}%
  \BibitemOpen
  \bibfield  {author} {\bibinfo {author} {\bibfnamefont {M.}~\bibnamefont
  {Wehrle}}, \bibinfo {author} {\bibfnamefont {S.}~\bibnamefont {Oberli}},\
  and\ \bibinfo {author} {\bibfnamefont {J.~J.~L.}\ \bibnamefont
  {Van\'{i}\v{c}ek}},\ }\href {https://doi.org/10.1021/acs.jpca.5b03907}
  {\bibfield  {journal} {\bibinfo  {journal} {J.~Phys.\ Chem.~A}\ }\textbf
  {\bibinfo {volume} {119}},\ \bibinfo {pages} {5685} (\bibinfo {year}
  {2015})}\BibitemShut {NoStop}%
\bibitem [{\citenamefont {Patoz}, \citenamefont {Begu\v{s}i{\'{c}}},\ and\
  \citenamefont {Van{\'{i}}{\v{c}}ek}(2018)}]{Patoz_Vanicek:2018}%
  \BibitemOpen
  \bibfield  {author} {\bibinfo {author} {\bibfnamefont {A.}~\bibnamefont
  {Patoz}}, \bibinfo {author} {\bibfnamefont {T.}~\bibnamefont
  {Begu\v{s}i{\'{c}}}},\ and\ \bibinfo {author} {\bibfnamefont {J.~J.~L.}\
  \bibnamefont {Van{\'{i}}{\v{c}}ek}},\ }\href
  {https://doi.org/10.1021/acs.jpclett.8b00827} {\bibfield  {journal} {\bibinfo
   {journal} {J.~Phys.\ Chem.\ Lett.}\ }\textbf {\bibinfo {volume} {9}},\
  \bibinfo {pages} {2367} (\bibinfo {year} {2018})}\BibitemShut {NoStop}%
\bibitem [{\citenamefont {Begu\v{s}i\'{c}}, \citenamefont {Tapavicza},\ and\
  \citenamefont {Van\'i\v{c}ek}(2022)}]{Begusic_Vanicek:2022}%
  \BibitemOpen
  \bibfield  {author} {\bibinfo {author} {\bibfnamefont {T.}~\bibnamefont
  {Begu\v{s}i\'{c}}}, \bibinfo {author} {\bibfnamefont {E.}~\bibnamefont
  {Tapavicza}},\ and\ \bibinfo {author} {\bibfnamefont {J.~J.~L.}\ \bibnamefont
  {Van\'i\v{c}ek}},\ }\href {https://doi.org/10.1021/acs.jctc.2c00030}
  {\bibfield  {journal} {\bibinfo  {journal} {J.~Chem.\ Theory Comput.}\
  }\textbf {\bibinfo {volume} {18}},\ \bibinfo {pages} {3065} (\bibinfo {year}
  {2022})}\BibitemShut {NoStop}%
\bibitem [{\citenamefont {Coalson}\ and\ \citenamefont
  {Karplus}(1990)}]{Coalson_Karplus:1990}%
  \BibitemOpen
  \bibfield  {author} {\bibinfo {author} {\bibfnamefont {R.~D.}\ \bibnamefont
  {Coalson}}\ and\ \bibinfo {author} {\bibfnamefont {M.}~\bibnamefont
  {Karplus}},\ }\href {https://doi.org/10.1063/1.458778} {\bibfield  {journal}
  {\bibinfo  {journal} {J.~Chem.\ Phys.}\ }\textbf {\bibinfo {volume} {93}},\
  \bibinfo {pages} {3919} (\bibinfo {year} {1990})}\BibitemShut {NoStop}%
\bibitem [{\citenamefont {Lasser}\ and\ \citenamefont
  {Lubich}(2020)}]{Lasser_Lubich:2020}%
  \BibitemOpen
  \bibfield  {author} {\bibinfo {author} {\bibfnamefont {C.}~\bibnamefont
  {Lasser}}\ and\ \bibinfo {author} {\bibfnamefont {C.}~\bibnamefont
  {Lubich}},\ }\href {https://doi.org/10.1017/S0962492920000033} {\bibfield
  {journal} {\bibinfo  {journal} {Acta Numerica}\ }\textbf {\bibinfo {volume}
  {29}},\ \bibinfo {pages} {229} (\bibinfo {year} {2020})}\BibitemShut
  {NoStop}%
\bibitem [{\citenamefont {Dirac}(1930)}]{Dirac:1930}%
  \BibitemOpen
  \bibfield  {author} {\bibinfo {author} {\bibfnamefont {P.~A.~M.}\
  \bibnamefont {Dirac}},\ }\href {https://doi.org/10.1017/S0305004100016108}
  {\bibfield  {journal} {\bibinfo  {journal} {Math.\ Proc.\ Camb.\ Phil.\
  Soc.}\ }\textbf {\bibinfo {volume} {26}},\ \bibinfo {pages} {376} (\bibinfo
  {year} {1930})}\BibitemShut {NoStop}%
\bibitem [{\citenamefont {Frenkel}(1934)}]{book_Frenkel:1934}%
  \BibitemOpen
  \bibfield  {author} {\bibinfo {author} {\bibfnamefont {J.}~\bibnamefont
  {Frenkel}},\ }\href@noop {} {\emph {\bibinfo {title} {Wave mechanics}}}\
  (\bibinfo  {publisher} {Clarendon Press},\ \bibinfo {address} {Oxford},\
  \bibinfo {year} {1934})\BibitemShut {NoStop}%
\bibitem [{\citenamefont {McLachlan}(1964)}]{McLachlan:1964}%
  \BibitemOpen
  \bibfield  {author} {\bibinfo {author} {\bibfnamefont {A.}~\bibnamefont
  {McLachlan}},\ }\href {https://doi.org/10.1080/00268976400100041} {\bibfield
  {journal} {\bibinfo  {journal} {Mol. Phys.}\ }\textbf {\bibinfo {volume}
  {8}},\ \bibinfo {pages} {39} (\bibinfo {year} {1964})}\BibitemShut {NoStop}%
\bibitem [{\citenamefont {Heller}(1976{\natexlab{b}})}]{Heller:1976}%
  \BibitemOpen
  \bibfield  {author} {\bibinfo {author} {\bibfnamefont {E.~J.}\ \bibnamefont
  {Heller}},\ }\href {https://doi.org/10.1063/1.431911} {\bibfield  {journal}
  {\bibinfo  {journal} {J.~Chem.\ Phys.}\ }\textbf {\bibinfo {volume} {64}},\
  \bibinfo {pages} {63} (\bibinfo {year} {1976}{\natexlab{b}})}\BibitemShut
  {NoStop}%
\bibitem [{\citenamefont {Kramer}\ and\ \citenamefont
  {Saraceno}(1980)}]{book_Kramer_Saraceno:1980}%
  \BibitemOpen
  \bibfield  {author} {\bibinfo {author} {\bibfnamefont {P.}~\bibnamefont
  {Kramer}}\ and\ \bibinfo {author} {\bibfnamefont {M.}~\bibnamefont
  {Saraceno}},\ }\href {https://doi.org/10.1007/3-540-10271-X_317} {\emph
  {\bibinfo {title} {Geometry of the time-dependent variational principle in
  quantum mechanics}}}\ (\bibinfo  {publisher} {Springer Berlin Heidelberg},\
  \bibinfo {address} {Berlin, Heidelberg},\ \bibinfo {year} {1980})\BibitemShut
  {NoStop}%
\bibitem [{\citenamefont {Hairer}, \citenamefont {Lubich},\ and\ \citenamefont
  {Wanner}(2006)}]{book_Hairer_Wanner:2006}%
  \BibitemOpen
  \bibfield  {author} {\bibinfo {author} {\bibfnamefont {E.}~\bibnamefont
  {Hairer}}, \bibinfo {author} {\bibfnamefont {C.}~\bibnamefont {Lubich}},\
  and\ \bibinfo {author} {\bibfnamefont {G.}~\bibnamefont {Wanner}},\ }\href
  {http://books.google.ch/books/about/Geometric_Numerical_Integration.html?id=T1TaNRLmZv8C&redir_esc=y}
  {\emph {\bibinfo {title} {Geometric Numerical Integration:
  Structure-Preserving Algorithms for Ordinary Differential Equations}}}\
  (\bibinfo  {publisher} {Springer Berlin Heidelberg New York},\ \bibinfo
  {year} {2006})\BibitemShut {NoStop}%
\bibitem [{\citenamefont {Buch}(2002)}]{Buch:2002}%
  \BibitemOpen
  \bibfield  {author} {\bibinfo {author} {\bibfnamefont {V.}~\bibnamefont
  {Buch}},\ }\href {https://doi.org/10.1063/1.1497968} {\bibfield  {journal}
  {\bibinfo  {journal} {J.~Chem.\ Phys.}\ }\textbf {\bibinfo {volume} {117}},\
  \bibinfo {pages} {4738} (\bibinfo {year} {2002})}\BibitemShut {NoStop}%
\bibitem [{\citenamefont {Hasegawa}(2014)}]{Hasegawa:2014}%
  \BibitemOpen
  \bibfield  {author} {\bibinfo {author} {\bibfnamefont {H.}~\bibnamefont
  {Hasegawa}},\ }\href {https://doi.org/10.1016/j.physleta.2013.12.035}
  {\bibfield  {journal} {\bibinfo  {journal} {Phys.\ Lett.~A}\ }\textbf
  {\bibinfo {volume} {378}},\ \bibinfo {pages} {691} (\bibinfo {year}
  {2014})}\BibitemShut {NoStop}%
\bibitem [{\citenamefont {Heather}\ and\ \citenamefont
  {Metiu}(1985)}]{Heather_Metiu:1985}%
  \BibitemOpen
  \bibfield  {author} {\bibinfo {author} {\bibfnamefont {R.}~\bibnamefont
  {Heather}}\ and\ \bibinfo {author} {\bibfnamefont {H.}~\bibnamefont
  {Metiu}},\ }\href {https://doi.org/10.1016/0009-2614(85)85353-7} {\bibfield
  {journal} {\bibinfo  {journal} {Chem.\ Phys.\ Lett.}\ }\textbf {\bibinfo
  {volume} {118}},\ \bibinfo {pages} {558} (\bibinfo {year}
  {1985})}\BibitemShut {NoStop}%
\bibitem [{\citenamefont {Poirier}(2017)}]{Poirier:DF_quantum_trajectories}%
  \BibitemOpen
  \bibfield  {author} {\bibinfo {author} {\bibfnamefont {B.}~\bibnamefont
  {Poirier}},\ }\href@noop {} {\enquote {\bibinfo {title} {{Dirac--Frenkel
  Variational Principle as Applied to Quantum Trajectories}},}\ } (\bibinfo
  {year} {2017})\BibitemShut {NoStop}%
\bibitem [{\citenamefont {Moghaddasi~Fereidani}\ and\ \citenamefont
  {Van\'i\v{c}ek}(2023)}]{Moghaddasi_Vanicek:2023a}%
  \BibitemOpen
  \bibfield  {author} {\bibinfo {author} {\bibfnamefont {R.}~\bibnamefont
  {Moghaddasi~Fereidani}}\ and\ \bibinfo {author} {\bibfnamefont {J.~J.~L.}\
  \bibnamefont {Van\'i\v{c}ek}},\ }\href@noop {} {\enquote {\bibinfo {title}
  {High-order geometric integrators for the local cubic variational thawed
  gaussian wavepacket dynamics},}\ } (\bibinfo {year} {2023})\BibitemShut
  {NoStop}%
\bibitem [{\citenamefont {Leimkuhler}\ and\ \citenamefont
  {Reich}(2004)}]{book_Leimkuhler_Reich:2004}%
  \BibitemOpen
  \bibfield  {author} {\bibinfo {author} {\bibfnamefont {B.}~\bibnamefont
  {Leimkuhler}}\ and\ \bibinfo {author} {\bibfnamefont {S.}~\bibnamefont
  {Reich}},\ }\href
  {http://books.google.ch/books/about/Simulating_Hamiltonian_Dynamics.html?id=tpb-tnsZi5YC&redir_esc=y}
  {\emph {\bibinfo {title} {Simulating Hamiltonian Dynamics}}}\ (\bibinfo
  {publisher} {Cambridge University Press},\ \bibinfo {year}
  {2004})\BibitemShut {NoStop}%
\bibitem [{\citenamefont {Yoshida}(1990)}]{Yoshida:1990}%
  \BibitemOpen
  \bibfield  {author} {\bibinfo {author} {\bibfnamefont {H.}~\bibnamefont
  {Yoshida}},\ }\href {https://doi.org/10.1016/0375-9601(90)90092-3} {\bibfield
   {journal} {\bibinfo  {journal} {Phys.\ Lett.~A}\ }\textbf {\bibinfo {volume}
  {150}},\ \bibinfo {pages} {262} (\bibinfo {year} {1990})}\BibitemShut
  {NoStop}%
\bibitem [{\citenamefont {Suzuki}(1990)}]{Suzuki:1990}%
  \BibitemOpen
  \bibfield  {author} {\bibinfo {author} {\bibfnamefont {M.}~\bibnamefont
  {Suzuki}},\ }\href {https://doi.org/10.1016/0375-9601(90)90962-n} {\bibfield
  {journal} {\bibinfo  {journal} {Phys.\ Lett.~A}\ }\textbf {\bibinfo {volume}
  {146}},\ \bibinfo {pages} {319} (\bibinfo {year} {1990})}\BibitemShut
  {NoStop}%
\bibitem [{\citenamefont {McLachlan}(1995)}]{McLachlan:1995}%
  \BibitemOpen
  \bibfield  {author} {\bibinfo {author} {\bibfnamefont {R.~I.}\ \bibnamefont
  {McLachlan}},\ }\href {https://doi.org/10.1137/0916010} {\bibfield  {journal}
  {\bibinfo  {journal} {SIAM J.\ Sci.\ Comp.}\ }\textbf {\bibinfo {volume}
  {16}},\ \bibinfo {pages} {151} (\bibinfo {year} {1995})}\BibitemShut
  {NoStop}%
\bibitem [{\citenamefont {Wehrle}, \citenamefont {\v{S}ulc},\ and\
  \citenamefont {Van\'{\i}\v{c}ek}(2011)}]{Wehrle_Vanicek:2011}%
  \BibitemOpen
  \bibfield  {author} {\bibinfo {author} {\bibfnamefont {M.}~\bibnamefont
  {Wehrle}}, \bibinfo {author} {\bibfnamefont {M.}~\bibnamefont {\v{S}ulc}},\
  and\ \bibinfo {author} {\bibfnamefont {J.~J.~L.}\ \bibnamefont
  {Van\'{\i}\v{c}ek}},\ }\href {https://doi.org/10.2533/chimia.2011.334}
  {\bibfield  {journal} {\bibinfo  {journal} {Chimia}\ }\textbf {\bibinfo
  {volume} {65}},\ \bibinfo {pages} {334} (\bibinfo {year} {2011})}\BibitemShut
  {NoStop}%
\bibitem [{\citenamefont {Sofroniou}\ and\ \citenamefont
  {Spaletta}(2005)}]{Sofroniou_Spaletta:2005}%
  \BibitemOpen
  \bibfield  {author} {\bibinfo {author} {\bibfnamefont {M.}~\bibnamefont
  {Sofroniou}}\ and\ \bibinfo {author} {\bibfnamefont {G.}~\bibnamefont
  {Spaletta}},\ }\href {https://doi.org/10.1080/10556780500140664} {\bibfield
  {journal} {\bibinfo  {journal} {Optim.\ Method Softw.}\ }\textbf {\bibinfo
  {volume} {20}},\ \bibinfo {pages} {597} (\bibinfo {year} {2005})}\BibitemShut
  {NoStop}%
\bibitem [{\citenamefont {Choi}\ and\ \citenamefont
  {Van\'{i}\v{c}ek}(2019)}]{Choi_Vanicek:2019}%
  \BibitemOpen
  \bibfield  {author} {\bibinfo {author} {\bibfnamefont {S.}~\bibnamefont
  {Choi}}\ and\ \bibinfo {author} {\bibfnamefont {J.~J.~L.}\ \bibnamefont
  {Van\'{i}\v{c}ek}},\ }\href {https://doi.org/10.1063/1.5092611} {\bibfield
  {journal} {\bibinfo  {journal} {J.~Chem.\ Phys.}\ }\textbf {\bibinfo {volume}
  {150}},\ \bibinfo {pages} {204112} (\bibinfo {year} {2019})}\BibitemShut
  {NoStop}%
\bibitem [{\citenamefont {Roulet}, \citenamefont {Choi},\ and\ \citenamefont
  {Van\'{i}\v{c}ek}(2019)}]{Roulet_Vanicek:2019}%
  \BibitemOpen
  \bibfield  {author} {\bibinfo {author} {\bibfnamefont {J.}~\bibnamefont
  {Roulet}}, \bibinfo {author} {\bibfnamefont {S.}~\bibnamefont {Choi}},\ and\
  \bibinfo {author} {\bibfnamefont {J.~J.~L.}\ \bibnamefont
  {Van\'{i}\v{c}ek}},\ }\href {https://doi.org/10.1063/1.5094046} {\bibfield
  {journal} {\bibinfo  {journal} {J.~Chem.\ Phys.}\ }\textbf {\bibinfo {volume}
  {150}},\ \bibinfo {pages} {204113} (\bibinfo {year} {2019})}\BibitemShut
  {NoStop}%
\bibitem [{\citenamefont {Born}\ and\ \citenamefont
  {Oppenheimer}(1927)}]{Born_Oppenheimer:1927}%
  \BibitemOpen
  \bibfield  {author} {\bibinfo {author} {\bibfnamefont {M.}~\bibnamefont
  {Born}}\ and\ \bibinfo {author} {\bibfnamefont {R.}~\bibnamefont
  {Oppenheimer}},\ }\href {https://doi.org/10.1002/andp.19273892002} {\bibfield
   {journal} {\bibinfo  {journal} {Ann.~d.~Phys.}\ }\textbf {\bibinfo {volume}
  {389}},\ \bibinfo {pages} {457} (\bibinfo {year} {1927})}\BibitemShut
  {NoStop}%
\bibitem [{\citenamefont {Heller}(1981{\natexlab{b}})}]{Heller:1981a}%
  \BibitemOpen
  \bibfield  {author} {\bibinfo {author} {\bibfnamefont {E.~J.}\ \bibnamefont
  {Heller}},\ }\href {https://doi.org/10.1021/ar00072a002} {\bibfield
  {journal} {\bibinfo  {journal} {Acc.\ Chem.\ Res.}\ }\textbf {\bibinfo
  {volume} {14}},\ \bibinfo {pages} {368} (\bibinfo {year}
  {1981}{\natexlab{b}})}\BibitemShut {NoStop}%
\bibitem [{\citenamefont {Roulet}\ and\ \citenamefont
  {Van\'{i}\v{c}ek}(2021)}]{Roulet_Vanicek:2021}%
  \BibitemOpen
  \bibfield  {author} {\bibinfo {author} {\bibfnamefont {J.}~\bibnamefont
  {Roulet}}\ and\ \bibinfo {author} {\bibfnamefont {J.~J.~L.}\ \bibnamefont
  {Van\'{i}\v{c}ek}},\ }\href {https://doi.org/10.1063/5.0050071} {\bibfield
  {journal} {\bibinfo  {journal} {J.~Chem.\ Phys.}\ }\textbf {\bibinfo {volume}
  {154}},\ \bibinfo {pages} {154106} (\bibinfo {year} {2021})}\BibitemShut
  {NoStop}%
\bibitem [{\citenamefont {Van\'i\v{c}ek}(2023)}]{Vanicek:2023}%
  \BibitemOpen
  \bibfield  {author} {\bibinfo {author} {\bibfnamefont {J.~J.~L.}\
  \bibnamefont {Van\'i\v{c}ek}},\ }\href
  {https://doi.org/10.48550/ARXIV.2302.10221} {\  (\bibinfo {year} {2023}),\
  10.48550/ARXIV.2302.10221}\BibitemShut {NoStop}%
\bibitem [{\citenamefont {Van{\'{i}}{\v{c}}ek}\ and\ \citenamefont
  {Begu{\v{s}}i{\'{c}}}(2021)}]{Vanicek_Begusic:2021}%
  \BibitemOpen
  \bibfield  {author} {\bibinfo {author} {\bibfnamefont {J.~J.~L.}\
  \bibnamefont {Van{\'{i}}{\v{c}}ek}}\ and\ \bibinfo {author} {\bibfnamefont
  {T.}~\bibnamefont {Begu{\v{s}}i{\'{c}}}},\ }in\ \href
  {https://doi.org/10.1016/B978-0-12-817234-6.00011-8} {\emph {\bibinfo
  {booktitle} {Molecular Spectroscopy and Quantum Dynamics}}},\ \bibinfo
  {editor} {edited by\ \bibinfo {editor} {\bibfnamefont {R.}~\bibnamefont
  {Marquardt}}\ and\ \bibinfo {editor} {\bibfnamefont {M.}~\bibnamefont
  {Quack}}}\ (\bibinfo  {publisher} {Elsevier},\ \bibinfo {year} {2021})\ pp.\
  \bibinfo {pages} {199--229}\BibitemShut {NoStop}%
\bibitem [{\citenamefont {Kramer}\ and\ \citenamefont
  {Saraceno}(1981)}]{Kramer_Saraceno:1981}%
  \BibitemOpen
  \bibfield  {author} {\bibinfo {author} {\bibfnamefont {P.}~\bibnamefont
  {Kramer}}\ and\ \bibinfo {author} {\bibfnamefont {M.}~\bibnamefont
  {Saraceno}},\ }\href@noop {} {\emph {\bibinfo {title} {Geometry of the
  time-dependent variational principle in quantum mechanics}}},\ \bibinfo
  {series} {Lecture notes in physics}, Vol.\ \bibinfo {volume} {140}\ (\bibinfo
   {publisher} {Springer-Verlag},\ \bibinfo {address} {Berlin},\ \bibinfo
  {year} {1981})\BibitemShut {NoStop}%
\bibitem [{\citenamefont {Beck}\ \emph {et~al.}(2000)\citenamefont {Beck},
  \citenamefont {J\"{a}ckle}, \citenamefont {Worth},\ and\ \citenamefont
  {Meyer}}]{Beck_Meyer:2000}%
  \BibitemOpen
  \bibfield  {author} {\bibinfo {author} {\bibfnamefont {M.}~\bibnamefont
  {Beck}}, \bibinfo {author} {\bibfnamefont {A.}~\bibnamefont {J\"{a}ckle}},
  \bibinfo {author} {\bibfnamefont {G.}~\bibnamefont {Worth}},\ and\ \bibinfo
  {author} {\bibfnamefont {H.-D.}\ \bibnamefont {Meyer}},\ }\href
  {https://doi.org/10.1016/S0370-1573(99)00047-2} {\bibfield  {journal}
  {\bibinfo  {journal} {Phys.\ Rep.}\ }\textbf {\bibinfo {volume} {324}},\
  \bibinfo {pages} {1} (\bibinfo {year} {2000})}\BibitemShut {NoStop}%
\bibitem [{\citenamefont {Habershon}(2012)}]{Habershon:2012}%
  \BibitemOpen
  \bibfield  {author} {\bibinfo {author} {\bibfnamefont {S.}~\bibnamefont
  {Habershon}},\ }\href {https://doi.org/10.1063/1.3671978} {\bibfield
  {journal} {\bibinfo  {journal} {J.~Chem.\ Phys.}\ }\textbf {\bibinfo {volume}
  {136}},\ \bibinfo {pages} {014109} (\bibinfo {year} {2012})}\BibitemShut
  {NoStop}%
\bibitem [{\citenamefont {Joubert-Doriol}\ and\ \citenamefont
  {Izmaylov}(2015)}]{Joubert-Doriol_Izmaylov:2015}%
  \BibitemOpen
  \bibfield  {author} {\bibinfo {author} {\bibfnamefont {L.}~\bibnamefont
  {Joubert-Doriol}}\ and\ \bibinfo {author} {\bibfnamefont {A.~F.}\
  \bibnamefont {Izmaylov}},\ }\href {https://doi.org/10.1063/1.4916384}
  {\bibfield  {journal} {\bibinfo  {journal} {J.~Chem.\ Phys.}\ }\textbf
  {\bibinfo {volume} {142}},\ \bibinfo {pages} {134107} (\bibinfo {year}
  {2015})}\BibitemShut {NoStop}%
\bibitem [{\citenamefont {Hackl}\ \emph {et~al.}(2020)\citenamefont {Hackl},
  \citenamefont {Guaita}, \citenamefont {Shi}, \citenamefont {Haegeman},
  \citenamefont {Demler},\ and\ \citenamefont {Cirac}}]{Hackl_Cirac:2020}%
  \BibitemOpen
  \bibfield  {author} {\bibinfo {author} {\bibfnamefont {L.}~\bibnamefont
  {Hackl}}, \bibinfo {author} {\bibfnamefont {T.}~\bibnamefont {Guaita}},
  \bibinfo {author} {\bibfnamefont {T.}~\bibnamefont {Shi}}, \bibinfo {author}
  {\bibfnamefont {J.}~\bibnamefont {Haegeman}}, \bibinfo {author}
  {\bibfnamefont {E.}~\bibnamefont {Demler}},\ and\ \bibinfo {author}
  {\bibfnamefont {J.~I.}\ \bibnamefont {Cirac}},\ }\href
  {https://doi.org/10.21468/SciPostPhys.9.4.048} {\bibfield  {journal}
  {\bibinfo  {journal} {SciPost Phys.}\ }\textbf {\bibinfo {volume} {9}},\
  \bibinfo {pages} {048} (\bibinfo {year} {2020})}\BibitemShut {NoStop}%
\bibitem [{\citenamefont {Kahan}\ and\ \citenamefont
  {Li}(1997)}]{Kahan_Li:1997}%
  \BibitemOpen
  \bibfield  {author} {\bibinfo {author} {\bibfnamefont {W.}~\bibnamefont
  {Kahan}}\ and\ \bibinfo {author} {\bibfnamefont {R.-C.}\ \bibnamefont {Li}},\
  }\href {https://doi.org/10.1090/s0025-5718-97-00873-9} {\bibfield  {journal}
  {\bibinfo  {journal} {Math.\ Comput.}\ }\textbf {\bibinfo {volume} {66}},\
  \bibinfo {pages} {1089} (\bibinfo {year} {1997})}\BibitemShut {NoStop}%
\bibitem [{\citenamefont {Choi}\ and\ \citenamefont
  {Van\'{i}\v{c}ek}(2021)}]{Choi_Vanicek:2021b}%
  \BibitemOpen
  \bibfield  {author} {\bibinfo {author} {\bibfnamefont {S.}~\bibnamefont
  {Choi}}\ and\ \bibinfo {author} {\bibfnamefont {J.~J.~L.}\ \bibnamefont
  {Van\'{i}\v{c}ek}},\ }\href {https://doi.org/10.1063/5.0061878} {\bibfield
  {journal} {\bibinfo  {journal} {J.~Chem.\ Phys.}\ }\textbf {\bibinfo {volume}
  {155}},\ \bibinfo {pages} {124104} (\bibinfo {year} {2021})}\BibitemShut
  {NoStop}%
\bibitem [{\citenamefont {Thorwart}, \citenamefont {Grifoni},\ and\
  \citenamefont {H{\"a}nggi}(2001)}]{Thorwart_Hanggi:2001}%
  \BibitemOpen
  \bibfield  {author} {\bibinfo {author} {\bibfnamefont {M.}~\bibnamefont
  {Thorwart}}, \bibinfo {author} {\bibfnamefont {M.}~\bibnamefont {Grifoni}},\
  and\ \bibinfo {author} {\bibfnamefont {P.}~\bibnamefont {H{\"a}nggi}},\
  }\href {https://doi.org/10.1006/aphy.2001.6174} {\bibfield  {journal}
  {\bibinfo  {journal} {Adv.\ Phys.}\ }\textbf {\bibinfo {volume} {293}},\
  \bibinfo {pages} {15} (\bibinfo {year} {2001})}\BibitemShut {NoStop}%
\bibitem [{\citenamefont {Razavy}(2003)}]{Book_Razavy:2003}%
  \BibitemOpen
  \bibfield  {author} {\bibinfo {author} {\bibfnamefont {M.}~\bibnamefont
  {Razavy}},\ }\href@noop {} {\emph {\bibinfo {title} {Quantum Theory of
  Tunneling}}}\ (\bibinfo  {publisher} {World Scientific Publishing},\ \bibinfo
  {address} {Singapore},\ \bibinfo {year} {2003})\BibitemShut {NoStop}%
\bibitem [{\citenamefont {Begu\v{s}i\'{c}}, \citenamefont {Cordova},\ and\
  \citenamefont {Van{\'{i}}{\v{c}}ek}(2019)}]{Begusic_Vanicek:2019}%
  \BibitemOpen
  \bibfield  {author} {\bibinfo {author} {\bibfnamefont {T.}~\bibnamefont
  {Begu\v{s}i\'{c}}}, \bibinfo {author} {\bibfnamefont {M.}~\bibnamefont
  {Cordova}},\ and\ \bibinfo {author} {\bibfnamefont {J.~J.~L.}\ \bibnamefont
  {Van{\'{i}}{\v{c}}ek}},\ }\href {https://doi.org/10.1063/1.5090122}
  {\bibfield  {journal} {\bibinfo  {journal} {J.~Chem.\ Phys.}\ }\textbf
  {\bibinfo {volume} {150}},\ \bibinfo {pages} {154117} (\bibinfo {year}
  {2019})}\BibitemShut {NoStop}%
\bibitem [{\citenamefont {Tannor}\ and\ \citenamefont
  {Heller}(1982)}]{Tannor_Heller:1982}%
  \BibitemOpen
  \bibfield  {author} {\bibinfo {author} {\bibfnamefont {D.~J.}\ \bibnamefont
  {Tannor}}\ and\ \bibinfo {author} {\bibfnamefont {E.~J.}\ \bibnamefont
  {Heller}},\ }\href {https://doi.org/10.1063/1.443643} {\bibfield  {journal}
  {\bibinfo  {journal} {J.~Chem.\ Phys.}\ }\textbf {\bibinfo {volume} {77}},\
  \bibinfo {pages} {202} (\bibinfo {year} {1982})}\BibitemShut {NoStop}%
\bibitem [{\citenamefont {Isserlis}(1918)}]{Isserlis:1918}%
  \BibitemOpen
  \bibfield  {author} {\bibinfo {author} {\bibfnamefont {L.}~\bibnamefont
  {Isserlis}},\ }\href {https://doi.org/10.1093/biomet/12.1-2.134} {\bibfield
  {journal} {\bibinfo  {journal} {Biometrika}\ }\textbf {\bibinfo {volume}
  {12}},\ \bibinfo {pages} {134} (\bibinfo {year} {1918})}\BibitemShut
  {NoStop}%
\bibitem [{\citenamefont {Brookes}(2011)}]{Brookes:2011}%
  \BibitemOpen
  \bibfield  {author} {\bibinfo {author} {\bibfnamefont {M.}~\bibnamefont
  {Brookes}},\ }\href@noop {} {\enquote {\bibinfo {title} {{The Matrix
  Reference Manual}},}\ }\bibinfo {howpublished} {[online]
  http://www.ee.imperial.ac.uk/hp/staff/dmb/matrix/intro.html} (\bibinfo {year}
  {2011})\BibitemShut {NoStop}%
\bibitem [{\citenamefont {Begu{\v{s}}i{\'{c}}}\ and\ \citenamefont
  {Van{\'{i}}{\v{c}}ek}(2020)}]{Begusic_Vanicek:2020a}%
  \BibitemOpen
  \bibfield  {author} {\bibinfo {author} {\bibfnamefont {T.}~\bibnamefont
  {Begu{\v{s}}i{\'{c}}}}\ and\ \bibinfo {author} {\bibfnamefont {J.~J.~L.}\
  \bibnamefont {Van{\'{i}}{\v{c}}ek}},\ }\href
  {https://doi.org/10.1063/5.0031216} {\bibfield  {journal} {\bibinfo
  {journal} {J.~Chem.\ Phys.}\ }\textbf {\bibinfo {volume} {153}},\ \bibinfo
  {pages} {184110} (\bibinfo {year} {2020})}\BibitemShut {NoStop}%
\bibitem [{\citenamefont {Petersen}\ and\ \citenamefont
  {Pedersen}(2012)}]{Petersen_Pedersen:2012}%
  \BibitemOpen
  \bibfield  {author} {\bibinfo {author} {\bibfnamefont {K.~B.}\ \bibnamefont
  {Petersen}}\ and\ \bibinfo {author} {\bibfnamefont {M.~S.}\ \bibnamefont
  {Pedersen}},\ }\href {http://www2.imm.dtu.dk/pubdb/p.php?3274} {\enquote
  {\bibinfo {title} {The matrix cookbook},}\ } (\bibinfo {year}
  {2012})\BibitemShut {NoStop}%
\bibitem [{\citenamefont {Marsden}\ and\ \citenamefont
  {Ratiu}(1999)}]{book_Marsden_Ratiu:1999}%
  \BibitemOpen
  \bibfield  {author} {\bibinfo {author} {\bibfnamefont {J.~E.}\ \bibnamefont
  {Marsden}}\ and\ \bibinfo {author} {\bibfnamefont {T.~S.}\ \bibnamefont
  {Ratiu}},\ }\href@noop {} {\emph {\bibinfo {title} {Introduction to mechanics
  and symmetry: a basic exposition of classical mechanical systems}}},\
  Vol.~\bibinfo {volume} {17}\ (\bibinfo  {publisher} {Springer Science \&
  Business Media},\ \bibinfo {year} {1999})\BibitemShut {NoStop}%
\bibitem [{\citenamefont {Vande~Linde}\ and\ \citenamefont
  {Hase}(1990)}]{Linde_Hase:1990}%
  \BibitemOpen
  \bibfield  {author} {\bibinfo {author} {\bibfnamefont {S.~R.}\ \bibnamefont
  {Vande~Linde}}\ and\ \bibinfo {author} {\bibfnamefont {W.~L.}\ \bibnamefont
  {Hase}},\ }\href {https://doi.org/10.1021/j100370a012} {\bibfield  {journal}
  {\bibinfo  {journal} {J.~Phys.\ Chem.}\ }\textbf {\bibinfo {volume} {94}},\
  \bibinfo {pages} {2778} (\bibinfo {year} {1990})}\BibitemShut {NoStop}%
\bibitem [{\citenamefont {Yagi}\ \emph {et~al.}(2004)\citenamefont {Yagi},
  \citenamefont {Hirao}, \citenamefont {Taketsugu}, \citenamefont {Schmidt},\
  and\ \citenamefont {Gordon}}]{Yagi_Gordon:2004}%
  \BibitemOpen
  \bibfield  {author} {\bibinfo {author} {\bibfnamefont {K.}~\bibnamefont
  {Yagi}}, \bibinfo {author} {\bibfnamefont {K.}~\bibnamefont {Hirao}},
  \bibinfo {author} {\bibfnamefont {T.}~\bibnamefont {Taketsugu}}, \bibinfo
  {author} {\bibfnamefont {M.~W.}\ \bibnamefont {Schmidt}},\ and\ \bibinfo
  {author} {\bibfnamefont {M.~S.}\ \bibnamefont {Gordon}},\ }\href
  {https://doi.org/10.1063/1.1764501} {\bibfield  {journal} {\bibinfo
  {journal} {J.~Chem.\ Phys.}\ }\textbf {\bibinfo {volume} {121}},\ \bibinfo
  {pages} {1383} (\bibinfo {year} {2004})}\BibitemShut {NoStop}%
\bibitem [{\citenamefont {Mandelli}, \citenamefont {Aieta},\ and\ \citenamefont
  {Ceotto}(2022)}]{Mandelli_Ceotto:2022}%
  \BibitemOpen
  \bibfield  {author} {\bibinfo {author} {\bibfnamefont {G.}~\bibnamefont
  {Mandelli}}, \bibinfo {author} {\bibfnamefont {C.}~\bibnamefont {Aieta}},\
  and\ \bibinfo {author} {\bibfnamefont {M.}~\bibnamefont {Ceotto}},\ }\href
  {https://doi.org/10.1021/acs.jctc.1c01143} {\bibfield  {journal} {\bibinfo
  {journal} {J.~Chem.\ Theory Comput.}\ }\textbf {\bibinfo {volume} {18}},\
  \bibinfo {pages} {623} (\bibinfo {year} {2022})}\BibitemShut {NoStop}%
\bibitem [{\citenamefont {Masoumi}, \citenamefont {Olum},\ and\ \citenamefont
  {Wachter}(2017)}]{Masoumi_Wachter:2017}%
  \BibitemOpen
  \bibfield  {author} {\bibinfo {author} {\bibfnamefont {A.}~\bibnamefont
  {Masoumi}}, \bibinfo {author} {\bibfnamefont {K.~D.}\ \bibnamefont {Olum}},\
  and\ \bibinfo {author} {\bibfnamefont {J.~M.}\ \bibnamefont {Wachter}},\
  }\href {https://doi.org/10.1088/1475-7516/2017/10/022} {\bibfield  {journal}
  {\bibinfo  {journal} {J. Cosmol. Astropart. Phys.}\ }\textbf {\bibinfo
  {volume} {2017}},\ \bibinfo {pages} {022} (\bibinfo {year}
  {2017})}\BibitemShut {NoStop}%
\end{thebibliography}%


%

\end{document}


\title{Supplementary material: High-order geometric integrators for the variational Gaussian approximation}
\author{Roya Moghaddasi Fereidani}
\email{roya.moghaddasifereidani@epfl.ch}
\author{Ji\v{r}\'{\i} Van\'{\i}\v{c}ek}
\email{jiri.vanicek@epfl.ch}
\affiliation{Laboratory of Theoretical Physical Chemistry, Institut des Sciences et Ing\'{e}nierie Chimiques, Ecole Polytechnique F\'{e}d\'{e}rale de Lausanne (EPFL), CH-1015 Lausanne, Switzerland}
\date{\today}

\begin{abstract}
This document provides information to support the main text. It includes the computational details of Fig. 1 of the main text, the convergence  and efficiency  of  the  high-order  symplectic  integrators obtained using the triple-jump, Suzuki-fractal, and optimal composition schemes, and the fourth-order Runge-Kutta method in both Heller's and  Hagedorn's parametrizations. It also contains the  separate convergence of individual  parameters  of  the  Gaussian wavepacket in both parametrizations and a discussion of the difference between the results of the two parametrizations.
\end{abstract}
\maketitle

\renewcommand{\theequation}{S\arabic{equation}}
\renewcommand{\thesection}{S\arabic{section}}
\renewcommand{\thefigure}{S\arabic{figure}}

\section{Computational details}

Figure 1 of the main text describes the dynamics of a Gaussian wavepacket propagated in a one-dimensional coupled Morse potential [Eq. (50) of the main text] with various methods. The potential has zero coupling term ($d_{e}=0$), energy $V_{\textrm{eq}}=10$ at the equilibrium position $q_{\textrm{eq}}=1.5$, dissociation energy $d^{\prime}_{e}=11.25$, and anharmonicity $\chi^{\prime}=0.02$. The initial wavepacket was a real Gaussian with zero position and momentum and a width matrix $A_{0}=i\omega_{0}$ corresponding to the ground vibrational state of a harmonic oscillator with frequency $\omega_{0}=1$. The wavepacket was propagated for $10000$ steps of $\Delta t =0.004$ with the second-order symplectic integrator. The position grid for the exact quantum dynamics was chosen to have $512$ points between $-5$ and $25$. Panel (a) shows the wavepacket evolving with the  effective potential of the VGA at the step $400$ of the simulation. 

\section{Convergence and efficiency of the high-order symplectic integrators}
\label{sec:sym_integrators}
\begin{figure}[htbp]
	\includegraphics[width=0.60\textwidth]{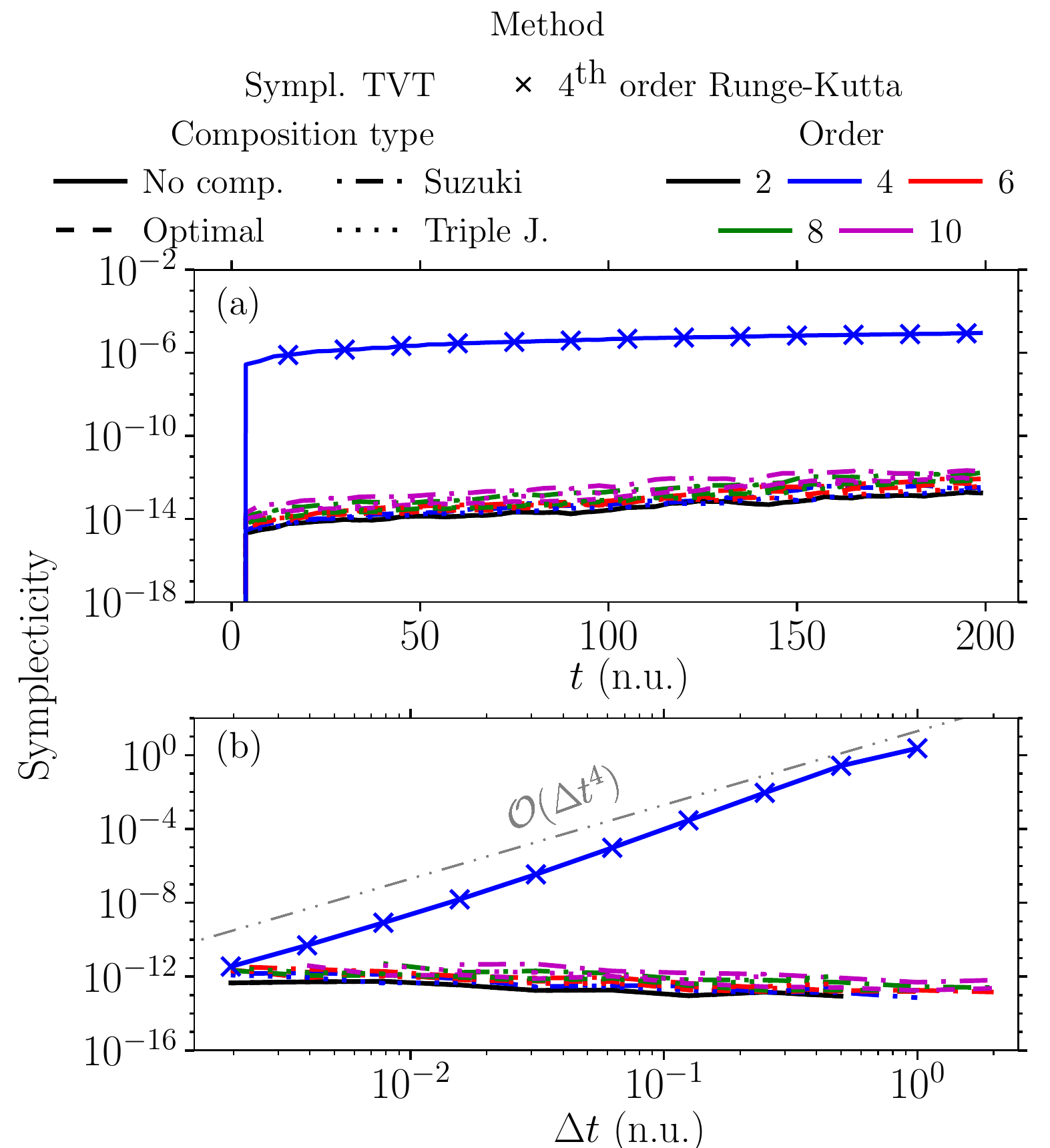} 
	\caption{Conservation of the symplectic structure of the Gaussian wavepackets by the symplectic integrators and its non-conservation by the fourth-order Runge-Kutta method. The system is the same two-dimensional coupled Morse potential as in Fig. 4 of the main text. The symplecticity $\mathcal{S}_{t}$, defined in Eq. (57) of the main text, is plotted (a) as a function of time $t$ for a fixed time step $\Delta t= 2^{-4}\, \textrm{n.u.}$ and (b) as a function of time step $\Delta t$ at the final time $t_{f}=200\, \textrm{n.u.}$}
	\label{Fig5}
\end{figure}
\begin{figure}[!htbp]
 	\includegraphics[width=1.01\textwidth]{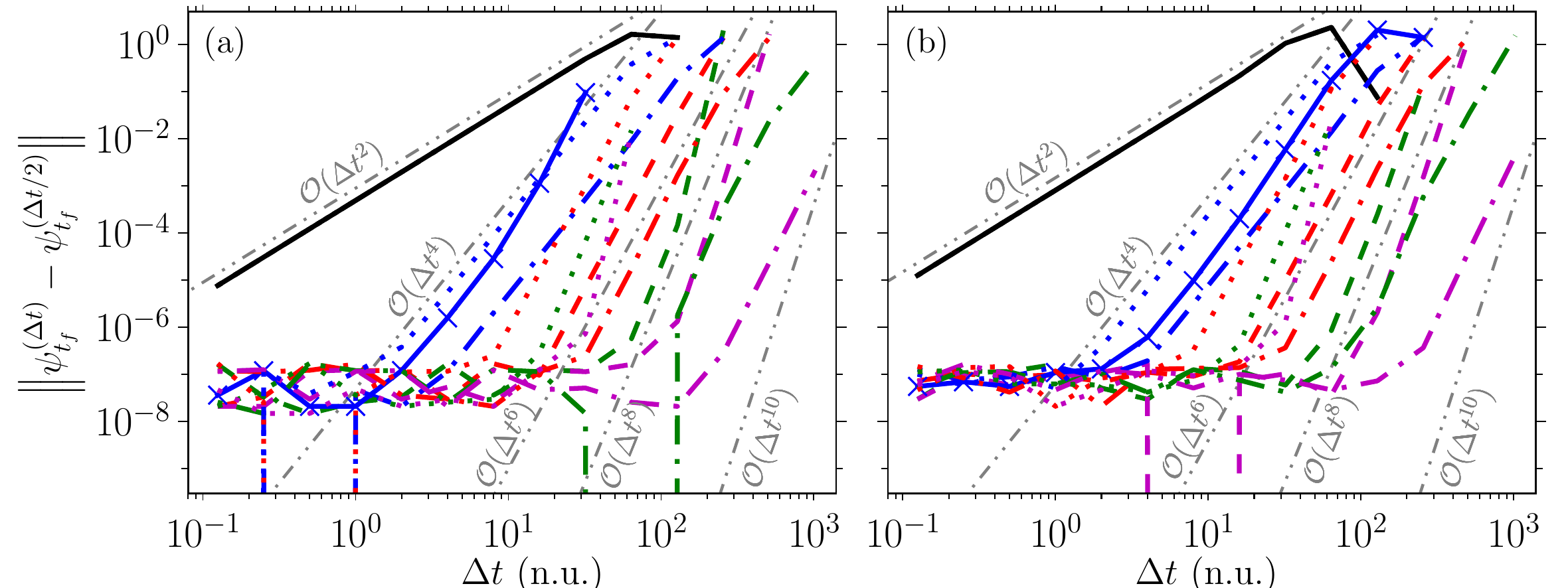}
	\caption{Convergence of the symplectic integrators and of the fourth-order Runge-Kutta method for the VGA, measured by the convergence error at the final time $t_{f} = 2^{16} \, \textrm{n.u.}=65536 \, \textrm{n.u.}$ as a function of the time step $\Delta t$.  The system is the same twenty-dimensional coupled Morse potential described in Sec.V.B.2 of the main text. The left and right panels show the values obtained using Heller's and Hagedorn's parametrizations, respectively. Line labels are the same as those in Fig. \ref{Fig5}.
	}
	\label{Fig7}
\end{figure}
\begin{figure}[!htbp]
	\includegraphics[width=1.0\textwidth]{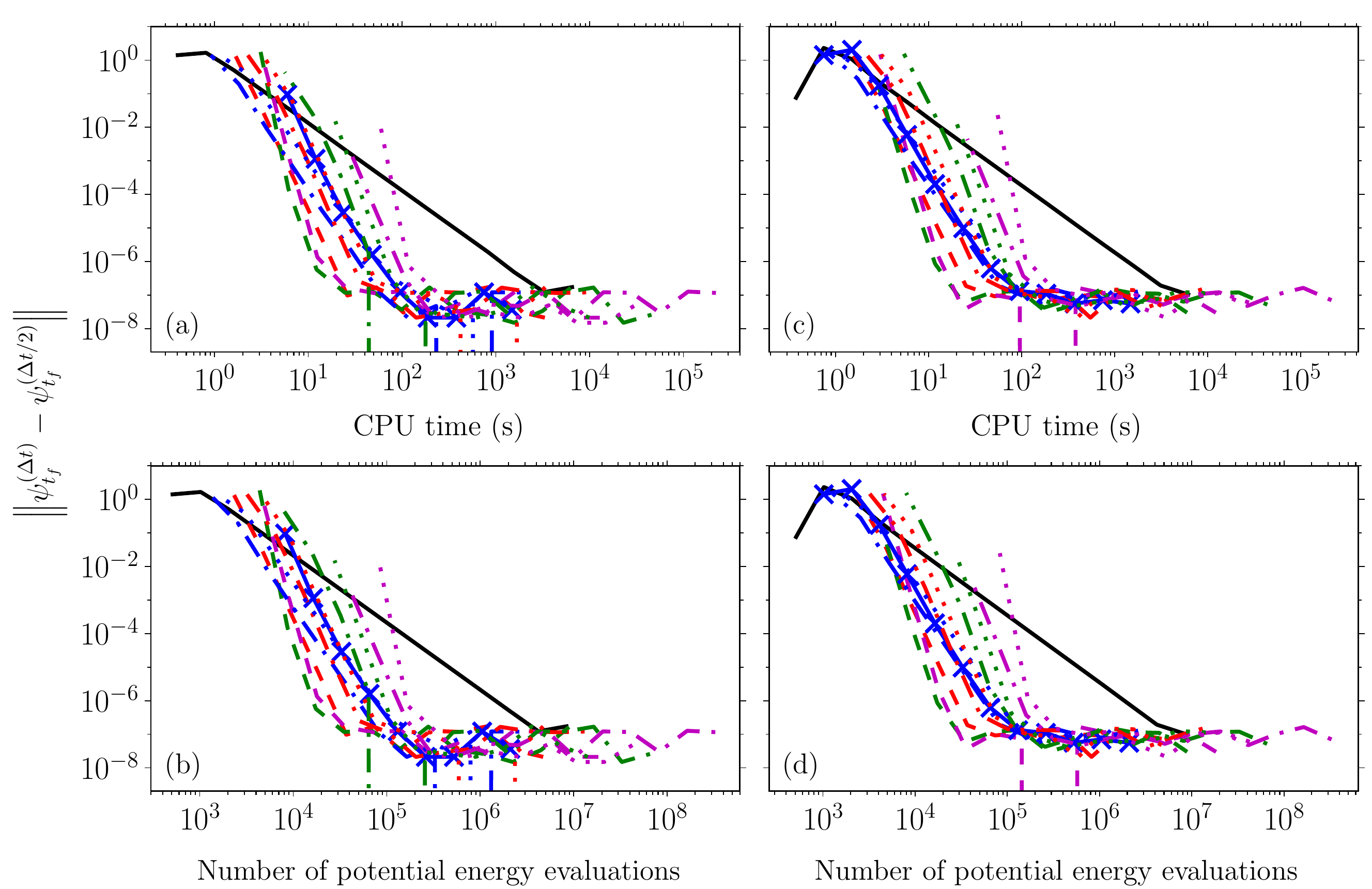} 
	\caption{Efficiency of the symplectic integrators and of the fourth-order Runge-Kutta method for the VGA in the twenty-dimensional coupled Morse potential described in the main text. Efficiency is measured by plotting the convergence error as a function of (a,c) the computational cost (CPU time) or (b,d) the number of potential energy evaluations. The left panels display the values obtained using Heller's parametrization, while the right panels show the results obtained using Hagedorn's parametrization. Line labels are the same as those in Fig. \ref{Fig5}.
	}
	\label{Fig8}
\end{figure}
\begin{figure}[ht]
	\includegraphics[width=0.98\textwidth]{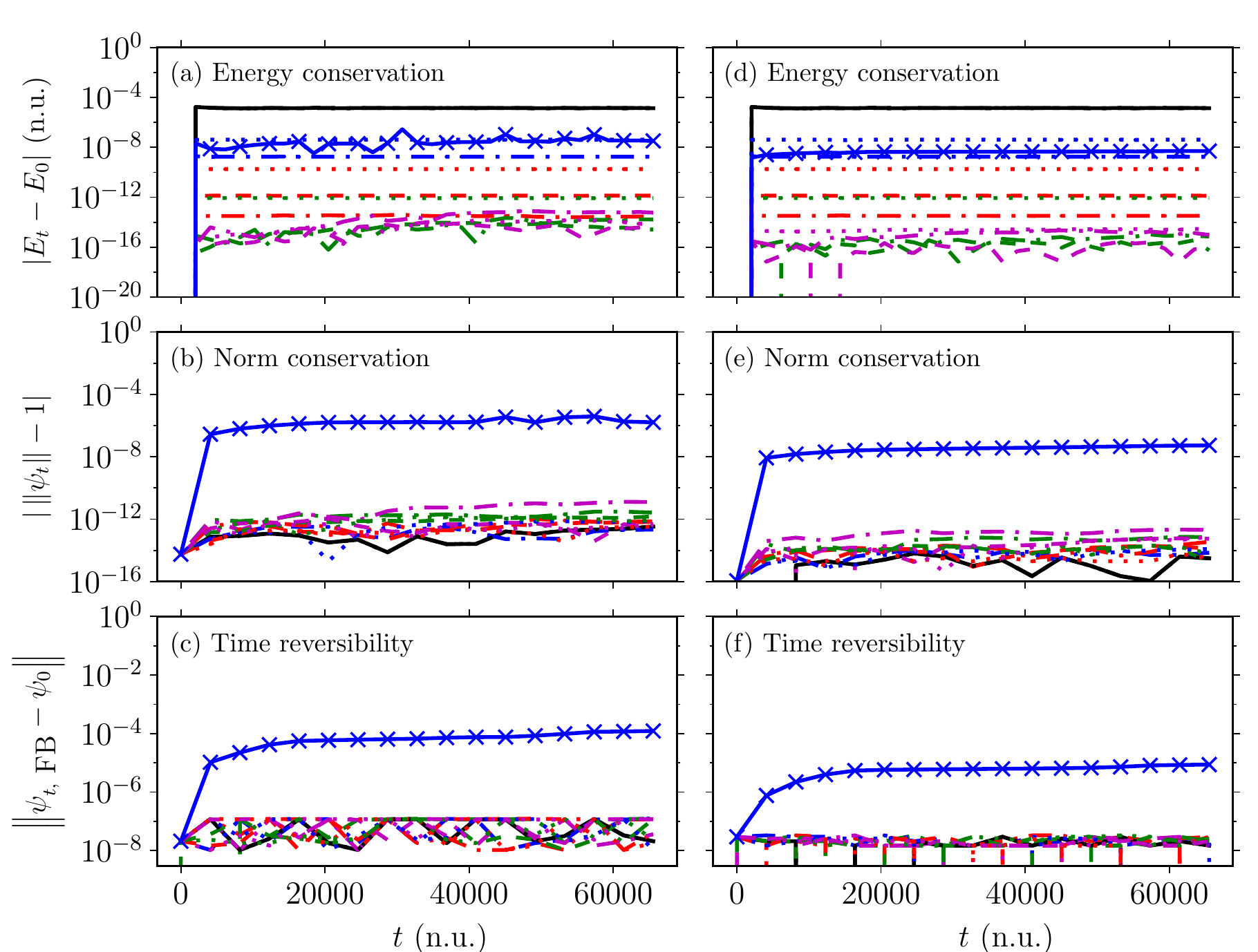} 
	\caption{Geometric properties of various integrators for the VGA as a function of time $t$ for a large time step $\Delta t = 8 \, \textrm{n.u.}$ (a,d) Energy, (b,e) norm, and (c,f) time reversibility [Eq. (58) of the main text] are shown. The left panels display the values obtained using Heller's parametrization, while the right panels display the values obtained using Hagedorn's parametrization. The system is the twenty-dimensional coupled Morse potential described in the main text. Line labels are the same as those in Fig. \ref{Fig5}.}
	\label{Fig9}
\end{figure}
\begin{figure}[!htbp]
	\hspace{-14mm}
	\includegraphics[width=0.95\textwidth]{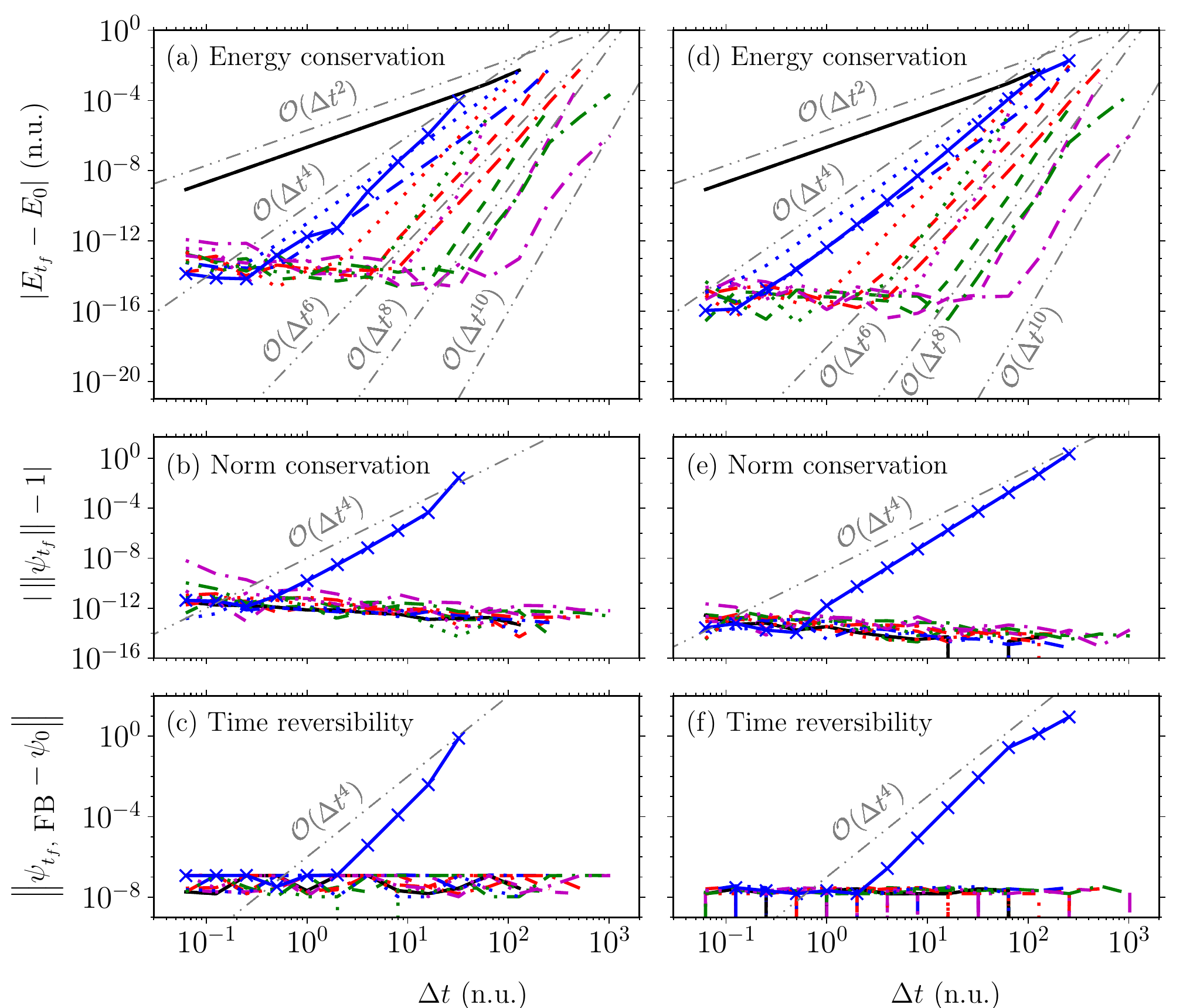} 
	\caption{Geometric properties of various integrators for the VGA as a function of the time step $\Delta t$ measured at the final time $t_{f} = 2^{16} \, \textrm{n.u.}=65536 \, \textrm{n.u.}$ (a,d) Energy, (b,e) norm, and (c,f) time reversibility [Eq. (58) of the main text] are shown. The left panels display the values obtained using Heller's parametrization, while the right panels show the values obtained using Hagedorn's parametrization. The system is the twenty-dimensional coupled Morse potential described in the main text. Line labels are the same as those in Fig. \ref{Fig5}.}
	\label{Fig10}
\end{figure}

In Figs.~\ref{Fig5}-\ref{Fig10}, we show the results presented in Fig. 5 and Figs. 7-10 of the main text for all the high-order symplectic integrators obtained using  the triple-jump, Suzuki-fractal, and optimal composition schemes. In addition, right panels of Figs.~\ref{Fig7}-\ref{Fig10} display the corresponding results obtained using Hagedorn's parametrization. Although the results of Heller's and Hagedorn's parametrizations are almost identical when using the symplectic integrators, they differ when employing the fourth-order Runge-Kutta method.~\cite{book_Leimkuhler_Reich:2004}  
This is due to the equations used to propagate the parameters $A_{t}$, $Q_{t}$ and $P_{t}$ at each time step. For the symplectic integrators, the equation evolving $A_{t}$ can be obtained exactly from those evolving $Q_{t}$ and $P_{t}$ since for the kinetic propagation substep, we have
\begin{align}
P_{t}\cdot Q^{-1}_{t} &= P_{0} \cdot(Q_{0}+t\, m^{-1} \cdot P_{0})^{-1}
\label{Eq:A_P_invQ_T1}\\
& = (Q_{0}\cdot P^{-1}_{0}+t\, m^{-1} )^{-1}
\label{Eq:A_P_invQ_T2}\\
&=(A^{-1}_{0}+t\, m^{-1} )^{-1}
\label{Eq:A_P_invQ_T3}\\
&=A_{t},
\label{Eq:A_P_invQ_T4}
\end{align}
where we used Eqs. (43) and (44) of the main text to write (\ref{Eq:A_P_invQ_T1}), the relation $A_{t}=P_{t}\cdot Q^{-1}_{t}$ to derive (\ref{Eq:A_P_invQ_T3}), and Eq. (34) of the main text to go from (\ref{Eq:A_P_invQ_T3}) to (\ref{Eq:A_P_invQ_T4}), and for the potential propagation substep, we have
\begin{align}
P_{t}\cdot Q^{-1}_{t} &= (P_{0}-t\,{\langle \hat{V}^{\prime\prime} \rangle} \cdot Q_{0})\cdot Q^{-1}_{0}
\label{Eq:A_P_invQ_V1} \\
&= P_{0}\cdot Q^{-1}_{0}-t\,{\langle \hat{V}^{\prime\prime} \rangle} 
\label{Eq:A_P_invQ_V2} \\
&=A_{0}-t\,{\langle \hat{V}^{\prime\prime} \rangle} 
\label{Eq:A_P_invQ_V3} \\
&=A_{t},
\label{Eq:A_P_invQ_V4}
\end{align}
where we used Eqs. (47) and (48) of the main text to write (\ref{Eq:A_P_invQ_V1}), the relation $A_{t}=P_{t}\cdot Q^{-1}_{t}$ to derive (\ref{Eq:A_P_invQ_V3}), and Eq. (38) of the main text to go from (\ref{Eq:A_P_invQ_V3}) to (\ref{Eq:A_P_invQ_V4}). 
In contrast, for the fourth-order Runge-Kutta approach,~\cite{book_Leimkuhler_Reich:2004} the equation propagating $A_{t}$ 
\begin{equation}
A_{t}=A_{0}-\big(A_{0}\cdot m^{-1}\cdot A_{0}+{\langle \hat{V}^{\prime\prime} \rangle} \big)\, \Delta t  
\end{equation}
cannot be obtained exactly from those propagating $Q_t$ and $P_t$  
\begin{align}
Q_{t}&= Q_{0} + m^{-1} \cdot P_0 \, \Delta t,\\
P_{t}&= P_{0} - {\langle \hat{V}^{\prime\prime} \rangle} \cdot Q_{0} \, \Delta t  
\end{align}
as
\begin{align}
P_{t}\cdot Q^{-1}_{t}&= \big(P_{0} - {\langle \hat{V}^{\prime\prime} \rangle} \cdot Q_{0} \, \Delta t\big) \cdot \big(Q_{0} + m^{-1} \cdot P_{0} \, \Delta t\big)^{-1}\\
&= \big(P_{0} - {\langle \hat{V}^{\prime\prime} \rangle} \cdot Q_{0} \, \Delta t\big) \cdot Q^{-1}_{0} \cdot \big(I + m^{-1} \cdot P_{0} \cdot Q^{-1}_{0} \, \Delta t\big)^{-1}\\
&= \big(P_{0} \cdot Q^{-1}_{0} - {\langle \hat{V}^{\prime\prime} \rangle} \, \Delta t\big)
 \cdot \big(I + m^{-1} \cdot P_{0} \cdot Q^{-1}_{0} \, \Delta t\big)^{-1}\\
&= \big(A_{0} - {\langle \hat{V}^{\prime\prime} \rangle} \, \Delta t\big)
 \cdot \big(1 + m^{-1} \cdot A_{0} \, \Delta t\big)^{-1}
\label{eq:step1}\\
&\overset{\Delta t \rightarrow 0}{\approx} (A_{0} - {\langle \hat{V}^{\prime\prime} \rangle} \, \Delta t\big)
\cdot \big(1 - m^{-1} \cdot A_{0} \, \Delta t\big)
\label{eq:step2}\\
&= A_{0} - A_{0} \cdot m^{-1} \cdot A_{0} \, \Delta t -{\langle \hat{V}^{\prime\prime} \rangle} \, \Delta t + \mathcal{O}(\Delta t^2)\\
&= A_{t} + \mathcal{O}(\Delta t^{2}).
\label{eq:AvsQP-RK4}
\end{align}
where $I$ is the identity matrix, and we used Taylor series of $\big(1 + m^{-1} \cdot A_{0} \, \Delta t\big)^{-1}$ going from Eq.~(\ref{eq:step1}) to Eq.~(\ref{eq:step2}). Thus, when evolving the Gaussian parameters by the fourth-order Runge-Kutta method, the results of Heller's and Hagedorn's parametrizations are nearly identical for small time steps, but as the time step increases, the values of the two parametrizations begin to diverge.

\begin{figure}[!htbp]
	\includegraphics[width=0.57\textwidth]{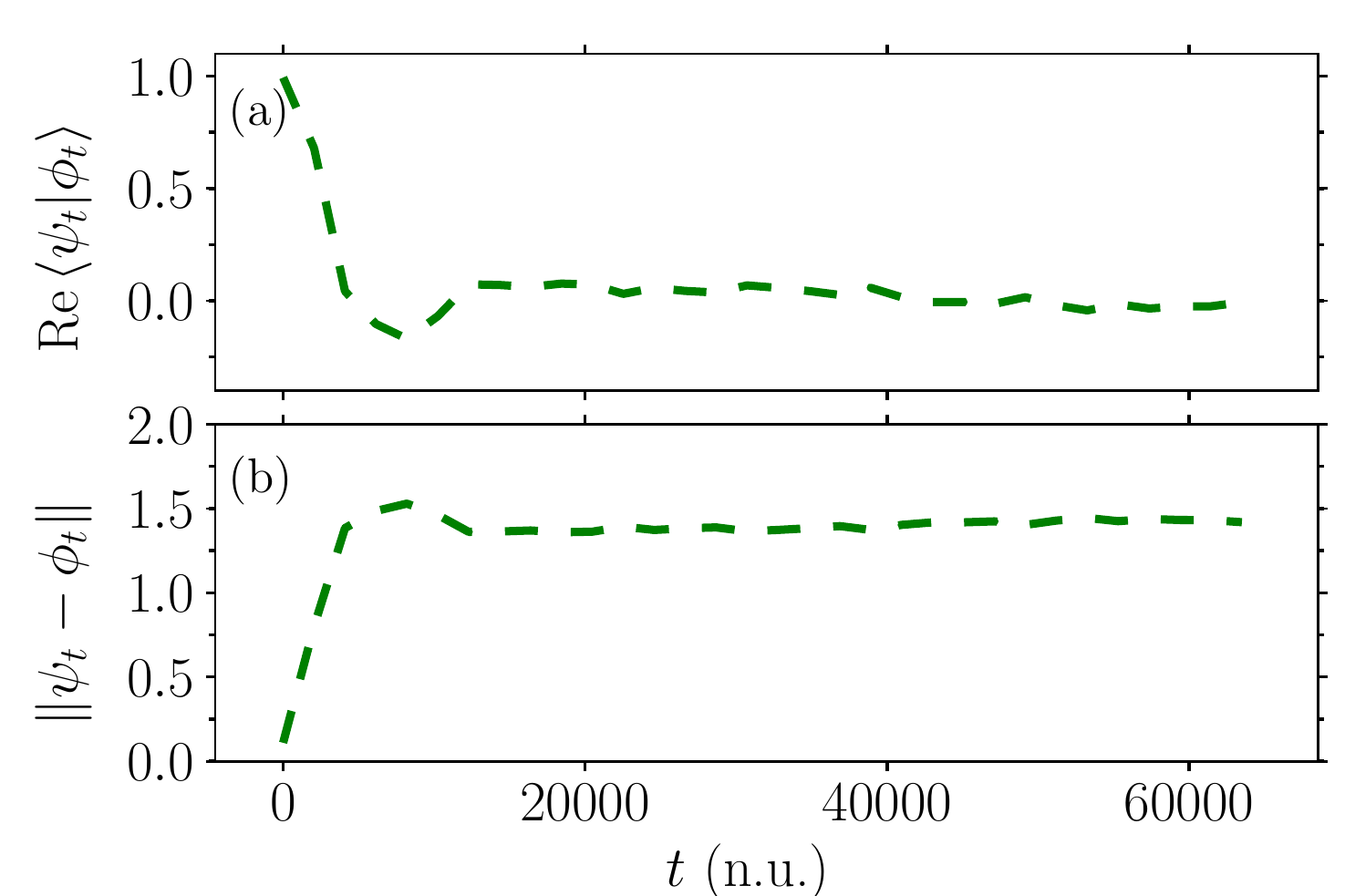} 
	\caption{Non-conservation of the (a) inner product [Eq. (B4) of the main text] and (b) distance between two states [Eq. (29) of the main text] by the VGA. The system is the twenty-dimensional coupled Morse potential [Eq. (50) of the main text], and state $\psi_{0}$ is the Gaussian wavepacket defined in Sec.V.B.2. State $\phi_{0}$ is $\psi_{0}$ displaced by $1\, \textrm{n.u}$ along all its twenty modes. The values are obtained using Hagedorn's parametrization.
	}
	\label{Fig11}
\end{figure}

\clearpage
\newpage
Figure~\ref{Fig11} represents the corresponding results of Fig. 11 of the main text obtained using Hagedorn's parametrization.

\begin{figure}[!htbp]
	\includegraphics[width=0.9\textwidth]{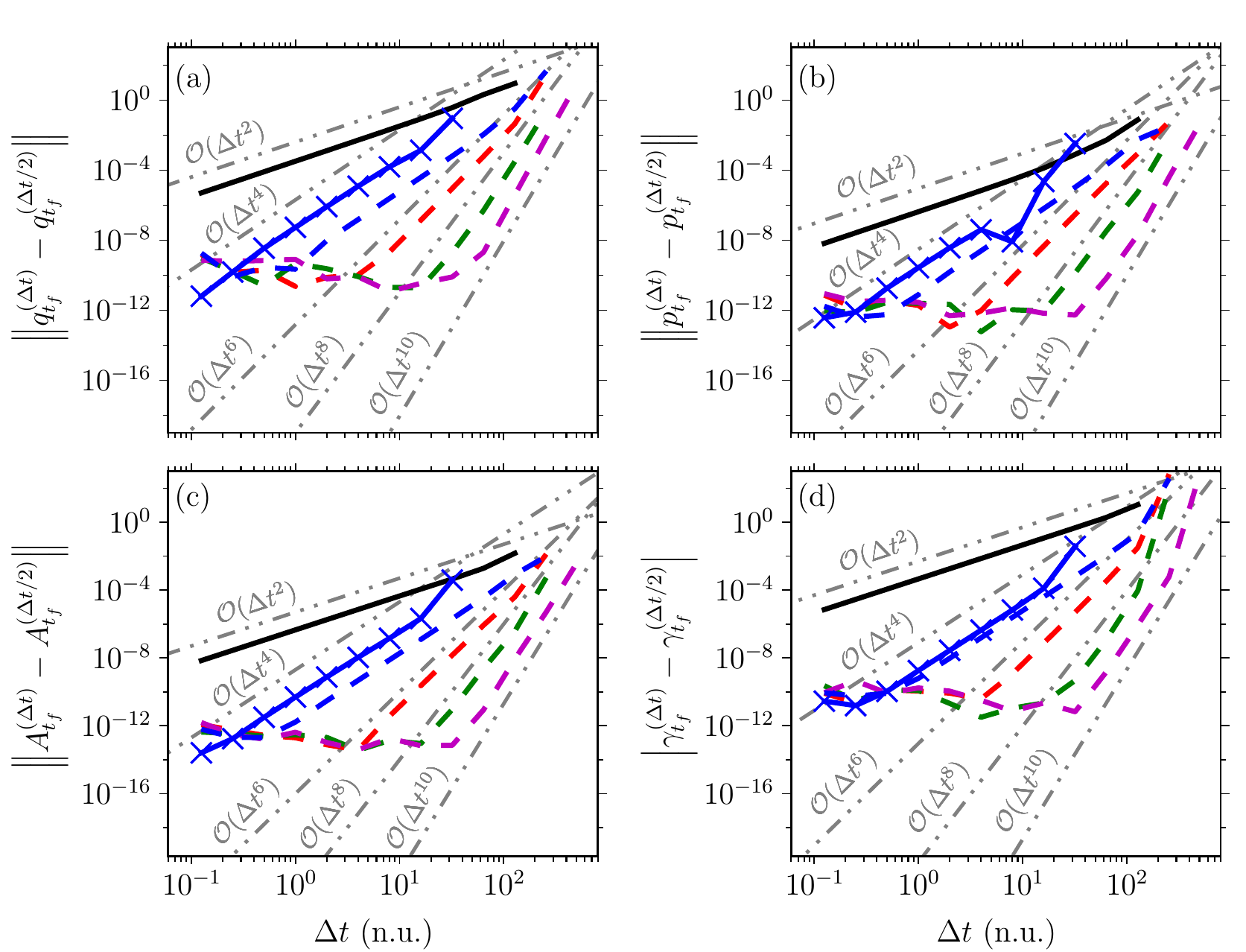} 
	\caption{Conservation of various integrators for the parameters (a) $q_{t}$, (b) $p_{t}$, (c) $A_{t}$ and (d) $\gamma_{t}$ of the Gaussian wavepacket [Eq. (4) of the main text], measured by the convergence error at the final time $t_{f} = 2^{16} \, \textrm{n.u.}=65536 \, \textrm{n.u.}$ as a function of the time step $\Delta t$. The system is the twenty-dimensional coupled Morse potential described in the main text. To avoid clutter, only the higher-order symplectic integrators obtained with the optimal composition are shown. Line labels are the same as those in Fig. \ref{Fig5}.
	}
	\label{Fig12}
\end{figure}

\begin{figure}[!htbp]
	\includegraphics[width=0.9\textwidth]{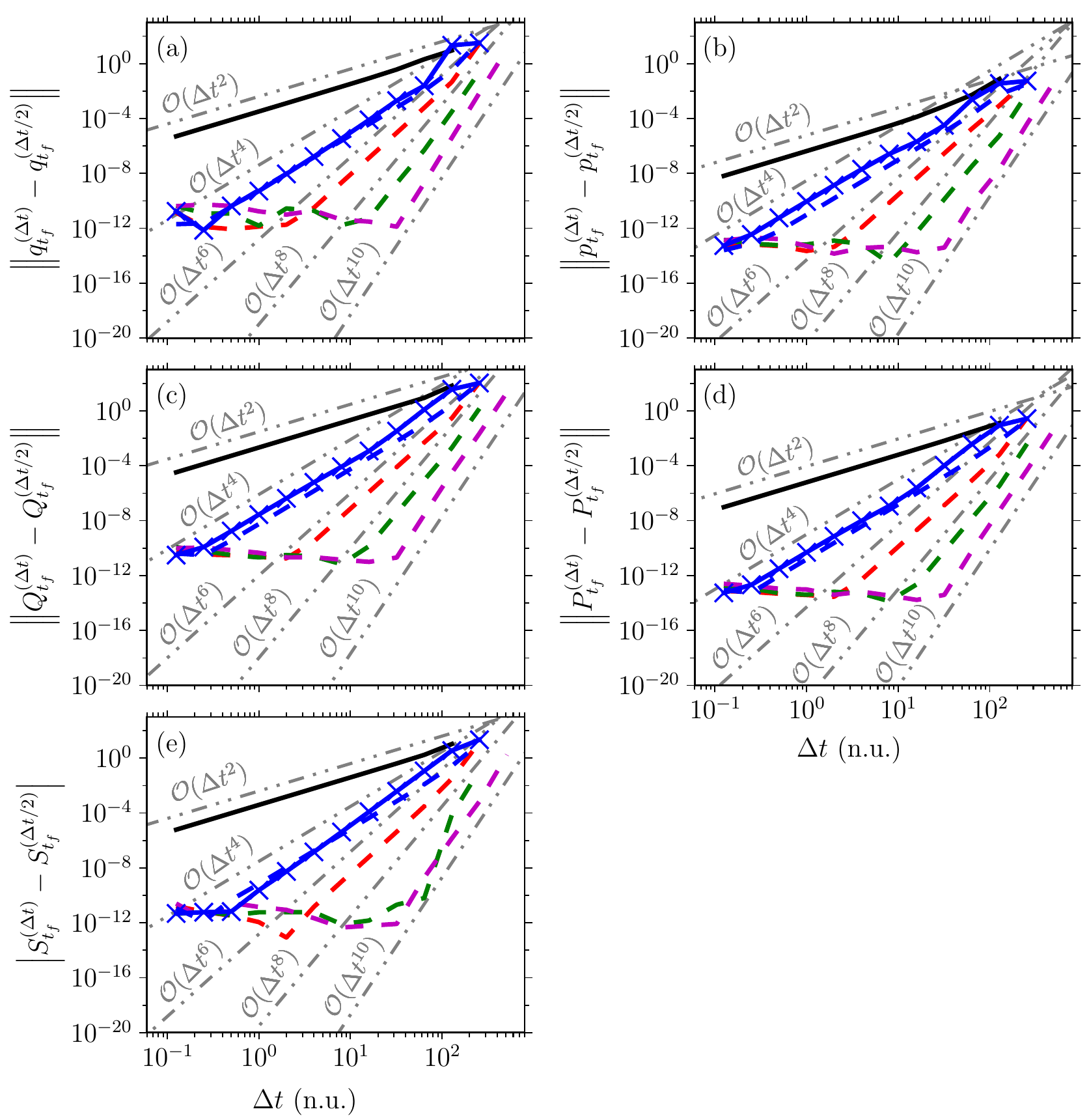} 
	\caption{Conservation of various integrators for the parameters (a) $q_{t}$, (b) $p_{t}$, (c) $Q_{t}$, (d) $P_{t}$ and (e) $S_{t}$ of the Gaussian wavepacket [Eq. (12) of the main text], measured by the convergence error at the final time $t_{f} = 2^{16} \, \textrm{n.u.}=65536 \, \textrm{n.u.}$ as a function of the time step $\Delta t$. The system is the twenty-dimensional coupled Morse potential described in the main text. To avoid clutter, only the higher-order symplectic integrators obtained with the optimal composition are shown. Line labels are the same as those in Fig. \ref{Fig5}.
	}
	\label{Fig13}
\end{figure}

Figures~\ref{Fig12} and~\ref{Fig13} show the convergence behaviour of the individual parameters of the Gaussian wavepacket in Heller's and Hagedorn's parametrizations. The convergence error of the Gaussian's parameter $\nu$, which can be the vector $q_{t}$ or $p_{t}$ or the matrix $A_{t}$ or $Q_{t}$ or $P_{t}$, is defined as the Frobenius distance 
\begin{align}
d(\nu^{(\Delta t)}_{t},\nu^{(\Delta t/2)}_{t}):= \lVert \nu^{(\Delta t)}_{t}-\nu^{(\Delta t/2)}_{t} \rVert,
\end{align}
where
\begin{align}
 \lVert A \rVert &:= \langle A, A \rangle^{1/2}, \\
 \langle A, B \rangle &:= \text{Tr}\big(A^{\dagger}B\big)
\end{align}
denote the Frobenius norm and the Frobenius inner product of matrices $A$ and $B$, and $\nu^{(\Delta t)}_{t}$ is the parameter at time $t$ obtained after propagation with the time step $\Delta t$.  The convergence error of the scalar parameters $\gamma_{t}$ and $S_{t}$ are measured by the absolute values $\lvert \gamma^{(\Delta t)}_{t}-\gamma^{(\Delta t/2)}_{t} \rvert$ and $\lvert S^{(\Delta t)}_{t}-S^{(\Delta t/2)}_{t} \rvert$.


\clearpage
\bibliographystyle{aipnum4-2}
\bibliography{biblio53,addition_VGA}